\documentclass[paper,notoc]{JHEP3}
\usepackage{epsfig,cite}
\usepackage{amsbsy} 
\usepackage{amsmath}
\usepackage{graphicx}
\usepackage{axodraw4j}

\def\beq{\begin{equation}}   
\def\eeq{\end{equation}}
\def\bea{\begin{eqnarray}}  
\def\eea{\end{eqnarray}} 

\def\eps{\epsilon}

\def\bsp#1\esp{\begin{split}#1\end{split}}
\def\n3lo{N$_3$LO}

\def\zbar{\bar{z}}

\newcommand{\mbint}{\int_{-i\infty}^{+i\infty}}
\newcommand{\cF}{{\mathcal{F}}}
\newcommand{\cR}{{\mathcal{R}}}
\newcommand{\cI}{{\mathcal{I}}}
\newcommand{\ord}{{\mathcal{O}}}

\title{Soft triple-real radiation for Higgs production at N3LO}

\author{Charalampos Anastasiou\\
  Institute for Theoretical Physics, ETH Zurich,
  8093 Zurich, Switzerland\\
  E-mail: \email{babis@phys.ethz.ch}}

\author{Claude Duhr\\
  Institute for Theoretical Physics, ETH Zurich,
  8093 Zurich, Switzerland\\
  Institute for Particle Physics Phenomenology, University of Durham,\\
  Durham, DH1 3LE, U.K.\\
  E-mail: \email{duhrc@itp.phys.ethz.ch}}

\author{Falko Dulat\\
  Institute for Theoretical Physics, ETH Zurich,
  8093 Zurich, Switzerland\\
  E-mail: \email{dulatf@itp.phys.ethz.ch}}

\author{Bernhard Mistlberger\\
  Institute for Theoretical Physics, ETH Zurich,
  8093 Zurich, Switzerland\\
  E-mail: \email{bmistlbe@itp.phys.ethz.ch}}

\abstract{
We  present the two first terms in the threshold expansion of Higgs 
production partonic cross-sections at hadron colliders for processes
with three partons in the final state. These are contributions to the
inclusive Higgs cross-section in gluon fusion at \n3lo.
We have developed a new technique  for the expansion of the squared
matrix-elements around 
the soft limit and for the reduction of the required phase-space
integrals to only ten single-scale master integrals.  We compute  
the master integrals building upon modern techniques for the
integration of multidimensional integrals in dimensional regularization.    
Our results constitute an important step towards a systematic
computation of the Higgs boson cross-section as an expansion 
around  the threshold limit.  
}


\preprint{IPPP/13/08\\DCPT/13/16}

\begin{document}

\catcode`\@=11
\font\manfnt=manfnt
\def\Watchout{\@ifnextchar [{\W@tchout}{\W@tchout[1]}}
\def\W@tchout[#1]{{\manfnt\@tempcnta#1\relax%
  \@whilenum\@tempcnta>\z@\do{%
    \char"7F\hskip 0.3em\advance\@tempcnta\m@ne}}}
\let\foo\W@tchout
\def\dubious{\@ifnextchar[{\@dubious}{\@dubious[1]}}
\let\enddubious\endlist
\def\@dubious[#1]{%
  \setbox\@tempboxa\hbox{\@W@tchout#1}
  \@tempdima\wd\@tempboxa
  \list{}{\leftmargin\@tempdima}\item[\hbox to 0pt{\hss\@W@tchout#1}]}
\def\@W@tchout#1{\W@tchout[#1]}
\catcode`\@=12

\def\YD#1{
\begin{picture}(80,60)(0,#1)
    \SetWidth{1.0}
\Text(4,8)[lb]{$2$}
\Text(4,48)[lb]{$1$}
\Text(67,8)[lb]{$1$}
\Text(67,48)[lb]{$2$}
\Line[arrow,arrowpos=0.5,arrowlength=5,arrowwidth=2,arrowinset=0.2,flip](20, 50)(10,50)
\Line[arrow,arrowpos=0.5,arrowlength=5,arrowwidth=2,arrowinset=0.2,flip](20,10)(10,10)
\Line[arrow,arrowpos=0.5,arrowlength=5,arrowwidth=2,arrowinset=0.2,flip](55,50)(65,50)
\Line[arrow,arrowpos=0.5,arrowlength=5,arrowwidth=2,arrowinset=0.2,flip](55,10)(65,10)
\Line(20,10)(55,10)
\Line(55,50)(55,10)
\Line(37.5,50)(37.5,10)
\Line(20,50)(20,10)
\Line(37.5, 50)(55, 50)
\Line[dash,dashsize=2](28.25,55)(28.25,5)
\Line[double,sep=2](20, 50)(37.5, 50)
\end{picture}
}


\def\FOnePic#1{
  \begin{picture}(0,0) (#1,-154)
    \SetWidth{1.0}
    \SetColor{Black}
    \Line[dash,dashsize=2](238,-133)(238,-169)
    \Line[arrow,arrowpos=0.5,arrowlength=5,arrowwidth=2,arrowinset=0.2,flip](250,-151)(262,-145)
    \Text(264,-147)[lb]{\small{\Black{$1$}}}
    \Line[arrow,arrowpos=0.5,arrowlength=5,arrowwidth=2,arrowinset=0.2,flip](250,-151)(262,-157)
    \Text(264,-160)[lb]{\small{\Black{$2$}}}
    \Arc[double,sep=2,clock](238,-151)(12,-180,-360)
    \Arc(238,-151)(12,-180,0)
    \Arc[clock](238,-160)(15,143.13,36.87)
    \Arc(238,-142)(15,-143.13,-36.87)
    \Line[arrow,arrowpos=0.5,arrowlength=5,arrowwidth=2,arrowinset=0.2](214,-145)(226,-151)
    \Text(209,-147)[lb]{\small{\Black{$1$}}}
    \Line[arrow,arrowpos=0.5,arrowlength=5,arrowwidth=2,arrowinset=0.2](214,-157)(226,-151)
    \Text(209,-160)[lb]{\small{\Black{$2$}}}
  \end{picture}
  }


\def\FTwoPic#1{
\begin{picture}(72,50) (#1,-26.5)
    \SetWidth{1.0}
    \SetColor{Black}
    \Line(79,-24)(55,-42)
    \Line[dash,dashsize=2](67,0)(67,-48)
    \Line(55,-6)(103,-24)
    \Line[double,sep=2](103,-24)(55,-42)
    \Line(79,-24)(103,-24)
    \Line(79,-24)(55,-6)
    \Line(55,-6)(55,-42)
    \Line[arrow,arrowpos=0.5,arrowlength=5,arrowwidth=2,arrowinset=0.2](43,0)(55,-6)
    \Text(38,-2)[lb]{\small{\Black{$1$}}}
    \Line[arrow,arrowpos=0.5,arrowlength=5,arrowwidth=2,arrowinset=0.2](43,-48)(55,-42)
    \Text(38,-50)[lb]{\small{\Black{$2$}}}
    \Line[arrow,arrowpos=0.5,arrowlength=5,arrowwidth=2,arrowinset=0.2,flip](103,-24)(115,-18)
    \Text(116,-20)[lb]{\small{\Black{$1$}}}
    \Line[arrow,arrowpos=0.5,arrowlength=5,arrowwidth=2,arrowinset=0.2,flip](103,-24)(115,-30)
    \Text(116,-33)[lb]{\small{\Black{$2$}}}
  \end{picture}
  }

  
  \def\FThreePic#1{
   \begin{picture}(72,50) (#1,-73)
     \SetWidth{1.0}
    \SetColor{Black}
    \Line(125,-52)(125,-88)
    \Line[dash,dashsize=2](137,-46)(137,-94)
    \Line(149,-70)(125,-52)
    \Line(149,-70)(173,-70)
    \Line(149,-70)(125,-88)
    \Line[arrow,arrowpos=0.5,arrowlength=5,arrowwidth=2,arrowinset=0.2,flip](173,-70)(185,-76)
    \Text(187,-80)[lb]{\small{\Black{$2$}}}
    \Line[arrow,arrowpos=0.5,arrowlength=5,arrowwidth=2,arrowinset=0.2,flip](173,-70)(185,-64)
    \Text(187,-66)[lb]{\small{\Black{$1$}}}
    \Line[arrow,arrowpos=0.5,arrowlength=5,arrowwidth=2,arrowinset=0.2](113,-94)(125,-88)
    \Text(108,-97)[lb]{\small{\Black{$2$}}}
    \Line[arrow,arrowpos=0.5,arrowlength=5,arrowwidth=2,arrowinset=0.2](113,-46)(125,-52)
    \Text(108,-48)[lb]{\small{\Black{$1$}}}
    \SetColor{White}
    \COval(143,-65)(4,4)(0){White}{White}
    \COval(143,-75)(3,4)(0){White}{White}
    \SetColor{Black}
    \Arc(149,-25)(51,-118.072,-61.928)
    \Arc[double,sep=2,clock](149,-115)(51,118.072,61.928)
  \end{picture}
  }

  
  \def\FFivePic#1{
   \begin{picture}(72,50) (#1,-182)
    \SetWidth{1.0}
    \SetColor{Black}
    \Line(263,-161)(263,-197)
    \Arc(266,-188)(-27.103,-144.006,-83.493)
        \Arc(284.929,-145.071)(27.103,-144.006,-83.493)
    \Line(311,-179)(287,-172)
    \Line(288,-172)(263,-198)
    \SetColor{White}
    \COval(281,-180)(4,4)(0){White}{White}
    \SetColor{Black}
    \Line[arrow,arrowpos=0.5,arrowlength=5,arrowwidth=2,arrowinset=0.2,flip](311,-179)(323,-185)
    \Text(325,-189)[lb]{\small{\Black{$2$}}}
    \Line[arrow,arrowpos=0.5,arrowlength=5,arrowwidth=2,arrowinset=0.2,flip](311,-179)(323,-173)
    \Text(325,-176)[lb]{\small{\Black{$1$}}}
    \Line[arrow,arrowpos=0.5,arrowlength=5,arrowwidth=2,arrowinset=0.2](251,-203)(263,-198)
    \Text(246,-207)[lb]{\small{\Black{$2$}}}
    \Line[arrow,arrowpos=0.5,arrowlength=5,arrowwidth=2,arrowinset=0.2](251,-155)(263,-161)
    \Text(246,-158)[lb]{\small{\Black{$1$}}}
    \Line[dash,dashsize=2](275,-155)(275,-203)
    \Line[double,sep=2](263,-179)(311,-179)
  \end{picture}
  }
  
  
  \def\FSixPic#1{
   \begin{picture}(72,50) (#1,-120)
    \SetWidth{1.0}
    \SetColor{Black}
    \Line(170,-97)(170,-133)
    \Line(218,-97)(170,-97)
    \Line(218,-133)(170,-133)
    \Line[dash,dashsize=2](182,-91)(182,-139)
    \Line(218,-97)(218,-133)
    \Line(206,-133)(170,-97)
    \SetColor{White}
    \COval(187,-113)(4,4)(0){White}{White}
    \SetColor{Black}
    \Line[double,sep=2](218,-97)(170,-121)
    \Line[arrow,arrowpos=0.5,arrowlength=5,arrowwidth=2,arrowinset=0.2,flip](218,-133)(230,-133)
    \Text(232,-136)[lb]{\small{\Black{$1$}}}
    \Line[arrow,arrowpos=0.5,arrowlength=5,arrowwidth=2,arrowinset=0.2,flip](218,-97)(230,-97)
    \Text(232,-100)[lb]{\small{\Black{$2$}}}
    \Line[arrow,arrowpos=0.5,arrowlength=5,arrowwidth=2,arrowinset=0.2](158,-97)(170,-97)
    \Text(153,-100)[lb]{\small{\Black{$1$}}}
    \Line[arrow,arrowpos=0.5,arrowlength=5,arrowwidth=2,arrowinset=0.2](158,-133)(170,-133)
    \Text(153,-136)[lb]{\small{\Black{$2$}}}
  \end{picture}
  }

  
  \def\FSevenPic#1{
    \begin{picture}(72,50) (#1,-133)
    \SetWidth{1.0}
    \SetColor{Black}
    \Line(218,-145)(170,-145)
    \Line[dash,dashsize=2](176,-103)(212,-151)
    \Line(194,-145)(194,-131)
    \Line(194,-122)(194,-109)
    \Line[arrow,arrowpos=0.5,arrowlength=5,arrowwidth=2,arrowinset=0.2](158,-109)(170,-109)
    \Text(153,-112)[lb]{\small{\Black{$1$}}}
    \Line[arrow,arrowpos=0.5,arrowlength=5,arrowwidth=2,arrowinset=0.2](158,-145)(170,-145)
    \Text(153,-148)[lb]{\small{\Black{$2$}}}
    \Line[arrow,arrowpos=0.5,arrowlength=5,arrowwidth=2,arrowinset=0.2,flip](218,-145)(230,-145)
    \Text(232,-148)[lb]{\small{\Black{$2$}}}
    \Line[arrow,arrowpos=0.5,arrowlength=5,arrowwidth=2,arrowinset=0.2,flip](218,-109)(230,-109)
    \Text(232,-112)[lb]{\small{\Black{$1$}}}
    \Line[double,sep=2](218,-109)(170,-145)
    \Line(218,-109)(170,-109)
    \Line(170,-109)(170,-145)
    \Line(218,-109)(218,-145)
  \end{picture}
  }

  \def\FEightPic#1{
    \begin{picture}(72,50) (#1,-119)
    \SetWidth{1.0}
    \SetColor{Black}
    \Line(277,-132)(229,-132)
    \Line(277,-96)(229,-96)
    \Line[double,sep=2](277,-96)(229,-132)
    \Line[dash,dashsize=2](271,-90)(235,-138)
    \Line(229,-96)(229,-132)
    \Line(277,-96)(277,-132)
    \Line(253,-132)(253,-118)
    \Line(253,-109)(253,-96)
    \Line[arrow,arrowpos=0.5,arrowlength=5,arrowwidth=2,arrowinset=0.2,flip](277,-132)(289,-132)
    \Text(292,-135)[lb]{\small{\Black{$2$}}}
    \Line[arrow,arrowpos=0.5,arrowlength=5,arrowwidth=2,arrowinset=0.2,flip](277,-96)(289,-96)
    \Text(292,-99)[lb]{\small{\Black{$1$}}}
    \Line[arrow,arrowpos=0.5,arrowlength=5,arrowwidth=2,arrowinset=0.2](217,-96)(229,-96)
    \Text(212,-99)[lb]{\small{\Black{$1$}}}
    \Line[arrow,arrowpos=0.5,arrowlength=5,arrowwidth=2,arrowinset=0.2](217,-132)(229,-132)
    \Text(212,-135)[lb]{\small{\Black{$2$}}}
  \end{picture}
  }

  
 \def\FNinePic#1{
   \begin{picture}(72,50) (#1,-195.5)
    \SetWidth{1.0}
    \SetColor{Black}
    \Line(240,-175)(240,-211)
    \Line(288,-193)(240,-175)
    \Line(266,-185)(240,-211)
    \SetColor{White}
    \COval(255,-198)(4,4)(0){White}{White}
    \SetColor{Black}
    \Line(262,-189)(240,-181)
    \Line[dash,dashsize=2](246,-169)(246,-217)
    \Line[arrow,arrowpos=0.5,arrowlength=5,arrowwidth=2,arrowinset=0.2](228,-169)(240,-175)
    \Text(223,-172)[lb]{\small{\Black{$1$}}}
    \Line[arrow,arrowpos=0.5,arrowlength=5,arrowwidth=2,arrowinset=0.2](228,-217)(240,-211)
    \Text(223,-221)[lb]{\small{\Black{$2$}}}
    \Line[double,sep=2](240,-199)(288,-193)
    \Line[arrow,arrowpos=0.5,arrowlength=5,arrowwidth=2,arrowinset=0.2,flip](288,-193)(300,-187)
    \Text(303,-190)[lb]{\small{\Black{$1$}}}
    \Line[arrow,arrowpos=0.5,arrowlength=5,arrowwidth=2,arrowinset=0.2,flip](288,-193)(300,-199)
    \Text(303,-203)[lb]{\small{\Black{$2$}}}
  \end{picture}
  }

  \def\XOnePic#1{
  \begin{picture}(0,0) (#1,-155)
    \SetWidth{1.0}
    \SetColor{Black}
    \Line(226,-151)(250,-151)
    \Arc[double,sep=2,clock](238,-151)(12,-180,-360)
    \Arc(238,-151)(12,-180,0)
    \Line[arrow,arrowpos=0.5,arrowlength=5,arrowwidth=2,arrowinset=0.2,flip](250,-151)(262,-145)
    \Line[arrow,arrowpos=0.5,arrowlength=5,arrowwidth=2,arrowinset=0.2,flip](250,-151)(262,-157)
    \Line[arrow,arrowpos=0.5,arrowlength=5,arrowwidth=2,arrowinset=0.2](214,-145)(226,-151)
    \Text(208,-148)[lb]{\small{\Black{$1$}}}
    \Line[arrow,arrowpos=0.5,arrowlength=5,arrowwidth=2,arrowinset=0.2](214,-157)(226,-151)
    \Text(208,-160)[lb]{\small{\Black{$2$}}}
    \Line[dash,dashsize=2](238,-133)(238,-169)
    \Text(265,-147)[lb]{\small{\Black{$1$}}}
    \Text(265,-160)[lb]{\small{\Black{$2$}}}
  \end{picture}
  }
  
    \def\XOnePicNoText#1{
  \begin{picture}(0,0) (#1,-155)
    \SetWidth{1.0}
    \SetColor{Black}
    \Line(226,-151)(250,-151)
    \Arc[double,sep=2,clock](238,-151)(12,-180,-360)
    \Arc(238,-151)(12,-180,0)
    \Line[arrow,arrowpos=0.5,arrowlength=5,arrowwidth=2,arrowinset=0.2,flip](250,-151)(262,-145)
    \Line[arrow,arrowpos=0.5,arrowlength=5,arrowwidth=2,arrowinset=0.2,flip](250,-151)(262,-157)
    \Line[arrow,arrowpos=0.5,arrowlength=5,arrowwidth=2,arrowinset=0.2](214,-145)(226,-151)
    \Line[arrow,arrowpos=0.5,arrowlength=5,arrowwidth=2,arrowinset=0.2](214,-157)(226,-151)
    \Line[dash,dashsize=2](238,-133)(238,-169)
  \end{picture}
  }

  
  \def\XEigthteenPic#1{
   \begin{picture}(20,50) (#1,-153)
        \SetWidth{1.0}
    \SetColor{Black}
    \Line(218,-126)(266,-150)
    \Line[arrow,arrowpos=0.5,arrowlength=5,arrowwidth=2,arrowinset=0.2](206,-119)(218,-126)
    \Text(200,-122)[lb]{\small{\Black{$1$}}}
    \Line[arrow,arrowpos=0.5,arrowlength=5,arrowwidth=2,arrowinset=0.2](206,-180)(218,-174)
    \Text(200,-183)[lb]{\small{\Black{$2$}}}
    \Line[arrow,arrowpos=0.5,arrowlength=5,arrowwidth=2,arrowinset=0.2,flip](266,-150)(278,-144)
    \Text(281,-146)[lb]{\small{\Black{$1$}}}
    \Line[arrow,arrowpos=0.5,arrowlength=5,arrowwidth=2,arrowinset=0.2,flip](266,-150)(278,-156)
    \Text(281,-160)[lb]{\small{\Black{$2$}}}
    \Line(242,-162)(218,-174)
    \Line[double,sep=2](242,-162)(266,-150)
    \Line(242,-138)(218,-174)
    \SetColor{White}
    \COval(233,-151)(5,2)(0){White}{White}
    \SetColor{Black}
    \Line(242,-162)(218,-126)
    \Line[dash,dashsize=2](220,-120)(256,-174)
    \SetColor{White}
    \COval(254,-172)(4,16)(0){White}{White}
  \end{picture}
  }

  \def\XOneNLOPic#1{
   \begin{picture}(0,0) (#1,-22)
    \SetWidth{1.0}
    \SetColor{Black}
    \Arc[dash,dashsize=1,clock](200,-18)(12,-0,-180)
    \SetColor{White}
    \Vertex(191,-27){10}
    \SetColor{Black}
    \Arc[clock](190.167,-27.833)(10.069,102.426,-12.426)
    \Arc(197.833,-20.167)(10.069,167.574,282.426)
    \Line[arrow,arrowpos=0.5,arrowlength=5,arrowwidth=2,arrowinset=0.2,flip](212,-18)(224,-12)
    \Line[arrow,arrowpos=0.5,arrowlength=5,arrowwidth=2,arrowinset=0.2,flip](212,-18)(224,-24)
    \Line[dash,dashsize=2](210,1)(186,-35)
    \Arc[double,sep=2,clock](200,-18)(12,-180,-360)
    \Vertex(200,-6){2.6}
    \Line[arrow,arrowpos=0.5,arrowlength=5,arrowwidth=2,arrowinset=0.2](176,-24)(188,-18)
    \Line[arrow,arrowpos=0.5,arrowlength=5,arrowwidth=2,arrowinset=0.2](176,-12)(188,-18)
  \end{picture}
  }
  
  \def\XOneNNLOPic#1{
    \begin{picture}(0,0) (#1,-22)
    \SetWidth{1.0}
    \SetColor{Black}
    \Arc[dash,dashsize=1,clock](200,-18)(12,-0,-180)
    \Vertex(209,-10){2.6}
    \Vertex(208,-27){2.6}
    \SetColor{White}
    \Vertex(191,-27){10}
    \SetColor{Black}
    \Arc[clock](190.167,-27.833)(10.069,102.426,-12.426)
    \Arc(197.833,-20.167)(10.069,167.574,282.426)
    \Line[arrow,arrowpos=0.5,arrowlength=5,arrowwidth=2,arrowinset=0.2,flip](212,-18)(224,-12)
    \Line[arrow,arrowpos=0.5,arrowlength=5,arrowwidth=2,arrowinset=0.2,flip](212,-18)(224,-24)
    \Line[dash,dashsize=2](210,1)(186,-35)
    \Arc[double,sep=2,clock](200,-18)(12,-180,-360)
    \Vertex(200,-6){2.6}
    \Line[arrow,arrowpos=0.5,arrowlength=5,arrowwidth=2,arrowinset=0.2](176,-24)(188,-18)
    \Line[arrow,arrowpos=0.5,arrowlength=5,arrowwidth=2,arrowinset=0.2](176,-12)(188,-18)
  \end{picture}
  }
  
\def\XEightteenLOPic#1{
 \begin{picture}(0,0) (#1,-162)
   \SetWidth{1.0}
    \SetColor{Black}
    \Line(198,-140)(234,-158)
        \Line(216,-149)(198,-176)
    \SetColor{White}
    \COval(210,-158)(3,3)(0){White}{White}
    \SetColor{Black}
    \Line[arrow,arrowpos=0.5,arrowlength=5,arrowwidth=2,arrowinset=0.2,flip](234,-158)(243,-163)
    \Line[arrow,arrowpos=0.5,arrowlength=5,arrowwidth=2,arrowinset=0.2,flip](234,-158)(243,-154)
    \Line[arrow,arrowpos=0.5,arrowlength=5,arrowwidth=2,arrowinset=0.2](189,-134)(198,-140)
    \Line[arrow,arrowpos=0.5,arrowlength=5,arrowwidth=2,arrowinset=0.2](189,-180)(198,-176)
    \Line(216,-167)(198,-176)
    \Line[double,sep=2](216,-167)(234,-158)
    \Line(216,-167)(198,-140)
    \Line[dash,dashsize=2](199,-135)(226,-173)
  \end{picture}
  }

 \def\XEightteenNLOPic#1{
 \begin{picture}(0,0) (#1,-160)
    \SetWidth{1.0}
    \SetColor{Black}
    \Line[double,sep=2](326,-155)(362,-155)
    \Line(326,-137)(362,-155)
    \Line[arrow,arrowpos=0.5,arrowlength=5,arrowwidth=2,arrowinset=0.2,flip](362,-155)(371,-160)
    \Line[arrow,arrowpos=0.5,arrowlength=5,arrowwidth=2,arrowinset=0.2,flip](362,-155)(371,-151)
    \Line[arrow,arrowpos=0.5,arrowlength=5,arrowwidth=2,arrowinset=0.2](317,-133)(326,-137)
    \Line[arrow,arrowpos=0.5,arrowlength=5,arrowwidth=2,arrowinset=0.2](317,-178)(326,-173)
    \Line(362,-155)(326,-173)
    \Line(326,-137)(326,-173)
    \Line[dash,dashsize=2](334,-131)(334,-182)
    \Vertex(343,-155){2.6}  \end{picture}
}

 \def\XEightteenNNLOPic#1{
 \begin{picture}(0,0) (#1,-275)
    \SetWidth{1.0}
        \Line(216,-261)(198,-288)
    \SetColor{Black}
    \Line[dash,dashsize=1](216,-261)(234,-270)
    \Line(198,-252)(216,-261)
    \Vertex(221,-276.5){2.6}
    \Vertex(228,-273){2.6}
    \SetColor{White}
    \COval(210,-270)(3,3)(0){White}{White}
    \SetColor{Black}
    \Line[arrow,arrowpos=0.5,arrowlength=5,arrowwidth=2,arrowinset=0.2,flip](234,-270)(243,-275)
    \Line[arrow,arrowpos=0.5,arrowlength=5,arrowwidth=2,arrowinset=0.2,flip](234,-270)(243,-266)
    \Line[arrow,arrowpos=0.5,arrowlength=5,arrowwidth=2,arrowinset=0.2](189,-246)(198,-252)
    \Line[arrow,arrowpos=0.5,arrowlength=5,arrowwidth=2,arrowinset=0.2](189,-292)(198,-288)
    \Line[double,sep=2](216,-279)(234,-270)
    \Line(216,-279)(198,-252)
    \Line[dash,dashsize=2](201,-246)(228,-284)
    \Line(216,-279)(198,-288)  \end{picture}
}

\def\FOneLOPic#1{
  \begin{picture}(0,0) (#1,-154)
    \SetWidth{1.0}
    \SetColor{Black}
    \Line[dash,dashsize=2](238,-133)(238,-169)
    \Line[arrow,arrowpos=0.5,arrowlength=5,arrowwidth=2,arrowinset=0.2,flip](250,-151)(262,-145)
    \Line[arrow,arrowpos=0.5,arrowlength=5,arrowwidth=2,arrowinset=0.2,flip](250,-151)(262,-157)
    \Arc[double,sep=2,clock](238,-151)(12,-180,-360)
    \Arc(238,-151)(12,-180,0)
    \Arc[clock](238,-160)(15,143.13,36.87)
    \Arc(238,-142)(15,-143.13,-36.87)
    \Line[arrow,arrowpos=0.5,arrowlength=5,arrowwidth=2,arrowinset=0.2](214,-145)(226,-151)
    \Line[arrow,arrowpos=0.5,arrowlength=5,arrowwidth=2,arrowinset=0.2](214,-157)(226,-151)
  \end{picture}
  }
  
  \def\FOneNLOPic#1{
   \begin{picture}(0,0) (#1,-22)
    \SetWidth{1.0}
    \SetColor{Black}
    \Arc[dash,dashsize=1,clock](200,-18)(12,-0,-180)
    \SetColor{White}
    \Vertex(191,-27){10}
    \SetColor{Black}
    \Arc[clock](190.167,-27.833)(10.069,102.426,-12.426)
    \Arc(197.833,-20.167)(10.069,167.574,282.426)
    \Line[arrow,arrowpos=0.5,arrowlength=5,arrowwidth=2,arrowinset=0.2,flip](212,-18)(224,-12)
    \Line[arrow,arrowpos=0.5,arrowlength=5,arrowwidth=2,arrowinset=0.2,flip](212,-18)(224,-24)
    \Line[dash,dashsize=2](210,1)(186,-35)
    \Arc[double,sep=2,clock](200,-18)(12,-180,-360)
    \Vertex(200,-6){2.6}
    \Line[arrow,arrowpos=0.5,arrowlength=5,arrowwidth=2,arrowinset=0.2](176,-24)(188,-18)
    \Line[arrow,arrowpos=0.5,arrowlength=5,arrowwidth=2,arrowinset=0.2](176,-12)(188,-18)
\Line(188,-18)(200,-30)
  \end{picture}
  }
  
  \def\FOneNNLOPic#1{
    \begin{picture}(0,0) (#1,-22)
    \SetWidth{1.0}
    \SetColor{Black}
    \Arc[dash,dashsize=1,clock](200,-18)(12,-0,-180)
    \Vertex(209,-10){2.6}
    \Vertex(208,-27){2.6}
    \SetColor{White}
    \Vertex(191,-27){10}
    \SetColor{Black}
    \Arc[clock](190.167,-27.833)(10.069,102.426,-12.426)
    \Arc(197.833,-20.167)(10.069,167.574,282.426)
    \Line[arrow,arrowpos=0.5,arrowlength=5,arrowwidth=2,arrowinset=0.2,flip](212,-18)(224,-12)
    \Line[arrow,arrowpos=0.5,arrowlength=5,arrowwidth=2,arrowinset=0.2,flip](212,-18)(224,-24)
    \Line[dash,dashsize=2](210,1)(186,-35)
    \Arc[double,sep=2,clock](200,-18)(12,-180,-360)
    \Vertex(200,-6){2.6}
    \Line[arrow,arrowpos=0.5,arrowlength=5,arrowwidth=2,arrowinset=0.2](176,-24)(188,-18)
    \Line[arrow,arrowpos=0.5,arrowlength=5,arrowwidth=2,arrowinset=0.2](176,-12)(188,-18)
\Line(188,-18)(200,-30)
  \end{picture}
  }
  
  \def\FTenPic#1{
  \begin{picture}(72,50) (#1,-146)
   \SetWidth{1.0}
    \SetColor{Black}
       \Line(216,-161)(168,-161)
    \Line(216,-125)(176,-161)
    \Line[double,sep=2](216,-125)(168,-125)
    \Line(216,-125)(216,-161)
    \Line[dash,dashsize=2](183,-117)(183,-171)
    \Line[arrow,arrowpos=0.5,arrowlength=5,arrowwidth=2,arrowinset=0.2](156,-125)(168,-125)
    \Text(150,-128)[lb]{\small{\Black{$1$}}}
    \Line[arrow,arrowpos=0.5,arrowlength=5,arrowwidth=2,arrowinset=0.2,flip](216,-161)(228,-161)
    \Text(231,-164)[lb]{\small{\Black{$2$}}}
    \Line[arrow,arrowpos=0.5,arrowlength=5,arrowwidth=2,arrowinset=0.2,flip](216,-125)(228,-125)
    \Text(231,-128)[lb]{\small{\Black{$1$}}}
    \SetColor{White}
    \Vertex(191,-147){4.472}
    \SetColor{Black}
    \Line(206,-161)(167,-125)
    \Line(168,-125)(168,-161)
    \Line[arrow,arrowpos=0.5,arrowlength=5,arrowwidth=2,arrowinset=0.2](156,-161)(168,-161)
    \Text(150,-164)[lb]{\small{\Black{$2$}}}
  \end{picture}
}

\def\FElevenPic#1{
  \begin{picture}(72,50) (#1,-155)
    \SetWidth{1.0}
    \SetColor{Black}
    \Line(203,-158)(177,-158)
    \Line[double,sep=2](215,-173)(167,-173)
    \Line(215,-137)(177,-158)
    \Line(215,-137)(215,-173)
    \Line[arrow,arrowpos=0.5,arrowlength=5,arrowwidth=2,arrowinset=0.2](155,-137)(167,-137)
    \Text(150,-140)[lb]{\small{\Black{$1$}}}
    \Line[arrow,arrowpos=0.5,arrowlength=5,arrowwidth=2,arrowinset=0.2,flip](215,-173)(227,-173)
    \Text(229,-176)[lb]{\small{\Black{$2$}}}
    \Line[arrow,arrowpos=0.5,arrowlength=5,arrowwidth=2,arrowinset=0.2,flip](215,-137)(227,-137)
    \Text(229,-140)[lb]{\small{\Black{$1$}}}
    \SetColor{White}
    \Vertex(190,-151){4.472}
    \SetColor{Black}
    \Line(203,-158)(167,-137)
    \Line(167,-137)(167,-173)
    \Line[arrow,arrowpos=0.5,arrowlength=5,arrowwidth=2,arrowinset=0.2](155,-173)(167,-173)
    \Text(150,-176)[lb]{\small{\Black{$2$}}}
    \Line(215,-173)(203,-158)
    \Line(177,-158)(167,-173)
    \Line[dash,dashsize=2](190,-128)(190,-182)
  \end{picture}
}

\section{Introduction}
\label{sec:introduction}

The first years of experiments at the Large Hadron Collider (LHC) have
probed decisively the physics of electroweak symmetry breaking,
demonstrating a great understanding of the theory at this energy scale. 
The success of particle theory must be attributed not only to the Standard 
Model (SM), but also to remarkable progress in perturbative calculations.  
Results in this field of research are outstanding and have
resulted in very accurate comparisons with data. 
Indeed, partonic cross-sections for a large variety of processes are now routinely computed
at next-to-leading order (NLO) in the strong coupling constant
$\alpha_s$~\footnote{See, for example, refs.~\cite{Bern:2011ep,Ita:2011wn,Berger:2010zx,Berger:2009ep,Berger:2009zg,KeithEllis:2009bu,Melia:2011dw,Melnikov:2009wh,Hirschi:2011pa,Bevilacqua:2011aa,Bevilacqua:2011xh,Bevilacqua:2009zn,Bevilacqua:2010ve,Frederix:2011ss}}.
In addition, modern Monte-Carlo programs have reached a high level of
sophistication with the development of methods to match and merge parton showers with fixed-order
calculations\footnote{See, for example, refs.~\cite{Catani:2001cc,Lonnblad:2001iq,Krauss:2002up,Frixione:2002ik,Nason:2004rx,Lavesson:2005xu,Mangano:2006rw,Alwall:2007fs,Frixione:2007vw,Alwall:2008qv,Hoeche:2009rj,Hamilton:2009ne,Hamilton:2010wh,Hoche:2010pf,Hoche:2010kg,Lonnblad:2011xx,Hoeche:2011fd,Hamilton:2012np,Hoeche:2012yf,Gehrmann:2012yg,Frederix:2012ps}}. 
Finally, basic processes at the LHC with a small final-state multiplicity can
now be computed fully differentially at next-to-next-to-leading order
(NNLO)\footnote{See, for example, refs.~\cite{Buehler:2012cu,Ridder:2013mf,Czakon:2012pz,Czakon:2012zr,Baernreuther:2012ws,Catani:2011qz,Catani:2009sm,Ferrera:2011bk,Grazzini:2008tf,Melnikov:2006kv,Anastasiou:2007mz,Anastasiou:2005qj,Anastasiou:2004xq}}.

A classic example where perturbative calculations have been crucial for the interpretation 
of LHC data is the production of a Higgs boson. Perturbative corrections 
for Higgs boson signal processes are generically large, especially so for production
via gluon fusion \cite{Graudenz:1992pv,Dawson:1990zj,Djouadi:1991tka,Spira:1995rr,Ravindran:2003um,Harlander:2002wh,Anastasiou:2002yz}.  For this process, the total and fully differential 
cross-sections are known through NNLO with a  typical uncertainty of about $10\%$ due to variations of the  factorization 
and renormalization scales \cite{Anastasiou:2007mz,Anastasiou:2011pi}. Indeed, without the knowledge of the NNLO corrections, predictions at LO or
NLO would be assigned
theory uncertainties which are larger than the already achieved
experimental uncertainties. With future LHC data,  a comparison with
theory at a level of precision of a few percent will be revealing the
fine details of the mechanism of electroweak symmetry breaking and 
potentially uncovering new fundamental laws of physics. 

Besides being of importance to Higgs physics,
NNLO predictions have promoted the Drell-Yan  process to a standard
candle for studies at hadron colliders, where 
observables in this process are computed with a typical precision of
better than $2\%$ as indicated by scale variations~\cite{Hamberg:1990np,Anastasiou:2003yy}. 
Measurements of the mass of the $W$ boson and the weak mixing angle
from Drell-Yan data are one of the most stringent constraints on the
Standard Model. Furthermore, Drell-Yan production data is an essential
input for the extraction of parton distribution functions from hadron collider processes. 
Last but not least, the clean detector signatures of
Drell-Yan events, combined with the excellent theoretical predictions of their rates, allow
for a precise determination of the luminosity. Indeed, luminosity
determination from Drell-Yan is relatively insensitive to high pile-up conditions, a fact which will be even more
important during the high-luminosity phase of the LHC.

We believe that it is possible to develop efficient methods for
computing Higgs boson and Drell-Yan hadroproduction cross-sections 
at the next perturbative order, ``\n3lo''. The benefits of such a
program are potentially very important. 

In Higgs production via gluon fusion the
uncertainty  due to scale variations at \n3lo is anticipated to be half
of what is found at NNLO~\cite{Moch:2005ky}. Scale variation estimates cannot be derived from first principles, but entail a certain degree of subjectiveness. With the knowledge of
more orders in the perturbative expansion, the remaining perturbative uncertainty can be estimated more reliably,
not only from scale variations but also from the progression of the
series. Improvements of the Higgs boson production cross-section uncertainty will
propagate into the determination of the couplings of the Higgs boson.   

Although existing NNLO calculations for 
the Drell-Yan process indicate that the cross-section is already known very precisely, an explicit \n3lo computation will educate us further by challenging or verifying the traditional prescriptions for uncertainty estimates. Furthermore, high-luminosity runs at the LHC require the triggering of leptons
at a higher transverse momentum, 
resulting in a deterioration of the NNLO scale uncertainty~\cite{Melnikov:2006kv}.  
An excellent theoretical precision can then be
recovered by including the \n3lo perturbative corrections.  

So far, there has been no \n3lo calculation for hadron collider processes performed in the literature.
In this paper, we attempt a first step towards computing inclusive
hadroproduction cross-sections at this perturbative order. We focus on
one of the most complicated \n3lo contributions to the inclusive Higgs boson cross-section in the gluon fusion 
channel, namely, the contributions coming from partonic cross-sections for
tree-level radiative processes with three additional partons emitted
in the final state.  A direct integration of the corresponding
matrix-elements over phase space is however tedious. To simplify the problem, we
perform a threshold expansion and devise a method to compute the
coefficients of the expansion analytically.  One of the main results of our paper
is the computation of
the first two terms of the threshold expansion for triple real corrections to inclusive Higgs production. 

Our expansion method builds upon {\it reverse unitarity}~\cite{Anastasiou:2002yz,Anastasiou:2002wq,Anastasiou:2002qz,Anastasiou:2003yy,Anastasiou:2003ds}, a technique
developed for the calculation of the inclusive Higgs cross-section 
and the rapidity distribution of electroweak gauge bosons at NNLO.  
The main idea is that phase-space integrals of matrix-elements are dual to loop integrals 
in their algebraic properties:  recurrence
identities in the powers of propagators and the number of dimensions
as well  as differential equations satisfied by loop
integrals also apply to phase-space integrals. 
This is achieved by associating on-shell conditions or other
phase-space constraints in the form of delta functions with differences
of otherwise identical propagators with opposite infinitesimal
imaginary parts, in the spirit of Cutkosky's rules. In this paper, we
observe and exploit that the duality rules of reverse unitarity can 
be expanded in kinematic parameters. 

After performing a threshold expansion of loop integrals dual to
 the cross-sections for Higgs plus three partons  processes, we apply
 automated reduction algorithms to reduce the coefficients of the
 expansion to master integrals. We have performed the reduction using
 existing public programs  and programs developed specifically for the
 purpose of this computation.  

The reduction yields ten master integrals for the first two
coefficients in the threshold expansion.  The master integrals themselves
are not specific to Higgs production and will appear in the threshold
expansion of other hadroproduction processes, such as Drell-Yan. 
For processes with a single colorless final state our set of  master integrals is complete 
and our expressions are universal for the partonic cross-sections with
the same initial state at leading order in the threshold expansion. 
  
For the computation of the {\it soft} master integrals we have employed a 
combination of methods. 
The master integrals are evaluated by using a parametrization of the phase space
in terms of the energies and angles of the momenta of the soft partons.
When needed, we perform the phase-space integrals by
introducing Mellin-Barnes representations.  We evaluate some of the 
Mellin-Barnes master integrals in a closed form as
hypergeometric functions or directly as a Laurent-expansion in the dimensional 
regulator $\epsilon$ with infinite nested sums as coefficients, which naturally evaluate to 
multiple zeta values in all cases. Some master integrals
cannot be evaluated easily from their Mellin-Barnes representations. 
We have developed a procedure to turn Mellin-Barnes integrals into 
integrals over positive real parameters, which are easy to expand in
$\epsilon$ whenever the integral is finite. The resulting parametric integrals
are then evaluated by integrating out the integration variables one-by-one
in terms of multiple polylogarithms. The last integration then again produces 
multiple zeta values in a natural way. 

We have restricted our computation of the partonic cross-sections 
to the first two terms  of the threshold expansion.  There are two obvious
ways to extend this work in the future. We could continue with 
the computation of subleading terms in the threshold expansion. This 
requires a more intensive but not prohibitive computer algebra and 
the evaluation of {\it soft} master integrals from topologies 
which contribute only to higher orders in the soft expansion. 
The soft expansion is rather fast converging for Higgs production but
rather slow for Drell-Yan, as has been noticed at NNLO. 
A second possibility is to perform the phase-space integrations 
for arbitrary kinematics, reducing them to a different set of master
integrals.  The {\it soft} master integrals which we 
present in this article can serve as boundary conditions for
solving the differential equations satisfied by the master integrals
in arbitrary kinematics.

This article is organized as follows: In Section~\ref{sec:runitarity} we present our general strategy based on reverse-unitarity 
to perform the threshold expansion for real-emission cross sections in terms of soft integrals, and we illustrate the method on some simple examples in Section~\ref{sec:validation}. In Sections~\ref{sec:DRR} and~\ref{sec:PS_soft} we study some properties of soft integrals in general, and we show that there is an easy and canonical way to derive dimensional recurrence relations and Mellin-Barnes integral representations for generic soft integrals. The soft master integrals contributing to the first two terms in the threshold expansion of the leading-order partonic cross sections for $p\,p\to H+3$ partons are discussed in Section~\ref{sec:masters_definition}, and the corresponding results for the cross sections are presented in Section~\ref{sec:amplitude}. Section~\ref{sec:masters_computation} contains technical details about the computation of the soft master integrals, and in Section~\ref{sec:conclusion} we draw our conclusions and give an outlook for future work. The paper contains several appendices discussing phase parametrisations and a generic formula for the phase-space volume for $H+n$ partons, as well as a new method to derive a parametric integral representation from a Mellin-Barnes integral and a description of an algorithmic way to perform the analytic integration of certain classes of parametric integrals.


\section{Reverse unitarity, threshold expansion and soft integrals}
\label{sec:runitarity}

We consider the production of a Higgs  boson in association with $j=3 \ldots N$ massless partons 
in the final state from two massless partons $i=1,2$ in the initial state,
\begin{equation}
1 + 2 \to H + 3 + \ldots + N.
\end{equation}
The inclusive cross-section for this process in dimensional regularization is given by a 
phase-space integral over the momenta $q_j$ of the final-state partons, 
\begin{equation}
\label{eq:runitstart1}
\sigma = \int d\Phi_{N-1}(q_H,q_3,\ldots,q_N;M^2;s;D)\left| {\cal A} \right|^2(\{ q_j\}, q_1, q_2;D).  
\end{equation}
We work in $D=4-2\eps$ dimensions and denote by $q_1, q_2$ the momenta of the initial-state partons and we also use the shorthand notation 
\[ 
q_{12} = q_1+q_2, \; q_{345} = q_3+q_4+q_5, \textrm{ etc.}
\]
In the following we will often drop the functional dependence on the dimension $D$ for clarity.
The mass of the Higgs boson is denoted by $M$, and we denote the (squared) center-of-mass energy by $s=(q_1+q_2)^2$. 
$|{\cal A}|^2$ represents the squared matrix-element multiplied with the appropriate flux and symmetry factors. The $D$-dimensional phase space measure is given by
\beq\bsp
\label{eq:psm}
d\Phi&_{N-1}(q_H,q_3,\ldots,q_N;M^2;s;D)\\
&\,= (2\pi)^D\,\delta^{(D)}(q_{12} - q_H - q_{3\ldots N})\,\frac{d^Dq_H}{(2\pi)^{D-1}}\delta_+(q_H^2-M^2)\prod_{j=3}^N\frac{d^Dq_j}{(2\pi)^{D-1}}\delta_+(q_j^2)\,,
\esp\eeq
with $\delta_+(q^2-m^2) = \delta(q^2-m^2)\Theta(q^0)$. Integrating out the momentum of the Higgs boson, we can rewrite eq.~\eqref{eq:runitstart1} as
\begin{equation}
\label{eq:runitstart}
\sigma = \frac{1}{2\pi}\int \left[ \prod_{j=3}^N  \frac{d^Dq_j}{(2\pi)^{D-1}} \delta_+(q_j^2) \right] \delta_+\left( \left[  q_{3\ldots N} - q_{12}\right]^2 - M^2 \right)
\left| {\cal A} \right|^2(\{ q_j\}, q_1, q_2).  
\end{equation}

 In this article, we restrict ourselves to the case of real-radiation matrix-elements without virtual corrections. 
We introduce the variables
\begin{equation}
\label{eq:zbar}
z =\frac{M^2}{s} {\rm~~and~~}\zbar = 1 -z\,.
\end{equation}
We now rescale the momenta of all the partons, 
\begin{equation}\label{eq:mom_scaling}
q_i = \left\{\begin{array}{ll}
\sqrt{s}\, p_i\,, & \textrm{if } i=1,2\,, \\
\sqrt{s}\,\zbar\, p_i\,, & \textrm{if } i=3\ldots N\,,
\end{array}\right.
\end{equation}
which captures  the scaling of the partonic momenta in the
final-state. We emphasize that this is not, as yet, an approximation,
but rather a convenient change of integration variables 
which captures the correct asymptotic behavior at threshold
as $\zbar \to 0$.  In the following we assume $s=1$, and we find 
\beq\bsp
\label{eq:runitasymptot}
\sigma =  \zbar^{(D-2)(N-2)-1}\, \frac{1}{2\pi}
\int& \left[ \prod_{j=3}^N  \frac{d^Dp_j}{(2\pi)^{D-1}} \delta_+(p_j^2) \right] 
\delta_+\left( \left[p_{3\ldots N} - p_{12}\right]^2 - z\, p_{3\ldots N}^2  \right)\\
&\,\times\left| {\cal A} \right|^2(\{ \zbar\, p_j\}, p_1, p_2). 
\esp\eeq
Note that the full $s$-dependence can easily be recovered from dimensional analysis.

The squared matrix-element  $\left| {\cal A} \right|^2$ consists 
of a rapidly growing number of terms with $N$, yielding a
correspondingly large number of phase-space integrals. 
The method of reverse unitarity, developed in 
refs.~\cite{Anastasiou:2002yz,Anastasiou:2002wq,Anastasiou:2002qz,Anastasiou:2003yy,Anastasiou:2003ds}, 
allows the reduction of phase-space integrals to a basis of fewer master
integrals by establishing a duality of phase-space and loop integrals,
where the latter are amenable to algebraic methods~\cite{Laporta:2001dd,Gehrmann:1999as} 
based on integration by parts~\cite{Tkachov:1981wb,Chetyrkin:1981qh}.  

Following the reverse unitarity approach, on-shell and other
phase-space constraints are dual to propagators: 
\begin{equation}
\delta_+(q^2) \to \left( \frac{1}{q^2}\right)_c \equiv \frac{1}{2\pi i}\,\mathrm{Disc}\,\frac{1}{q^2} = \frac{1}{2\pi i}\left(\frac 1
{q^2 + i0} - \frac{1}{q^2-i0}\right)\,.
\end{equation}
``Cut'' propagators can be differentiated in a similar way to ordinary
propagators with respect to their momenta, 
\begin{equation}
\frac \partial {\partial q_\mu }
\left[ \left( \frac{1}{q^2}\right)_c \right]^\nu  
= - \nu  
\left[ \left( \frac{1}{q^2}\right)_c \right]^{\nu+1} 
 2 q^\mu\,,  
\end{equation}
leading to identical integration-by-parts (IBP) identities for phase-space
integrals as for their dual loop integrals. Solving the system of IBP 
identities for phase-space integrals proceeds in the same way as for
loop integrals, with the exception that for cut-propagators we can use
the simplifying identity: 
\begin{equation}
\left[ \left( \frac{1}{q^2}\right)_c \right]^{-\nu} \to 0, \quad \forall
\; \nu=0,1,2,\ldots  
\end{equation}

According to the reverse unitarity method, we find a dual forward
scattering loop-amplitude with $N-1$ cut-propagators for the real
radiation contribution of eq.~\eqref{eq:runitasymptot}, namely, 
\beq\bsp
\sigma =  \zbar^{(D-2)(N-2)-1} \,\frac{1}{2\pi}
\int& \left[ \prod_{j=3}^N  \frac{d^Dp_j}{(2\pi)^{D-1}} \left( \frac 1 {p_j^2} \right)_c \right]  
\left[ \frac 1 { \left[p_{3\ldots N} - p_{12}\right]^2 - z \,p_{3\ldots N}^2 } \right]_c\\
&\times
\left| {\cal A} \right|^2(\{ \zbar p_j\}, p_1, p_2)\,.
\esp\eeq

In this article, we take one further step and expand cut-propagators
and the squared matrix-elements  around $z=1$, 
\begin{equation}
\left| {\cal A} \right|^2(\{ \zbar \,p_j\}, p_1, p_2)
= \zbar^{-2(N-2)} 
\sum_{k=0}^\infty 
\left| {\cal A} \right|_k^2(\{ p_j\}, p_1, p_2) \zbar^k, 
\end{equation}
and 
\begin{eqnarray}\label{eq:prop_expansion}
&& 
 \frac 1 {\left[p_{3\ldots N}-p_{12}\right]^2-z\,p_{3\ldots N}^2 } 
= \sum_{k=0}^\infty
\zbar^k  
\frac{
\left(- p_{3\ldots N}^2 \right)^k
}{
\left[ p_{12}\cdot\left(p_{12}-2p_{3\ldots N}\right) \right]^{k+1}
} \,.
\end{eqnarray}
In this approximation, the cross section can be expanded into a power series in $\zbar$,
\beq
\sigma = \zbar^{(D-4)(N-2)-1}\,\sum_{k=0}^\infty\,\zbar^k\,\sigma^{S(k)}\,.
\eeq
The coefficients of the power series are given by
\beq
\sigma^{S(k)}=\sum_{l=0}^k\,(-1)^l\int d\Phi_{N-1}^S\,\left[\frac{1}{p_{12}\cdot\left(p_{12}-2p_{3\ldots N}\right)}\right]_c^l\,\left(p_{3\ldots N}^2\right)^l\,\left| {\cal A} \right|_{k-l}^2(\{ p_j\}, p_1, p_2)\,,
\eeq
where $d\Phi_{N-1}^S$ denotes the ``soft'' phase space measure
\beq\bsp\label{eq:soft_measure}
d\Phi_{N-1}^S &\,\equiv \frac{1}{2\pi}\,\left[\frac{1}{p_{12}\cdot\left(p_{12}-2p_{3\ldots N}\right)}\right]_c\,\prod_{j=3}^N \frac{d^Dp_j}{(2\pi)^{D-1}} \left( \frac 1 {p_j^2} \right)_c\\
&\,=\frac{1}{2\pi}\,\delta_{+}(p_{12}^2-2p_{12}\cdot p_{3\ldots N})\,\prod_{j=3}^N \frac{d^Dp_j}{(2\pi)^{D-1}} \delta_{+}(p_j^2)
\,.
\esp\eeq
The integrals which emerge after the $\zbar$ expansion depend
trivially on one dimensionful parameter  $p_{12}^2=s$. If we put $s=1$, the integrals
are number integrals whose only functional dependence is through the space-time dimension $D=4-2\eps$. We will refer to such integrals as \emph{soft (phase space) integrals}, and they are the main subject of this paper. We note that, apart from the cut Higgs boson propagator, the integrands of soft phase space integrals are homogeneous functions under a simultaneous rescaling of the final-state momenta. In addition, a soft integral can 
be reduced to a set of ``soft'' master integrals using IBP
identities by exploiting the duality to loop integrals via reverse-unitarity.
We will illustrate this property in the next section where we check our method on several examples.


\section{Validation of the method and examples} 
\label{sec:validation}

In this section, we study the validity of the method described in
the previous section at NLO and NNLO -- two perturbative orders that are
well studied in the literature and so we can compare our results readily
with known results. In particular, we show that our method reproduces the correct 
results for the leading behavior of NLO and NNLO real emission amplitudes in the soft limit,
as well as for the subleading terms in the expansion of the phase space volume up to \n3lo and for a non-trivial double real emission master integral at NNLO.

At NLO, all phase space integrals that contribute to the real emission amplitude in general kinematics can be reduced to the phase space volume for $H+1\,\text{parton}$,
\beq
\Phi_2(\zbar;\eps) = \frac{1}{2(4\pi)^{1-\eps}}\,\frac{\Gamma(1-\eps)}{\Gamma(2-2\eps)}\,\zbar^{1-2\eps}\,.
\eeq
As there is only one master integral which is a monomial in $\zbar$, our method trivially gives the correct answer at NLO.

At NNLO all double real emission phase space integral can be reduced in general kinematics to a linear combination of 18 master integrals~\cite{Anastasiou:2002yz}. The leading contribution of all master integrals in the soft limit to all orders in $\eps$ was computed in ref.~\cite{Anastasiou:2012kq}, and it was observed that in this limit 17 master integrals are proportional to the soft limit of the phase space volume for $H+2$ partons,
\beq\bsp
\label{eq:NNLOpsvolume}
\Phi_3(\zbar;\eps) &\, = \frac{1}{2(4\pi)^{3-2\eps}}\,\zbar^{3-4\epsilon}  \frac{\Gamma(1-\eps)^2}{\Gamma(4-4\eps)}  {_2F_1}(1-\epsilon,2-2\epsilon;4-4\epsilon;\zbar)\\
&\, = \zbar^{3-4\eps}\,\Phi_3^S(\eps) + \ord(\zbar^4)\, ,
\esp\eeq
where we defined
\beq
\Phi_3^S(\eps) =  \frac{1}{2(4\pi)^{3-2\eps}}\, \frac{\Gamma(1-\eps)^2}{\Gamma(4-4\eps)}\,.
\eeq
More precisely, it was shown in ref.~\cite{Anastasiou:2012kq} that if ${\bf X}^S_i(\zbar;\eps)$ denotes the 
leading term  in the soft limit of the double real emission master integrals, then we can write\footnote{Note that the normalization differs slightly from the normalization of ref.~\cite{Anastasiou:2012kq}.}
\beq\bsp\label{eq:NNLOmasters}
{\bf X}^S_i(\zbar;\eps) &\,= {\bf S}_i(\zbar;\eps)\,\Phi_3^S(\eps)\,,\qquad 1\le i\le 17\,,\\
{\bf X}^S_{18}(\zbar;\eps)&\, = -4\,\zbar^{-1-4\eps}\,\frac{(1-2\eps)(3-4\eps)(1-4\eps)}{\eps^3}\,{_3 F_2}(1,1,-\eps;1-\eps,1-2\eps;1)\,\Phi_3^S(\eps)\,,
\esp\eeq
where ${\bf S}_i(\zbar;\eps)$ are monomials in $\zbar$ and rational functions of $\eps$. 
Using the method described in the previous section, we can easily explain the structure of eq.~\eqref{eq:NNLOmasters}. Indeed, we observe that in the soft limit all the double real emission phase space integrals can be reduced to only two master integrals. In particular, the IBP identities in the soft limit allow us to express all but one of the ${\bf X}_i^S$ in terms of the phase space volume, and the coefficients appearing in the reduction are precisely the functions ${\bf S}_i$. In other words, in the soft limit all double real emission phase space integrals can be reduced to linear combinations of the following  two soft master integrals

\begin{eqnarray}
\XOnePic{280} & = &\int d\Phi_3^S\,,\\
\XEigthteenPic{270} & =&\int\frac{d\Phi_3^S}{s_{14}s_{23}s_{34}}\,.
\end{eqnarray}
\vskip0.7cm\noindent
Our method thus provides the correct leading soft behavior of the double real emission contribution at NNLO. We emphasize that all the diagrams in this paper represent soft phase space integrals, i.e., all the diagrams represent integrals with respect to the soft phase space measure of eq.~\eqref{eq:soft_measure}. In addition, the invariants appearing in the integrands of the soft integrals are defined with respect to the rescaled momenta defined in eq.~\eqref{eq:mom_scaling},
\beq
s_{ij}=(\tau_ip_i+\tau_jp_j)^2\,,\qquad \tau_i=\left\{\begin{array}{ll}
-1\,,& \textrm{if } i=1,2\,,\\
+1\,,&\textrm{if } i=3\ldots N\,.
\end{array}\right.
\eeq

Our method does not only allow us to compute the leading soft behavior, but we can consistently expand around the soft limit $\zbar=0$. In the following we show that we can correctly reproduce the first few terms in the soft expansion of double and triple emission phase space volumes, as well as for the NNLO master integral ${\bf X}_{18}$ of refs.~\cite{Anastasiou:2002yz,Anastasiou:2012kq,Pak:2011hs}.

Let us start with the phase space volume for $H+2$ partons in the limit where the two partons are soft. On the one hand, from eq.~\eqref{eq:NNLOpsvolume} we immediately see that $\Phi_3$ admits the expansion
\beq\bsp\label{eq:X1_expansion_direct}
\Phi_3(\zbar;\eps) &\, = \zbar^{3-4\eps}\,\Phi_3^S(\eps) \sum_{n=0}^\infty \frac{(1-\epsilon)_n(2-2\epsilon)_n}{(4-4\epsilon)_n}\,\zbar^n\\
&\, = \zbar^{3-4\eps}\,\Phi_3^S(\eps) \Big[1+\frac{1-\eps}{2}\,\zbar+ \frac{(1-\eps)(2-\eps)(3-2\eps)}{4(5-4\eps)}\,\zbar^2 +\ord(\zbar^3)\Big]\,.
\esp\eeq
On the other hand, using eq.~\eqref{eq:prop_expansion} we obtain the diagrammatic expansion

\beq\label{eq:X1_expansion_diagrams}
\Phi_3(\zbar;\eps)  =\zbar^{3-4\eps}\Bigg[\phantom{aaaaaaaaa}\,\,\,-\zbar\phantom{aaaaaaaaa}\,\,\,+\zbar^2\phantom{aaaaaaaaa}\,\,\,+\ord(\zbar^3)\Bigg]\,, \XOnePicNoText{479}  \XOneNLOPic{364} \XOneNNLOPic{284}
\eeq
\vskip0.5cm\noindent
where the dashed lines indicate numerator factors and dots represent additional powers of the propagators or the numerators.
The diagrams appearing in eq.~\eqref{eq:X1_expansion_diagrams} are in one-to-one correspondence with the terms in the expansion~\eqref{eq:X1_expansion_direct}. Indeed, IBP reduction of the integrals in eq.~\eqref{eq:X1_expansion_diagrams} reveals

\begin{eqnarray}
\XOneNLOPic{228}& = & -\frac{1-\eps}{2}\XOnePicNoText{208}\phantom{aaaaaaaaaa}\,,\\
&&\phantom{A}\nonumber\\
\XOneNNLOPic{228}& = & \frac{(1-\eps)(2-\eps)(3-2\eps)}{4(5-4\eps)}\XOnePicNoText{209}\phantom{aaaaaaaaaa}\,,
\end{eqnarray}
\vskip0.5cm\noindent
in perfect agreement with eq.~\eqref{eq:X1_expansion_direct}. We checked explicitly that our method reproduces correctly the first ten terms of the soft expansion of the phase space volume for $H+2$ partons.

As a second example we derive the subleading terms in the soft expansion of the double real emission master integral ${\bf X}_{18}$. Unlike for the phase space volume, no result is known for ${\bf X}_{18}$ valid to all orders in $\eps$ in general kinematics, but the integral was evaluated explicitly up to $\ord(\eps)$ in terms of harmonic polylogarithms~\cite{Remiddi:1999ew} in ref.~\cite{Anastasiou:2002yz,Anastasiou:2012kq,Pak:2011hs}. We can thus compare the result of our method order by order in $\eps$ to the expansion of the harmonic polylogarithms around $z=1$. Using eq.~\eqref{eq:prop_expansion} we obtain

\beq
\int\frac{d\Phi_3}{q_{14}^2q_{23}^2q_{34}^2} = \zbar^{-1-4\eps}\Bigg[\phantom{aaaaaaaaaa}\,\,\,-\zbar\phantom{aaaaaaaaaa}\,\,\,+\zbar^2\phantom{aaaaaaaaaa}\,\,\,+\ord(\zbar^3)\Bigg]\,, \XEightteenLOPic{468}  \XEightteenNLOPic{516} \XEightteenNNLOPic{305}
\eeq
\vskip0.5cm\noindent
IBP reduction of the diagrams appearing in the subleading terms gives

\begin{eqnarray}
\XEightteenNLOPic{377} & = & -\frac{2 (1-4 \epsilon ) (3-4 \epsilon ) (1-2 \epsilon )}{\epsilon ^2} \XOnePicNoText{205} \phantom{aaaaaaaaaa}\,\,\,\,,\\
\phantom{A}&&\nonumber\\
\phantom{A}&&\nonumber\\
\XEightteenNNLOPic{250} & = & \frac{(3-4 \epsilon ) (1-2 \epsilon ) \left(2 \epsilon ^2-2 \epsilon +1\right)}{\epsilon ^2}\XOnePicNoText{205} \phantom{aaaaaaaaaa}\,\,\,\,.
\end{eqnarray}
\vskip0.5cm\noindent
We checked that using these identities we can reproduce correctly the first five terms in the soft expansion of ${\bf X}_{18}$.

The aim of this paper is to compute the leading terms in the soft expansion of the triple real emission amplitude for inclusive Higgs production. In order to test our method at \n3lo, we verified that we can reproduce the correct soft expansion of the phase space volume for $H+3$ partons. The phase space volume for $H+3$ partons in general kinematics can be written in the form (see Appendix~\ref{sec:psv})
\beq\bsp
\label{eq:N3LOpsvolume}
\Phi_4(\zbar;\eps) &\, = \frac{1}{2(4\pi)^{5-3\eps}}\,\zbar^{5-6\epsilon}  \frac{\Gamma(1-\eps)^3}{\Gamma(6-6\eps)}  {_2F_1}(2-2\epsilon,3-3\epsilon;6-6\epsilon;\zbar)\\
&\, = \zbar^{5-6\eps}\,\Phi_4^S(\eps)\left[1+(1-\eps)\,\zbar+\frac{(1-\eps)(3-2\eps)(4-3\eps)}{2(7-6\eps)}\,\zbar^2+\ord(\zbar^3)\right]\, ,
\esp\eeq
where we defined
\beq\label{eq:soft_Phi_4}
\Phi_4^S(\eps) =  \frac{1}{2(4\pi)^{5-3\eps}}\, \frac{\Gamma(1-\eps)^3}{\Gamma(6-6\eps)}\,.
\eeq
Using our method, we obtain the following diagrammatic expansion

\beq\label{eq:Phi4_expansion_diagrams}
\Phi_4(\zbar;\eps)  =\zbar^{5-6\eps}\Bigg[\phantom{aaaaaaaaa}\,\,\,-\zbar\phantom{aaaaaaaaa}\,\,\,+\zbar^2\phantom{aaaaaaaaa}\,\,\,+\ord(\zbar^3)\Bigg]\,. \FOneLOPic{477}  \FOneNLOPic{361} \FOneNNLOPic{281}
\eeq
\vskip0.5cm\noindent
All the diagrams in the expansion can be reduced to the soft phase space volume, as expected,

\begin{eqnarray}
\FOneNLOPic{228}& = & -(1-\eps)\FOneLOPic{208}\phantom{aaaaaaaaaa}\,,\\
&&\phantom{A}\nonumber\\
\FOneNNLOPic{228}& = & \frac{(1-\eps)(3-2\eps)(4-3\eps)}{2(7-6\eps)}\FOneLOPic{209}\phantom{aaaaaaaaaa}\,.
\end{eqnarray}
\vskip0.5cm\noindent

To summarize, our method provides a systematic way to perform the threshold expansion of phase space integrals for the production of a heavy colorless state. Every term in the expansion corresponds to a soft integral, as defined in Section~\ref{sec:runitarity}, which can be reduced to a small set of soft master integral using IBP reduction. In the next two sections we study some additional properties of soft integrals in general, before applying our method to compute the threshold expansion of the triple real emission contribution to inclusive Higgs production.

\section{Relations between phase space integrals in different dimensions}
\label{sec:DRR}
It is well known that loop integrals in different space-time dimensions are related by so-called dimensional shift identities~\cite{Tarasov:1996br}. 
After reduction to master integrals, the dimensional shift identities reduce to recurrence relations in the space-time dimension $D$ for the master integrals themselves~\cite{Smirnov:1999wz,Anastasiou:1999bn,Anastasiou:2000mf,Anastasiou:2000kp,Lee:2009dh}. 
Reverse unitarity rules show that the same dimensional shift identities hold for the dual 
phase-space integrals (see also ref.~\cite{Lee:2012te}). 

In this section we present an easy way to derive the dimensional shift identities for phase space integrals.
We stress that the results of this section are generic and apply to phase space integrals in general kinematics. To start, let us consider a phase space integral in $D$ dimensions,
\beq
F(D;\nu_1,\dots,\nu_n) = \int d\Phi_{N-1}(D) f(\nu_1,\dots,\nu_n)\,,
\eeq
where we explicitly indicate the dependence on the space-time dimension $D$.
The integrand $f$ can be written as a product,
\beq
f(\nu_1,\dots,\nu_N) = \prod_{l=1}^N P_l^{-\nu_l},
\eeq
where the $P_l$ are polynomials in the rescaled kinematic invariants $s_{ij}$, raised to some power $\nu_l$. In Appendix~\ref{app:PSparam} we show that the phase space measure $d\Phi_{N-1}(D)$ for $\mbox{parton} + \mbox{parton} \rightarrow H + (N-2)\,\mbox{partons}$  can be parametrized solely in terms of kinematic invariants,
\beq\bsp
\label{eq:invParam}
d\Phi_{N-1}(D)&= \mathcal{N}_{N-2}(D)\,\bar{z}^{(N-2)(D-2)-1}\left(\prod_{\substack{1\leq i,j \leq N\\i\neq j,(i,j)\neq(1,2)}}\!\!\!\!ds_{ij}\right)\\
&\times\delta\left(1-\sum_{i=3}^N(s_{1i}+s_{2i})+\bar{z}\sum_{i=3}^{N}\sum_{j=3}^{i-1}s_{ij}\right)G_N^{\frac{D-N-1}{2}}(\{s_{ij}\})\Theta[G_N(\{s_{ij}\})]\,,
\esp\eeq
where we have defined the normalisation factor
\beq\bsp
\label{eq:normN}
\mathcal{N}_{N-2}(D) &= (-1)^{\frac{(N-2)(N-3)}{2}}2^{-(N-2)\frac{D}{2}}(2\pi)^{(N-1)-(N-2)D} \prod_{i=3}^{N}\Omega_{D-i+1},\\
\Omega_D &= \frac{2\pi^{\frac{D}{2}}}{\Gamma(\frac{D}{2})}\,,
\esp\eeq
and the Gram determinant $G_N$  
\beq
G_N(\{s_{ij}\}) = \det(s_{1i}s_{2j}+s_{1j}s_{2i}-s_{ij})_{3\leq i,j \leq n}\,.
\eeq
Obviously, the $D$-dependent constants factor out of the integral, and so the actual integral depends on $D$ only through the exponent of the Gram determinant.
It is then easy to see that the phase space measures in shifted dimensions are related by
\beq
\label{eq:facParam}
d\Phi_{N-1}(D+2) = \frac{\mathcal{N}_{N-2}(D+2)}{\mathcal{N}_{N-2}(D)}\bar{z}^{2(N-2)}\,d\Phi_{N-1}(D)\, G_N(\{s_{ij}\})\,.
\eeq
We can thus express a phase space integral in $D+2$ dimensions as
\beq
\label{eq:dplustwo}
F(D+2;\nu_1,\dots,\nu_n)= \frac{\mathcal{N}_{N-2}(D+2)}{\mathcal{N}_{N-2}(D)}\bar{z}^{2(N-2)}\int d\Phi_{N-1}(D)\, G_N(\{s_{ij}\})\, f(\nu_1,\dots,\nu_n)\,.
\eeq

For a given set of polynomials $\{P_l\}$, the form of the integrand $f$ depends only on the exponents $\{\nu_i\}$.
One therefore introduces the operators $I_i^+$ and $I_i^-$ acting on $f$ as 
\beq\bsp
\label{eq:ops}
(I_i^+f)(\nu_1,\dots,\nu_i,\dots,\nu_n) &= f(\nu_1,\dots,\nu_i+1,\dots,\nu_n)\,, \\
(I_i^-f)(\nu_1,\dots,\nu_i,\dots,\nu_n) &= f(\nu_1,\dots,\nu_i-1,\dots,\nu_n)\,.
\esp\eeq
If we assume that the $\{P_l\}$ are linearly independent, we can express the invariants $s_{ij}$ as linear combinations of the $I^+_i$ and $I^-_i$.
This allows us to rewrite the extra power of the Gram determinant in eq.~\eqref{eq:facParam} as a polynomial of degree $N$ in the $I_i^-$
\beq\bsp
G_N(\{s_{ij}\})f(\nu_1,\dots,\nu_{n}) &= G_N(\{I_i^-\})\,f(\nu_1,\dots,\nu_n)\,.
\esp\eeq
We thus obtain the following compact formula relating phase space integrals in different dimensions,
\beq
F(D+2;\nu_1,\dots,\nu_n) = \frac{\mathcal{N}_{N-2}(D+2)}{\mathcal{N}_{N-2}(D)}\bar{z}^{2(N-2)} G_N(\{I_i^-\})\,F(D;\nu_1,\dots,\nu_n)\,.
\eeq
Every term in the polynomial can be evaluated according to the action of the $I^-_i$ operators, yielding a superposition of modified integrals in $D$ dimensions.
By applying this method to a master integral, we can express the master integral in $D+2$ dimensions as a linear combination of integrals in $D$ dimensions. Using IBP identities, we can reduce the integrals in $D$ dimensions to master integrals and thus we find a relation between the master integral in $D+2$ dimensions and $D$ dimensions. This dimensional recurrence relation can formally be written as
\beq
F_i(D+2) = \sum_j c_{ij}(D) F_j(D)\,,
\eeq
with coefficients $c_{ij}(D)$ that are determined from the IBP reduction.
\section{From soft integrals to angular integrals and Mellin-Barnes integrals}
\label{sec:PS_soft}
In this section we show that there is a canonical way to derive a Mellin-Barnes (MB) representation for the soft integrals that appear in the threshold expansion of phase space integrals. 
The soft integrals we need to consider have the form
\beq
F(\eps)=\int d\Phi_{N-1}^S\,f(\{p_j\};p_1,p_2)\,,
\eeq
where $f(\{p_j\};p_1,p_2)$ is a ratio of products of multi-particle invariants that is homogeneous under a simultaneous rescaling of the final-state momenta, i.e.,
\beq
f(\{\lambda\,p_j\};p_1,p_2)=f(\{p_j\};p_1,p_2)\,\lambda^a\,,
\eeq
for some $a$.  Note that this is true only if the (cut) Higgs boson propagator it not raised to an additional power. In that case, however, we can always reduce the integrand to a homogeneous function using IBP identities.

Next, we note that there is a subclass of soft integrals that have an additional property: they are homogeneous with respect to \emph{individual} rescalings of the final-state momenta, i.e.,
\beq\label{eq:scaling}
f(\{\lambda_j\,p_j\};p_1,p_2)=f(\{p_j\};p_1,p_2)\,\prod_{j=3}^N\lambda_j^{a_j}\,,
\eeq 
for some $a_j$. This subclass of soft integrals is precisely the one where the integrand consists of products of power of \emph{two}-particle invariants,
\beq\label{eq:2part}
f(\{p_j\};p_1,p_2)=\prod_{k=1}^ms_{i_kj_k}^{-\alpha_k}=\prod_{k=1}^m(2p_{i_k}\cdot p_{j_k})^{-\alpha_k}\,,
\eeq
where the index $k$ runs over all the two-particle invariants appearing in $f$.
Every soft integral can be converted into an integral of this type, to the price of introducing additional MB integrations. Indeed, if we write every multi-particle invariant as a sum of two-particle invariants, then we can convert sums into product by using the usual formula
\beq\label{eq:mb_int}
\frac{1}{(A+B)^\lambda} = \frac{1}{\Gamma(\lambda)}\mbint\frac{dz}{2\pi i}\Gamma(-z)\Gamma(\lambda+z)\frac{A^z}{B^{z+\lambda}}\,,
\eeq
where the contour separates the poles at $z=n$ from those in $z=\lambda-n$, $n\in\mathbb{N}$. Without loss of generality we can thus assume that our soft integral is homogeneous with respect to individual rescalings of the final state momenta.

If we concentrate on soft integral that satisfy eq.~\eqref{eq:scaling}, it is natural to choose a parametrization of the soft phase space that makes the homogeneity explicit. One possible parametrization with this property is the so-called 
 `energies and angles' parametrization, where the final-state momenta are parametrized as\footnote{We work with the rescaled momenta, eq.~\eqref{eq:mom_scaling}.}
\beq\bsp\label{eq:energies_and_angles}
p_1 &\,= \frac{1}{2}(1,1,0,\ldots)\,,\\
p_2 &\,= \frac{1}{2}(1,-1,0,\ldots)\,,\\
p_i &\,= \frac{1}{2}\,E_i\,\beta_i\,,\qquad 3\le i\le N\,.
\esp\eeq
The $E_i$ parametrize the energies of the final-state partons and $\beta_i$ is the $D$-velocity in the direction $p_i$. 
In this parametrization the phase space measure for each final-state parton takes the form
\beq
d^Dp_i\,\delta_+(p_i^2) = 2^{-(D-1)}\,\Theta(E_i)\,E_i^{D-3}\,dE_i\,d\Omega_i^{(D-1)}\,,
\eeq
where $d\Omega_i^{(D-1)}$ is the measure on the unit sphere parametrizing the solid angle of particle $i$. Furthermore, the on-shell condition for the Higgs boson, coming from the cut propagator in eq.~\eqref{eq:soft_measure}, can be rewritten as
\beq\bsp\label{eq:H_on_shell}
\left[\frac{1}{p_{12}\cdot\left(p_{12}-2p_{3\ldots N}\right)}\right]_c = \delta_+\left(p_{12}\cdot\left(p_{12}-2p_{3\ldots N}\right)\right)
=\delta\left(1-\sum_{i=3}^NE_i\right)\,.
\esp\eeq
Thus, the soft phase space measure can be parametrized as
\beq
d\Phi_{N-1}^{S} = (2\pi)^{N-1-(N-2)D}\,2^{-(N-2)(D-1)}\,\delta\left(1-\sum_{i=3}^NE_i\right)\,\prod_{i=3}^NE_i^{D-3}\,dE_i\,d\Omega_i^{(D-1)}\,.
\eeq
Using this parametrization, we see that every soft integral with an integrand of the form~\eqref{eq:2part} can be written as
\beq\label{eq:toy_integral}
F(\eps)=\int d\Phi^S_{N-1}\prod_{k=1}^ms_{i_kj_k}^{-\alpha_k} = 2^{\alpha}\int d\Phi^S_{N-1}\prod_{k=1}^m(E_{i_k}E_{j_k})^{-\alpha_k}\,(\beta_{i_k}\cdot\beta_{j_k})^{\alpha_k}\,,\qquad \alpha=\sum_{k=1}^m\alpha_k\,.
\eeq 
Both the measure and the integrand can be written in a factorized form, and so we can integrate out the energies in terms of a generalized Beta function,
\beq
\int_0^1\left(\prod_{k=3}^NdE_k\,E_k^{a_k-1}\right)\,\delta\left(1-\sum_{k=3}^NE_k\right) =\frac{\Gamma(a_1)\ldots\Gamma(a_m)}{\Gamma(a_1+\ldots +a_m)}\,.
\eeq
Hence, the only non-trivial integration is a multiple angular integration over the solid angles of the final-state partons. Angular integrals can be written in the general form
\beq\label{eq:generic_angular_integral}
\Omega_{D-1}^{(\alpha_1,\ldots,\alpha_m)}\left(\{\beta_{j_a}\cdot\beta_{j_b}\}\right)=\int\frac{d\Omega^{(D-1)}_i}{(\beta_{j_1}\cdot \beta_i)^{\alpha_1}\ldots(\beta_{j_m}\cdot \beta_i)^{\alpha_m}}\,.
\eeq
In ref.~\cite{Somogyi:2011ir} it was shown that such integrals fall into a class of generalized hypergeometric functions known as ${\bf H}$ functions, and an MB representation for the most general angular integral of this type was derived. We have thus a general recipe to derive MB representations for generic soft integrals.

Although the previous technique allows us to derive a multifold MB representation for every soft integral we need to consider, it can sometimes be useful to insert, if existent, explicit closed expressions for the angular integrals.
Indeed, for small values of $m$ the integrals~\eqref{eq:generic_angular_integral} are very simple and can be evaluated in closed form. In the following we briefly review some results for angular integrals which will be useful in our case. 

The case $m=0$ corresponds to the volume of the solid angle
\beq
\Omega_{D-1} = \int d\Omega^{(D-1)}_i = \frac{2\pi^{(D-1)/2}}{\Gamma\left(\frac{D-1}{2}\right)}\,.
\eeq
Note that this integral is sufficient to compute the soft phase space volume, which corresponds to putting $m=0$ in eq.~\eqref{eq:toy_integral}: we simply obtain a factor $\Omega_{D-1}$ for each final-state parton. Thus, we obtain
\beq\label{eq:soft_volume}
\Phi_{N-1}^S(\eps) = (2\pi)^{-2N+5+(N-2)\eps}\,2^{-(N-2)(2-\eps)}\,\frac{\Gamma(1-\eps)^{N-2}}{\Gamma(2(N-2)(1-\eps))}\,,
\eeq
in agreement with the results of Appendix~\ref{sec:psv}.

As we are only interested in massless momenta, $\beta_j^2=0$, Lorentz invariance implies that the angular integral with one propagator must evaluate to a constant. Indeed, we have
\beq
\Omega_{D-1}^{(\alpha)} = \int \frac{d\Omega_i^{(D-1)}}{(\beta_j\cdot \beta_i)^{\alpha}}=2^{2-\alpha-2 \epsilon}\,\pi ^{1-\epsilon } \,\frac{ \Gamma (1-\epsilon -\alpha)}{\Gamma (2-2 \epsilon -\alpha)}\,.
\eeq
For angular integrals with two massless propagators one obtains~\cite{vanNeerven:1985xr}
\beq\bsp\label{eq:Omega_j_k}
\Omega&_{D-1}^{(\alpha_{j_1},\alpha_{j_2})}(\beta_{j_1}\cdot\beta_{j_2}) = \int \frac{d\Omega_i^{(D-1)}}{(\beta_{j_1}\cdot \beta_i)^{\alpha_1}\,(\beta_{j_2}\cdot \beta_i)^{\alpha_2}}\\
&\,=2^{2-\alpha_1-\alpha_2-2 \epsilon}\,\pi ^{1-\epsilon }  \,\frac{\Gamma (1-\epsilon -\alpha_1) \Gamma (1-\epsilon -\alpha_2) }{\Gamma (1-\epsilon ) \Gamma (2-2 \epsilon -\alpha_1-\alpha_2)}\\
&\,\times{_2F_1}\left(\alpha_1,\alpha_2;1-\epsilon ;1-\frac{\beta_{j_1}\cdot \beta_{j_2}}{2}\right)\,.
\esp\eeq
To our knowledge, there are no closed formulas for angular integrals with three or more massless propagators. We will however need later on the angular integral with three massless propagators, which admits the MB representation~\cite{Somogyi:2011ir},
\beq\bsp\label{eq:Omega_i_j_k}
\Omega&_{D-1}^{(\alpha_{j_1},\alpha_{j_2},\alpha_{j_3})}(\beta_{j_1}\cdot\beta_{j_2},\beta_{j_2}\cdot\beta_{j_3},\beta_{j_1}\cdot\beta_{j_3})\\
& = \frac{2^{2-\alpha_1-\alpha_2-\alpha_3-2\eps}\,\pi^{1-\eps}}{\Gamma(\alpha_1)\Gamma(\alpha_2)\Gamma(\alpha_3)\Gamma(2-\alpha_1-\alpha_2-\alpha_3-2\eps)}\\
&\times\mbint\frac{dz_{12}dz_{13}dz_{23}}{(2\pi i)^3}\Gamma(-z_{12})\Gamma(-z_{13})\Gamma(-z_{23})
\Gamma(\alpha_1+z_{12}+z_{13})\Gamma(\alpha_2+z_{12}+z_{23})\\
&\qquad\times\Gamma(\alpha_3+z_{13}+z_{23})
\Gamma(1-\alpha_1-\alpha_2-\alpha_3-\eps-z_{12}-z_{13}-z_{23})\\
&\qquad\times\left(\frac{\beta_{j_1}\cdot\beta_{j_2}}{2}\right)^{z_{12}}
\left(\frac{\beta_{j_1}\cdot\beta_{j_3}}{2}\right)^{z_{13}}\left(\frac{\beta_{j_2}\cdot\beta_{j_3}}{2}\right)^{z_{23}}\,,
\esp\eeq
with $z_{ij}=z_i+z_j$, and where the contours separate the poles coming from $\Gamma$ functions of the form $\Gamma(\ldots-z_{ij})$ from those coming from $\Gamma(\ldots+z_{ij})$.

\section{Triple real emission phase space integrals in the soft limit}
\label{sec:masters_definition}
\subsection{Triple real soft master integrals for Higgs production}
After having studied some properties of soft integral in the previous sections, we will use the technology developed in the previous sections to compute the threshold expansion of the leading-order cross sections for $H$ plus five partons. More details about the construction of the amplitude in this limit will be given in Section~\ref{sec:amplitude}. Here it suffices to say that we have computed the squared amplitude and we have checked that in the limit where we only keep the first two terms in the threshold expansion, all the phase space integrals can be reduced to linear combinations of the following ten soft master integrals,

\begin{eqnarray}
\FOnePic{283}&=&\int {d\Phi_4^S } = \Phi_4^S(\eps)\,,\\
\label{eq:integral_F2}
\FTwoPic{55}&=&\int \frac{d\Phi_4^S}{(s_{13}+s_{15})s_{34}} = \Phi_4^S(\eps)\,\cF_2(\eps)\,,\\
\label{eq:integral_F3}
\FThreePic{125}&=&\int \frac{d\Phi_4^S}{s_{14}s_{23}s_{34}}= \Phi_4^S(\eps)\,\cF_3(\eps)\,,\\
\label{eq:integral_F11}
\FElevenPic{167}&=&\int \frac{d\Phi_4^S}{s_{13}s_{15}s_{34}s_{45}}= \Phi_4^S(\eps)\,\cF_{4}(\eps)\,,\\
\label{eq:integral_F5}
\FFivePic{262}&=&\int \frac{d\Phi_4^S}{(s_{14}+s_{15})s_{23}s_{345}}= \Phi_4^S(\eps)\,\cF_5(\eps)\,,\\
\label{eq:integral_F6}
\FSixPic{169}&=&\int \frac{d\Phi_4^S}{(s_{13}+s_{14})(s_{14}+s_{15})s_{23}s_{34}}= \Phi_4^S(\eps)\,\cF_6(\eps)\,,\\
\label{eq:integral_F7}
\FSevenPic{169}&=&\int \frac{d\Phi_4^S}{s_{15}s_{24}s_{34}s_{35}}= \Phi_4^S(\eps)\,\cF_7(\eps)\,,\\
\label{eq:integral_F8}
\FEightPic{228}&=&\int \frac{d\Phi_4^S}{(s_{13}+s_{15})(s_{23}+s_{24})s_{34}s_{35}}= \Phi_4^S(\eps)\,\cF_8(\eps)\,,\\
\label{eq:integral_F9}
\FNinePic{239}&=&\int \frac{d\Phi_4^S}{s_{15}(s_{14}+s_{15})s_{23}s_{34}s_{345}}= \Phi_4^S(\eps)\,\cF_9(\eps)\,,
\end{eqnarray}
\begin{eqnarray}
\label{eq:integral_F10}
\FTenPic{167}&=&\int \frac{d\Phi_4^S}{(s_{23}+s_{24})(s_{24}+s_{25})s_{34}s_{45}}= \Phi_4^S(\eps)\,\cF_{10}(\eps)\,.
\end{eqnarray}
\vskip0.7cm
\noindent

We have normalized all the integrals to the soft phase space volume for $H+3g$ defined in eq.~\eqref{eq:soft_Phi_4}. In the remainder of this section we give the dimensional recurrence relations satisfied by the master integrals and present the analytic results for each master integral as a Laurent expansion in the dimensional regulator $\eps$. Technical details about how to compute the master integrals analytically will be given in Section~\ref{sec:masters_computation}.

\subsection{Dimensional recurrence relations}
Using the technique described in Section~\ref{sec:DRR}, we can derive dimensional recurrence relations for all the master integrals defined in the previous section. The knowledge of these recurrence relations provides us with a strong check on our results. In addition, it turns out that the master integral $\mathcal{F}_9(D)$ is easier to compute in $D=6-2\eps$ dimensions, where it is finite, and the dimensional recurrence relations allow us to relate the six-dimensional and four-dimensional results in an easy way.

The recurrence relation for the soft phase space volume is trivial to obtain from the recurrence relation for the $\Gamma$ function,
\beq
\Phi_4^S(D+2) = \frac{(D-4) (D-3) (D-2)^3}{72 (D-1) (3 D-5) (3 D-4) (3 D-2) (3 D-1)} \frac{\Gamma(D-4)}{64 \pi^3 \Gamma(D-1)} \Phi_4^S(D)\,.
\eeq
As we have defined all our master integrals relative to the phase space volume $\Phi_4^S$, we can simplify their recurrence relations by factoring out the above result. We therefore define the ratio
\beq
\mathcal{R} = \frac{\mathcal{N}_3^{D}}{\mathcal{N}_3^{D+2}} \frac{\Phi_4^S(D+2)}{\Phi_4^S(D)} =\frac{(D-4) (D-3) (D-2)^3}{72 (D-1) (3 D-5) (3 D-4) (3 D-2) (3 D-1)}\,,
\eeq
where $\mathcal{N}$ was defined in eq.~\eqref{eq:normN}. We give the results for the remaining master integrals relative to $\mathcal{R}$. The dimensional recurrence relations for the non-trivial master integrals are
\begin{eqnarray}
\mathcal{F}_2(D+2)\mathcal{R} &=& -\frac{(D-4) (7 D-18)}{3 (3 D-5) (3 D-4)}
-\frac{(D-4)^2 (3 D-10)}{24 (3 D-7) (3 D-5) (3 D-4)}\mathcal{F}_2(D)\,,\\
\nonumber\\
\mathcal{F}_3(D+2)\mathcal{R} &=& \frac{\left(38-28 D+5 D^2\right)}{3 (D-4) (3 D-5)} \nonumber\\
&&-\frac{(D-4)^3 (D-3)}{18 (3 D-10) (3 D-8) (3 D-7) (3 D-5)}\mathcal{F}_3(D)\,,\\
\nonumber\\
\mathcal{F}_{4}(D+2)\mathcal{R} &=&-\frac{4 \left(386-387 D+128 D^2-14 D^3\right)}{(D-4)^2 (D-3)}\nonumber\\
&&-\frac{(D-4)^2 (3 D-14) }{24 (3 D-11) (3 D-8) (3 D-7)}\mathcal{F}_{4}(D)\,,\\
\nonumber\\
\mathcal{F}_5(D+2)\mathcal{R} &=&-\frac{(D-4) \left(4752-9636 D+6706 D^2-1962 D^3+207 D^4\right)}{72 (D-3) (D-1) (3 D-10) (3 D-8) (3 D-5)} \nonumber \\
&&+\frac{(D-4)^2 (D-2)}{96 (D-1) (3 D-7) (3 D-5)}\mathcal{F}_5(D)\,,\\
\nonumber\\
\mathcal{F}_6(D+2)\mathcal{R} &=& \frac{\left(4256-6684 D+4224 D^2-1345 D^3+216 D^4-14 D^5\right)}{3 (D-4)^2 (D-3) (D-2)^2} \nonumber\\
&&+\frac{(D-4) (3 D-10)}{9 (D-2)^2 (3 D-7)}\mathcal{F}_2(D)\\
&&-\frac{(D-4)^3}{24 (D-2) (3 D-11) (3 D-7)}\mathcal{F}_6(D)\,,\nonumber\\
\nonumber\\
\mathcal{F}_7(D+2)\mathcal{R} &=& -\frac{4 (2 D-7)}{(D-4) (D-3)} \nonumber\\
&&+\frac{(D-4)^4}{72 (3 D-11) (3 D-10) (3 D-8) (3 D-7)}\mathcal{F}_7(D)\,,\\
\nonumber\\
\mathcal{F}_8(D+2)\mathcal{R} &=&\frac{2 \left(231-114 D+14 D^2\right)}{3 (D-4) (D-3)}+\frac{2 (D-4)^2 (7 D-24)}{9 (D-3) (3 D-8) (3 D-7)}\mathcal{F}_2(D)\nonumber\\
&&+\frac{(D-4)^4}{72 (3 D-11) (3 D-10) (3 D-8) (3 D-7)}\mathcal{F}_8(D)\,, \\
\nonumber\\
\mathcal{F}_9(D+2)\mathcal{R} &=&\frac{2 (3 D-7) \left(6672-7824 D+3460 D^2-684 D^3+51 D^4\right)}{3 (D-4)^2 (D-3)^2 (3 D-10)}\nonumber\\
&&+\frac{(D-4) (3 D-10) (5 D-17)}{12 (D-3)^2 (3 D-8)}\mathcal{F}_2(D)+\frac{(D-4)}{6 (D-3) (3 D-8)}\mathcal{F}_5(D)\nonumber\\
&&+\frac{(D-4)^3 (3 D-14)}{96 (D-3) (3 D-13) (3 D-11) (3 D-8)}\mathcal{F}_9(D)\,,\\
\nonumber\\
\mathcal{F}_{10}(D+2)\mathcal{R} &=&-\frac{4 \left(26-39 D+16 D^2-2 D^3\right)}{(D-4)^2 (D-3)}-\frac{(D-4)^2 (3 D-10)}{3 (D-3) (3 D-8) (3 D-7)} \mathcal{F}_2(D)\nonumber\\
&&-\frac{(D-4)^2 (3 D-14)}{24 (3 D-11) (3 D-8) (3 D-7)} \mathcal{F}_{10}(D)\,.
\end{eqnarray}
\vskip0.7cm

\subsection{Analytic results for the soft master integrals}
\label{sec:masters_results}
In this section we present the analytical results for the master integrals contributing to hadronic Higgs production in the leading and next-to-leading soft approximation. As the explicit evaluation of the master integrals is rather long and technical, we delay all details about the computation to Section~\ref{sec:masters_computation} and only summarize the results at this point. The first master integral, the soft phase space volume, was already given in eq.~\eqref{eq:soft_Phi_4} and will not be repeated here. All the remaining master integrals have been evaluated as a Laurent series in the dimensional regulator up to terms involving zeta values of weight at most six. We have checked that our results agree numerically with the MB integral representation for soft integrals derived in Section~\ref{sec:PS_soft}. In addition, the results satisfy the dimensional recurrence relations for the master integrals given in the previous section (integrals in the shifted dimension have been evaluated numerically using the MB representation).
Finally, we make an intriguing observation in our results: if we express all the zeta values up to weight six in the basis $\{\zeta_2,\zeta_3,\zeta_4,\zeta_2\,\zeta_3,\zeta_5,\zeta_3^2,\zeta_6\}$, the coefficients in front of the values are integers in all cases. We note that this statement is only true in the specific basis of zeta values that we chose. The results for the master integrals are listed in the rest of this section.

\beq\bsp\label{eq:F2_result}
\cF_2(\eps)&\,=\frac{1}{\Phi_4^S(\eps)}\int \frac{d\Phi_4^S}{(s_{13}+s_{15})s_{34}} \\
&\,=\frac{\Gamma (6-6 \epsilon ) \Gamma (1-2 \epsilon )^2}{\epsilon\,  \Gamma (3-6 \epsilon ) \Gamma (2-2 \epsilon )^2}\, _3F_2(1,1,1-\epsilon ;2-2 \epsilon ,2-2 \epsilon ;1)\\
&\,=\frac{60}{\eps}\,\zeta_2 + 420\,\zeta_3-282\,\zeta_2 + \eps \Big( 1800\,\zeta_4-1974\,\zeta_3+432\,\zeta_2 \Big) \\
&\,+ \eps^{2} \Big( 5580\,\zeta_5+480\,\zeta_2\,\zeta_3-8460\,\zeta_4+3024\,\zeta_3-216\,\zeta_2 \Big)\\
&\, + \eps^{3} \Big( 19260\,\zeta_6+1680\,\zeta_3^2-26226\,\zeta_5-2256\,\zeta_2\,\zeta_3+12960\,\zeta_4-1512\,\zeta_3 \Big)\\
&\,+\ord(\eps^4)\,.
\esp\eeq
\beq\bsp\label{eq:F3_result}
\cF_3(\eps)&\,=\frac{1}{\Phi_4^S(\eps)}\int \frac{d\Phi_4^S}{s_{14}s_{23}s_{34}} \\
&\,=\frac{90}{\eps^{4}} - \frac{693}{\eps^{3}} + \frac{1}{\eps^{2}} \Big( -60\,\zeta_2+1917 \Big) + \frac{1}{\eps} \Big( -300\,\zeta_3+462\,\zeta_2-2268 \Big) \\
&\,-930\,\zeta_4+2310\,\zeta_3-1278\,\zeta_2+972 + \eps \Big( -2220\,\zeta_5-120\,\zeta_2\,\zeta_3+7161\,\zeta_4\\
&\,-6390\,\zeta_3+1512\,\zeta_2 \Big) + \eps^{2} \Big( -5555\,\zeta_6-300\,\zeta_3^2+17094\,\zeta_5+924\,\zeta_2\,\zeta_3\\
&\,-19809\,\zeta_4+7560\,\zeta_3-648\,\zeta_2 \Big)+\ord(\eps^3)\,.
\esp\eeq
\beq\bsp\label{eq:F11_result}
\cF_{4}(\eps)&\,=\frac{1}{\Phi_4^S(\eps)}\int \frac{d\Phi_4^S}{s_{13}s_{15}s_{34}s_{45}} \\
&\,=-\frac{3\, \Gamma (6-6 \epsilon ) \Gamma (1-2 \epsilon ) }{2 \epsilon ^4 \Gamma (1-6 \epsilon )}\\
&\,\times\Bigg[\frac{3\, \Gamma (1-2 \epsilon ) \Gamma (\epsilon +1)}{(1+3 \epsilon) \Gamma (1-3 \epsilon )}\,{_3F_2}(-3 \epsilon -1,-2 \epsilon ,-\epsilon ;-3 \epsilon ,-3 \epsilon ;1)\\
&\,+\frac{1}{(1+\epsilon ) \Gamma (1-2 \epsilon )}\,{_4F_3}(1,1,1-\epsilon ,-2 \epsilon ;1-2 \epsilon ,1-2 \epsilon ,2+\epsilon;1)\Bigg]\\
&\,= -\frac{600}{\epsilon^4} + \frac{10020}{\epsilon^3}-\frac{70560}{\epsilon ^2}-\frac{1}{\epsilon}\Big(480\, \zeta_3-303480\Big)-3600\, \zeta_4+8016\, \zeta_3\\
&\,-1007640-\epsilon\Big(17280\, \zeta _5-60120\, \zeta_4+56448\, \zeta _3-3061800\Big)-\epsilon^2\Big(66000\, \zeta_6\\
&\,+1920\, \zeta_3^2-288576\, \zeta_5+423360\, \zeta_4-242784\, \zeta_3+9185400\Big) + \ord(\epsilon^3)\,.
\esp\eeq
\beq\bsp\label{eq:F5_result}
\cF_5(\eps)&\,=\frac{1}{\Phi_4^S(\eps)}\int \frac{d\Phi_4^S}{(s_{14}+s_{15})s_{23}s_{345}} \\
&\,= -\frac{120}{\eps} \,\zeta_2 -960\,\zeta_3+684\,\zeta_2 + \eps \Big( -4620\,\zeta_4+5472\,\zeta_3-1188\,\zeta_2 \Big)\\
&\, + \eps^{2} \Big( -17160\,\zeta_5-720\,\zeta_2\,\zeta_3+26334\,\zeta_4-9504\,\zeta_3+648\,\zeta_2 \Big)\\
&\, + \eps^{3} \Big( -64110\,\zeta_6-2880\,\zeta_3^2+97812\,\zeta_5+4104\,\zeta_2\,\zeta_3-45738\,\zeta_4+5184\,\zeta_3 \Big)\\
&\,+\ord(\eps^4)\,.
\esp\eeq
\beq\bsp\label{eq:F6_result}
\cF_6(\eps)&\,=\frac{1}{\Phi_4^S(\eps)}\int \frac{d\Phi_4^S}{(s_{13}+s_{14})(s_{14}+s_{15})s_{23}s_{34}} \\
&\,= \frac{10}{\eps^{5}} - \frac{137}{\eps^{4}} + \frac{1}{\eps^{3}} \Big( 40\,\zeta_2+675 \Big) + \frac{1}{\eps^{2}} \Big( 320\,\zeta_3-548\,\zeta_2-1530 \Big)\\
&\, + \frac{1}{\eps} \Big( 1500\,\zeta_4-4384\,\zeta_3+2700\,\zeta_2+1620 \Big) + 5160\,\zeta_5+320\,\zeta_2\,\zeta_3-20550\,\zeta_4\\
&\,+21600\,\zeta_3-6120\,\zeta_2-648 + \eps \Big( 18340\,\zeta_6+1280\,\zeta_3^2-70692\,\zeta_5-4384\,\zeta_2\,\zeta_3\\
&\,+101250\,\zeta_4-48960\,\zeta_3+6480\,\zeta_2 \Big)+\ord(\eps^2)\,.
\esp\eeq
\beq\bsp\label{eq:F7_result}
\cF_7(\eps)&\,=\frac{1}{\Phi_4^S(\eps)}\int \frac{d\Phi_4^S}{s_{15}s_{24}s_{34}s_{35}}\\
&\,= -\frac{3}{2}\,\frac{\Gamma (6-6 \epsilon) }{ \eps^5\, \Gamma (1-6 \epsilon )}\, _3F_2(1,1,-2 \epsilon ;1-2 \epsilon ,1-2 \epsilon ;1)\\
&\,=-\frac{360}{\eps^{5}}  + \frac{4932}{\eps^{4}} + \frac{1}{\eps^{3}} \Big( 720\,\zeta_2-24300 \Big) + \frac{1}{\eps^{2}} \Big( 4320\,\zeta_3-9864\,\zeta_2+55080 \Big) \\
&\,+ \frac{1}{\eps} \Big( 15120\,\zeta_4-59184\,\zeta_3+48600\,\zeta_2-58320 \Big) + 43200\,\zeta_5-207144\,\zeta_4\\
&\,+291600\,\zeta_3-110160\,\zeta_2+23328 + \eps \Big( 111600\,\zeta_6-591840\,\zeta_5+1020600\,\zeta_4\\
&\,-660960\,\zeta_3+116640\,\zeta_2 \Big)+\ord(\eps^2)\,.
\esp\eeq
\beq\bsp\label{eq:F8_result}
\cF_8(\eps)&\,=\frac{1}{\Phi_4^S(\eps)}\int \frac{d\Phi_4^S}{(s_{13}+s_{15})(s_{23}+s_{24})s_{34}s_{35}}\\
&\,= -\frac{60}{\eps^{5}} + \frac{822}{\eps^{4}} + \frac{1}{\eps^{3}} \Big( 240\,\zeta_2-4050 \Big) + \frac{1}{\eps^{2}} \Big( 2400\,\zeta_3-3288\,\zeta_2+9180 \Big)\\
&\, + \frac{1}{\eps} \Big( 13320\,\zeta_4-32880\,\zeta_3+16200\,\zeta_2-9720 \Big) + 51840\,\zeta_5+3360\,\zeta_2\,\zeta_3\\
&\,-182484\,\zeta_4+162000\,\zeta_3-36720\,\zeta_2+3888 + \eps \Big( 207600\,\zeta_6+11760\,\zeta_3^2\\
&\,-710208\,\zeta_5-46032\,\zeta_2\,\zeta_3+899100\,\zeta_4-367200\,\zeta_3+38880\,\zeta_2 \Big)+\ord(\eps^2)\,.
\esp\eeq
\beq\bsp\label{eq:F9_result}
\cF_9(\eps)&\,=\frac{1}{\Phi_4^S(\eps)}\int \frac{d\Phi_4^S}{s_{15}(s_{14}+s_{15})s_{23}s_{34}s_{345}} \\
&\,= \frac{160}{\eps^{5}}  - \frac{1712}{\eps^{4}} + \frac{1}{\eps^{3}} \Big( -120\,\zeta_2+2784 \Big) + \frac{1}{\eps^{2}} \Big( -120\,\zeta_3+1284\,\zeta_2+31968 \Big) \\
&\,+ \frac{1}{\eps} \Big( 2520\,\zeta_4+1284\,\zeta_3-2088\,\zeta_2-216864 \Big) + 15720\,\zeta_5+1920\,\zeta_2\,\zeta_3\\
&\,-26964\,\zeta_4-2088\,\zeta_3-23976\,\zeta_2+795744 + \eps \Big( 82520\,\zeta_6+9600\,\zeta_3^2\\
&\,-168204\,\zeta_5-20544\,\zeta_2\,\zeta_3+43848\,\zeta_4-23976\,\zeta_3+162648\,\zeta_2-2449440 \Big)\\
&\,+\ord(\eps^2)\,.
\esp\eeq
\beq\bsp\label{eq:F10_result}
\cF_{10}(\eps)&\,=\frac{1}{\Phi_4^S(\eps)}\int \frac{d\Phi_4^S}{(s_{23}+s_{24})(s_{24}+s_{25})s_{34}s_{45}} \\
&\,=-\frac{120}{\epsilon ^4}+\frac{2004}{\epsilon ^3}-\frac{14112}{\epsilon ^2}+\frac{1}{\epsilon }\Big(240\, \zeta _3+60696\Big)+1980\, \zeta _4-4008\, \zeta _3 \\
&\,-201528 
+\epsilon\Big(6960\, \zeta _5+1680\, \zeta _2\, \zeta _3-33066\, \zeta _4+28224\, \zeta _3+612360\Big)\\
&\,+\eps^2\Big(32700\, \zeta _6+6840\, \zeta _3^2-116232\, \zeta _5
-28056\, \zeta _2\, \zeta _3+232848\, \zeta _4-121392\, \zeta _3\\
&\,-1837080\Big)+\ord(\eps^3)\,.
\esp\eeq
\section{The threshold approximation to the partonic $2\rightarrow{}H+3$ parton cross-section}
\label{sec:amplitude}

\subsection{General setup}
The production of a Higgs boson in the collision of two hadrons $h_1,\, h_2$ is dominated by QCD processes.
The hadronic cross-section is related to the partonic cross-section by the general factorisation formula
\beq
\sigma_{h_1+h_2\rightarrow H+X}=\sum\limits_{i,j}\int\limits_0^1dx_1 dx_2 f_i^{h_1}(x_1)f_j^{h_2}(x_2) \sigma_{i+j\rightarrow H}(M^2,x_1x_2 S).
\eeq
Here $f_i^h(x)$ are the parton-distribution functions for the parton $i$ inside of the hadron $h$, $\sigma_{i+j\rightarrow H}$ is the partonic cross-section and ${S}$ is the square of the total centre-of-mass energy of the hadronic system. The centre-of-mass energy squared of the partonic system is consequently given by $s=x_i x_j S$.
The partonic cross-section is expanded perturbatively in the strong coupling constant $\alpha_S$.
The leading QCD contribution arises in the  SM via a top-quark loop at $\mathcal{O}(\alpha_S^2)$.

We consider the Higgs boson to be relatively light compared to the top quark. This justifies to work in the limit of an infinite top quark mass and consider $N_f$ light quarks.  
We describe the interaction of the Higgs boson with gluons by introducing the effective Lagrangian
\beq
\mathcal{L}_{eff}=-\frac{1}{4}c_HG_{\mu\nu}^aG^{a\mu\nu} H.
\eeq
Here $G^a_{\mu\nu}$ denotes the gluon field strength tensor and $H$ is the Higgs field. The Wilson coefficient $c_H$ can be found, e.g. in ref.~\cite{Chetyrkin:2005ia,Schroder:2005hy,Furlan:2011uq}.
The next-to-next-to-leading (NNLO) order correction to the inclusive Higgs cross-section has been computed in the past by employing this effective theory. 
In this work we present a part of the next term in in the perturbative series (\n3lo) in the effective theory.

At every order in perturbative QCD, the cross-section receives contributions from various real and virtual radiation processes. 
We consider only tree-level processes with three real-emission partons in the final state.
A direct integration over the phase-space of the corresponding matrix elements is challenging. 
As we have discussed in the introduction, we pave the way towards the full computation 
by performing an expansion of the phase-space integrals around  the kinematic limit where  the 
Higgs boson is produced at threshold, $s\sim M^2$. In this limit,  all partons emitted in the final state are soft and their momenta vanish as $\bar z  \to 0$.
We expand the cross section in  the small parameter $\bar{z}$ defined in~\eqref{eq:zbar}. In the following, we shall present the leading and subleading terms in the threshold expansion.
\beq
\sigma_{i\,j\rightarrow H + X}(s,\bar{z})=\bar{z}^{-1-6\epsilon}s^{3\epsilon}\sum\limits_{k=0}^\infty \bar{z}^k \sigma^{S(k)}_{i\,j\rightarrow H + X}.
\eeq
Although we considered in our calculation the Higgs boson $H$ as a final state, we would like to stress the universality of our result for 
the leading term in the soft expansion for any other colorless final state produced by gluons in the initial state~\cite{Catani:1999ss}. 
The subleading term in the soft expansion is no longer universal.

\subsection{Calculation}
To obtain the real-emission cross-section we generate Feynman diagrams using QGRAF~\cite{Nogueira:1991ex} and compute squared matrix-elements using programs based on \textsc{GiNaC}~\cite{Bauer:2000cp} or \textsc{FORM}~\cite{FORM} and our own \textsc{C++} code for color and spin algebra. 
We perform our calculation in Feynman gauge with $D$ gluon polarizations in order to maintain a simple structure for the denominators of the squared amplitudes. To recover the 
result for physical gluon polarizations, we add matrix-elements with Faddeev-Popov ghosts as external states. 
We compare our results with a set of numeric cross-sections for different phase-space points obtained with {\sc MadGraph}~\cite{Alwall:2011uj} and find perfect agreement. 

In a next step we use reverse-unitarity and interpret the phase-space integrals as three-loop integrals as described in Section~\ref{sec:runitarity} and we expand them into a Laurent series in $\bar{z}$. The leading and next-to-leading terms in this series are then referred to as the soft and next-to-soft limit of the real emission cross-section. 
We then derive integration-by-parts (IBP) identities for the scalar soft phase-space integrals. We reduce all integrals to a set of 10 master integrals by employing the  Laporta algorithm~\cite{Laporta:2001dd}. We implemented this algorithm in a \textsc{C++} code that was developed by us specifically for this project, as well as the program AIR~\cite{Anastasiou:2004vj}. The calculation of the remaining 10 integrals is discussed in Section~\ref{sec:masters_computation}. 

\subsection{Leading soft contribution}
In this section we present the leading soft contributions to the $i\,j\rightarrow H+X$ real-emission cross-sections, where $X$ where represents the three final-state partons. To obtain a $\rm{\overline{MS}}$-renormalised quantity we redefine the strong coupling constant
\beq
\alpha_S=\alpha_S^R(\mu_R)e^{\epsilon \gamma_E}(4\pi \mu_R)^{-\epsilon},
\eeq
where $\gamma_E =\Gamma'(1)$ is the Euler-Mascheroni constant and $\mu_R$ is the renormalisation scale. In our calculation we consider $N_f$ massless quark flavours.
We include a flux factor of $\frac{1}{2s}$ and
average over incoming colours and spins, leading to a factor of $\frac{1}{2(1-\epsilon)(N_c^2-1)}$ for gluons and $\frac{1}{2N_c}$ for (anti-) quarks.
 We abbreviate $C_F=\frac{Nc^2-1}{2 N_c}$ and $C_A=N_c$. In the case that the phase-space integration contains $n$ identical particles we include a symmetry factor of $\frac{1}{n!}$.
For cross-sections with 4 (anti-) quarks we distinguish between the case of only one or two quark flavours. 
The two-flavour case is indicated by explicitly labeling the (anti-) quarks $q_i$ or $\bar{q}_i$. If the (anti-) quarks do not carry any label they are considered all to be of the same flavour. We present our results first in terms of the soft master integrals of Section~\ref{sec:masters_definition}. The resulting expressions are valid to all orders in the dimensional regulator $\epsilon$. Next, we insert the results for the soft master integrals and write the cross-section as a Laurent series in $\epsilon$ up to terms containing zeta values of weight six, keeping a factor corresponding to the volume of the soft phase-space $\Phi_4^S(\epsilon)$, eq.~\eqref{eq:soft_Phi_4}, unexpanded.

The cross-section containing only gluons is given by
\bea
\sigma^{S(0)}_{g\,g\rightarrow H+g\,g\,g}&=&\frac{1}{3!}\frac{1}{8(1-\epsilon)^2(N_c^2-1)^2}(4\pi\alpha_S)^3\Phi_4^S(\epsilon) C_A^4 C_F c_H^2  \\
\times\Bigg\{&&\mathcal{F}_1(\epsilon )32 \text{} \frac{1}{(3-2 \epsilon )^2 \epsilon ^5 \left(2 \epsilon ^2+\epsilon -1\right)}\nonumber\\
&&\times\left(33696 \epsilon ^{11}-270000 \epsilon ^{10}+1341936 \epsilon ^9-3650136 \epsilon ^8\right.\nonumber\\
&&+4913370 \epsilon ^7-2001211 \epsilon ^6-3045896 \epsilon ^5+5040807 \epsilon ^4\nonumber\\
&&\left.-3323131 \epsilon ^3+1144330 \epsilon ^2-197535 \epsilon +13050\right)\nonumber\\
&+&\mathcal{F}_2(\epsilon )\frac{64}{3} \text{} \frac{432 \epsilon ^7+288 \epsilon ^6-372 \epsilon ^5-188 \epsilon ^4+10 \epsilon ^3+317 \epsilon ^2-142 \epsilon +15}{\epsilon ^2 \left(4 \epsilon ^3-4 \epsilon ^2-5 \epsilon +3\right)}\nonumber\\
&-&\mathcal{F}_3(\epsilon )288 \text{} \left(6 \epsilon +\frac{1}{\epsilon }-7\right)\nonumber\\
&-&\mathcal{F}_5(\epsilon )32 \text{} \frac{12 \epsilon ^5+10 \epsilon ^4+4 \epsilon ^3-49 \epsilon ^2+26 \epsilon -3}{\epsilon ^2 (\epsilon +1) (2 \epsilon -3)}\nonumber\\
&-&\mathcal{F}_7(\epsilon )\frac{64}{3} \text{} (\epsilon -1)\nonumber\\
&-&\mathcal{F}_8(\epsilon )\frac{64}{3} \text{} (\epsilon -1)\nonumber\\
&+&\mathcal{F}_9(\epsilon )32 \text{} \frac{3 \epsilon ^2-2 \epsilon -1}{6 \epsilon +1}\Bigg\}\nonumber\\
&=&\frac{2^5}{3^4}  \frac{1}{3!}\frac{1}{8(N_c^2-1)^2}(4\pi\alpha_S)^3 \Phi_4^S(\epsilon) C_A^4 C_F c_H^2  \nonumber\\
\times\Bigg\{&&-\frac{218700}{\epsilon ^5}+\frac{2554740}{\epsilon ^4}+\frac{1}{\epsilon ^3}\big(131220 \zeta _2-9709605\big)\nonumber\\
&+&\frac{1}{\epsilon ^2}\big(782460 \zeta _3-1630854 \zeta _2+14950359\big)+\frac{1}{\epsilon }\big(2869830 \zeta _4-9687762 \zeta _3\nonumber\\
&+&+6810588 \zeta _2-8547924\big)+8373780 \zeta _5+301320 \zeta _2 \zeta _3-35377641 \zeta _4\nonumber\\
&+&40216932 \zeta _3-11741904 \zeta _2+107996+\epsilon\big(24995385 \zeta _6+763020 \zeta _3^2\nonumber\\
&-&103032486 \zeta _5-3541644 \zeta _2 \zeta _3+145858644 \zeta _4-68849712 \zeta _3\nonumber\\
&+&7687776 \zeta _2-455984\big)+\ord(\eps^2)\Bigg\}\nonumber.
\label{eq:softgluons}
\eea

Note that the only colour coefficient of the tree-level matrix element for a Higgs plus five gluons is $C_A^4C_F$ (see, e.g.,  ref.~\cite{Mangano:1990by}). Obviously this fact is left unchanged by the soft limit.

\newpage
In the case that two final-state partons are a quark anti-quark pair we sum over all possible quark flavours and obtain a factor $N_f$. The corresponding cross-section is then given by

\bea
\sigma^{S(0)}_{g\,g\rightarrow H+g\,q\,\bar{q}}&=&\frac{1}{8(1-\epsilon)^2(N_c^2-1)^2}(4\pi\alpha_S)^3 \Phi_4^S(\epsilon) C_A C_F c_H^2 N_f\\
 \times\Bigg\{ C_A^2 &\Big[&\mathcal{F}_1(\epsilon )32 \text{} \frac{1}{(3-2 \epsilon )^2 (\epsilon -1) \epsilon ^4 (\epsilon +1) (2 \epsilon -1)}\nonumber\\
 &&\times\left(288 \epsilon ^{11}-18288 \epsilon ^{10}+129712 \epsilon ^9-347128 \epsilon ^8+408738 \epsilon ^7-107919 \epsilon ^6\right.\nonumber\\
 &&\left.-271140 \epsilon ^5+359705 \epsilon ^4-210605 \epsilon ^3+66680 \epsilon ^2-10853 \epsilon +690\right)\nonumber\\
&+&\mathcal{F}_2(\epsilon )\frac{32}{3} \text{} \left(-\frac{108 \epsilon ^6+180 \epsilon ^5-201 \epsilon ^4-14 \epsilon ^3+67 \epsilon ^2-22 \epsilon +2}{-4 \epsilon ^5+8 \epsilon ^4+\epsilon ^3-8 \epsilon ^2+3 \epsilon }\right)\nonumber\\
&-&\mathcal{F}_5(\epsilon )32 \text{} \frac{\epsilon -6 \epsilon ^2}{-2 \epsilon ^2+\epsilon +3}\Big]\nonumber\\ 
-&&\mathcal{F}_1(\epsilon )64 \text{} \frac{72 \epsilon ^7-396 \epsilon ^6+982 \epsilon ^5-1377 \epsilon ^4+1134 \epsilon ^3-527 \epsilon ^2+122 \epsilon -10}{\epsilon ^4 (2 \epsilon -3)}
\Bigg\}\nonumber\\
&=& \frac{2^5}{3^7} \frac{1}{8(N_c^2-1)^2}(4\pi\alpha_S)^3 \Phi_4^S(\epsilon) C_A C_F c_H^2 N_f\nonumber\\
\times\Bigg\{&&\frac{153090}{\epsilon ^4}-\frac{1604043}{\epsilon ^3}+\frac{1}{\epsilon ^2}\big(-29160 \zeta _2+4903902\big)\nonumber\\
&+&\frac{1}{\epsilon }\big(-204120 \zeta _3+321732 \zeta _2-4833675\big)-874800 \zeta _4+2252124 \zeta _3-911088 \zeta _2\nonumber\\
&+&203535+\epsilon\big(-2711880 \zeta _5-233280 \zeta _2 \zeta _3+9651960 \zeta _4-6290136 \zeta _3-492210 \zeta _2\nonumber\\
&+&1667109\big)+\epsilon ^2\big(-9360360 \zeta _6-816480 \zeta _3^2+29921076 \zeta _5+2573856 \zeta _2 \zeta _3\nonumber\\
&-&26589060 \zeta _4-4323186 \zeta _3+4693212 \zeta _2+1294731\big) \nonumber\\
&+&2C_A C_F\Bigg[\frac{167670}{\epsilon ^4}-\frac{1743039}{\epsilon ^3}+\frac{1}{\epsilon ^2}\big(-29160 \zeta _2+5267592\big)+\frac{1}{\epsilon }\big(-204120 \zeta _3\nonumber\\
&+&321732 \zeta _2-5183163\big)-874800 \zeta _4+2252124 \zeta _3-911088 \zeta _2+337959\nonumber\\
&+&\epsilon\big(-2711880 \zeta _5-233280 \zeta _2 \zeta _3+9651960 \zeta _4-6290136 \zeta _3-492210 \zeta _2+1651749\big)\nonumber\\
&+&\epsilon ^2\big(-9360360 \zeta _6-816480 \zeta _3^2+29921076 \zeta _5+2573856 \zeta _2 \zeta _3-26589060 \zeta _4\nonumber\\
&-&4323186 \zeta _3+4693212 \zeta _2+1284491\big)\Bigg]+\ord(\eps^3)\Bigg\} \nonumber.
\eea

In the soft limit only the processes with gluons in the initial state contribute. All other contributions containing at least one (anti-) quark in the initial state vanish in this limit. This can be understood in terms of soft factorisation: there is no born-level process with a massless initial state fermion in Higgs production.
\newpage

\subsection{Next-to-leading soft contribution}
In this section we present our results for the next-to-leading term in the soft expansion of the real-emission cross sections. Only partonic subprocesses that have at least one gluon in the initial state give a non-zero contribution at this order. The results are

\bea
\sigma^{S(1)}_{g\,g\rightarrow H+g\,g\,g}&=&\frac{1}{3!}\frac{1}{8(1-\epsilon)^2(N_c^2-1)^2}(4\pi\alpha_S)^3 \Phi_4^S(\epsilon)C_A^4C_Fc_H^2 \\
\times\Bigg\{&-&\mathcal{F}_1(\epsilon )16 \text{} \frac{1}{(3-2 \epsilon )^2 \epsilon ^5 \left(6 \epsilon ^3+\epsilon ^2-4 \epsilon +1\right)}\nonumber\\
&&\times\left( 202176 \epsilon ^{13}-2464992 \epsilon ^{12}+12587184 \epsilon ^{11}-25936632 \epsilon ^{10}+9778008 \epsilon ^9 \right.\nonumber\\
&&\left.44147940 \epsilon ^8-70980864 \epsilon ^7+27845080 \epsilon ^6+29415875 \epsilon ^5-41617041 \epsilon ^4\right.\nonumber\\
&&\left.+22513771 \epsilon ^3-6360715 \epsilon ^2+915690 \epsilon -52200\right) \nonumber\\
&-&\mathcal{F}_2(\epsilon )\frac{64}{3} \text{} \frac{1}{\epsilon ^2 (\epsilon +1) (2 \epsilon -1) \left(6 \epsilon ^2-11 \epsilon +3\right)}\nonumber\\
&& \times\left(432 \epsilon ^9+4824 \epsilon ^8-1008 \epsilon ^7-5758 \epsilon ^6+2213 \epsilon ^5\right.\nonumber\\
&&\left.+968 \epsilon ^4+872 \epsilon ^3-1180 \epsilon ^2+347 \epsilon -30\right)\nonumber\\
&+&\mathcal{F}_3(\epsilon )192 \text{} \left(6 \epsilon ^2+11 \epsilon +\frac{3}{\epsilon }-20\right)\nonumber\\
&+&\mathcal{F}_5(\epsilon )64 \text{} \frac{72 \epsilon ^8-168 \epsilon ^7+74 \epsilon ^6+208 \epsilon ^5-132 \epsilon ^4-170 \epsilon ^3+151 \epsilon ^2-38 \epsilon +3}{\epsilon ^2 (\epsilon +1) \left(6 \epsilon ^2-11 \epsilon +3\right)}\nonumber\\
&+&\mathcal{F}_7(\epsilon )\frac{8}{3} \text{} \frac{24 \epsilon ^4+119 \epsilon ^3-242 \epsilon ^2+115 \epsilon -16}{(1-3 \epsilon )^2}\nonumber\\
&+&\mathcal{F}_8(\epsilon )\frac{8}{3} \text{} \frac{24 \epsilon ^4+119 \epsilon ^3-242 \epsilon ^2+115 \epsilon -16}{(1-3 \epsilon )^2}\nonumber\\
&-&\mathcal{F}_9(\epsilon )32 \text{} \frac{3 \epsilon ^4+16 \epsilon ^3-19 \epsilon ^2-2 \epsilon +2}{18 \epsilon ^2-3 \epsilon -1}\Bigg\}\nonumber\\
&=& \frac{2^5}{3^4} \frac{1}{3!}\frac{1}{8(N_c^2-1)^2}(4\pi\alpha_S)^3\Phi_4^S(\epsilon)C_A^4C_Fc_H^2\nonumber\\
\Bigg\{&&\frac{437400}{\epsilon ^5}-\frac{4934520}{\epsilon ^4}+\frac{1}{\epsilon ^3}\big(-262440 \zeta _2+17287938\big)\nonumber\\
&+&\frac{1}{\epsilon ^2}\big(-1564920 \zeta _3+3242268 \zeta _2-21475314\big)+\frac{1}{\epsilon }\big(-5739660 \zeta _4\nonumber\\
&+&19317204 \zeta _3-13267368 \zeta _2+4182774\big)-16747560 \zeta _5-602640 \zeta _2 \zeta _3\nonumber\\
&+&70752852 \zeta _4-78931800 \zeta _3+21368412 \zeta _2+7340216+\epsilon\big(-49990770 \zeta _6\nonumber\\
&-&1526040 \zeta _3^2+206978652 \zeta _5+6976368 \zeta _2 \zeta _3-288414837 \zeta _4+126945108 \zeta _3\nonumber\\
&-&10614588 \zeta _2+1548816\big)+\ord(\eps^2)\Bigg\}\nonumber.
\eea
\newpage

\bea
\sigma^{S(1)}_{g\,g\rightarrow H+g\,q\,\bar{q}}&=&\frac{1}{8(1-\epsilon)^2(N_c^2-1)^2}(4\pi\alpha_S)^3\Phi_4^S(\epsilon) C_FC_A c_H^2 N_f\\
\times\Bigg\{C_A^2&\Big[-&\mathcal{F}_1(\epsilon )16 \text{} \frac{1}{(3-2 \epsilon )^2 (\epsilon -1) \epsilon ^4 (\epsilon +1) (2 \epsilon -1) (3 \epsilon -1)}\nonumber\\
&&\times\left(1728 \epsilon ^{13}-117792 \epsilon ^{12}+595824 \epsilon ^{11}-489880 \epsilon ^{10}-2308824 \epsilon ^9\right.\nonumber\\
&&+5971472 \epsilon ^8-5007932 \epsilon ^7-440720 \epsilon ^6+4236833 \epsilon ^5-3757178 \epsilon ^4\nonumber\\
&&\left.+1686250 \epsilon ^3-424272 \epsilon ^2+56321 \epsilon -3030\right)\nonumber\\
&+&\mathcal{F}_2(\epsilon )\frac{4}{9} \text{} \left(\frac{1}{-4 \epsilon ^5+8 \epsilon ^4+\epsilon ^3-8 \epsilon ^2+3 \epsilon }\right)\nonumber\\
&&\times\left(720 \epsilon ^7+11784 \epsilon ^6+5072 \epsilon ^5-16002 \epsilon ^4+3569 \epsilon ^3+3603 \epsilon ^2-1729 \epsilon +183\right)\nonumber\\
&+&\mathcal{F}_5(\epsilon )\frac{4}{3} \text{} \frac{144 \epsilon ^6+96 \epsilon ^5+640 \epsilon ^4-368 \epsilon ^3+199 \epsilon ^2-80 \epsilon +9}{6 \epsilon ^4-5 \epsilon ^3-8 \epsilon ^2+3 \epsilon }\nonumber\\
&+&\mathcal{F}_7(\epsilon )8 \text{} \frac{(\epsilon -1) \epsilon }{9-27 \epsilon }\nonumber\\
&+&\mathcal{F}_8(\epsilon )8 \text{} \frac{(\epsilon -1) \epsilon }{9-27 \epsilon }\nonumber\\
&+&\mathcal{F}_9(\epsilon )2 \text{} \frac{\epsilon  \left(-3 \epsilon ^2+2 \epsilon +1\right)}{-54 \epsilon ^2+9 \epsilon +3}\Big]\nonumber\\
+&\Big[&\mathcal{F}_1(\epsilon )32 \text{} \frac{1}{\epsilon ^4 (2 \epsilon -3) (2 \epsilon -1) (3 \epsilon -1)}\nonumber\\
&&\times\left(864 \epsilon ^{10}-2736 \epsilon ^9+192 \epsilon ^8+14320 \epsilon ^7-37150 \epsilon ^6+49403 \epsilon ^5\right.\nonumber\\
&&\left.-38954 \epsilon ^4+18385 \epsilon ^3-4992 \epsilon ^2+708 \epsilon -40\right)\nonumber\\
&+&\mathcal{F}_2(\epsilon )16 \text{} \frac{12 \epsilon ^3-8 \epsilon ^2-5 \epsilon +1}{6 \epsilon -3}\Big] \Bigg\}\nonumber\\
&=& \frac{2^3}{3^7} \frac{1}{8(N_c^2-1)^2}(4\pi\alpha_S)^3\Phi_4^S(\epsilon) C_FC_A c_H^2 N_f\nonumber\\
\times\Bigg\{&-&\frac{1224720}{\epsilon ^4}+\frac{12657384}{\epsilon ^3}+\frac{1}{\epsilon ^2}\big(58320 \zeta _2-35969184\big)+\frac{1}{\epsilon }\big(408240 \zeta _3\nonumber\\
&-&526824 \zeta _2+21935124\big)+1837080 \zeta _4-3658608 \zeta _3-1156032 \zeta _2+20102076\nonumber\\
&+&\epsilon\big(2361960 \zeta _5+2303640 \zeta _2 \zeta _3-16405416 \zeta _4-8880516 \zeta _3+21218868 \zeta _2\nonumber\\
&-&16893276\big)+\epsilon ^2\big(10114875 \zeta _6+8193960 \zeta _3^2-17837172 \zeta _5-22092588 \zeta _2 \zeta _3\nonumber\\
&-&39133206 \zeta _4+154780416 \zeta _3-51706116 \zeta _2-19146300\big)\nonumber\\
&+&2C_AC_F\Bigg[-\frac{1341360}{\epsilon ^4}+\frac{13827672}{\epsilon ^3}+\frac{1}{\epsilon ^2}\big(58320 \zeta _2-39026448\big)+\frac{1}{\epsilon }\big(408240 \zeta _3\nonumber\\
&-&439344 \zeta _2+23445720\big)+1837080 \zeta _4-3046248 \zeta _3-1654668 \zeta _2+19514088\nonumber\\
&+&\epsilon\big(2361960 \zeta _5+2303640 \zeta _2 \zeta _3-13781016 \zeta _4-12370968 \zeta _3+20772720 \zeta _2\nonumber\\
&-&17445648\big)+\epsilon ^2\big(10114875 \zeta _6+8193960 \zeta _3^2-9701532 \zeta _5-21392748 \zeta _2 \zeta _3\nonumber\\
&-&54092286 \zeta _4+151657380 \zeta _3-49947768 \zeta _2-18496864\big)\Bigg]+\ord(\eps^3)\Bigg\}\nonumber.
\eea

\newpage
\bea
\sigma^{S(1)}_{g\,q\rightarrow H+g\,g\,q }&=&\sigma^{S(1)}_{g\,\bar{q}\rightarrow H+g\,g\,\bar{q} }=\frac{1}{2!}\frac{1}{8(1-\epsilon)N_c(N_c^2-1)}(4\pi\alpha_S)^3\Phi_4^S(\epsilon) C_F  c_H^2(C_A-2C_F)\nonumber\\
\times\Bigg\{C_A^4&\Big[&\mathcal{F}_1(\epsilon )\frac{2}{3} \text{} \frac{1}{\epsilon ^5 \left(4 \epsilon ^2-8 \epsilon +3\right)}\\
&&\times\left(864 \epsilon ^{10}-28512 \epsilon ^9+240744 \epsilon ^8-952620 \epsilon ^7+2144508 \epsilon ^6\right.\nonumber\\
&&\left.-2951235 \epsilon ^5+2534353 \epsilon ^4-1341935 \epsilon ^3+417659 \epsilon ^2-68326 \epsilon +4380\right)\nonumber\\
&+&\mathcal{F}_2(\epsilon )\frac{2}{9} \text{} \frac{324 \epsilon ^5-72 \epsilon ^4-567 \epsilon ^3+617 \epsilon ^2-201 \epsilon +19}{\epsilon ^2 (2 \epsilon -1)}\nonumber\\
&+&\mathcal{F}_3(\epsilon )12 \text{} \left(6 \epsilon +\frac{1}{\epsilon }-7\right)\nonumber\\
&+&\mathcal{F}_5(\epsilon )\text{} \frac{24 \epsilon ^5-112 \epsilon ^4+162 \epsilon ^3-48 \epsilon ^2-32 \epsilon +6}{9 \epsilon ^2-6 \epsilon ^3}\nonumber\\
&+&\mathcal{F}_7(\epsilon )\frac{4}{9} \text{} (\epsilon -1)\nonumber\\
&+&\mathcal{F}_8(\epsilon )\frac{4}{9} \text{} (\epsilon -1)\nonumber\\
&+&\mathcal{F}_9(\epsilon )\text{} \frac{-3 \epsilon ^2+2 \epsilon +1}{6 \epsilon +1}\nonumber \Big]\\
+C_A^2&\Big[-&\mathcal{F}_1(\epsilon )\frac{2}{3} \text{} \frac{1}{\epsilon ^5 \left(4 \epsilon ^2-8 \epsilon +3\right)}\nonumber\\
&&\times\left(1296 \epsilon ^{10}-68256 \epsilon ^9+531432 \epsilon ^8-1871976 \epsilon ^7+3715893 \epsilon ^6\right.\nonumber\\
&&\left.-4517352 \epsilon ^5+3465908 \epsilon ^4-1668835 \epsilon ^3+482104 \epsilon ^2-74714 \epsilon +4620\right.\nonumber\\
&+&\mathcal{F}_2(\epsilon )\frac{2}{3} \text{} \frac{108 \epsilon ^5-129 \epsilon ^3+75 \epsilon ^2-15 \epsilon +1}{\epsilon ^2 (2 \epsilon -1)}\nonumber\\
&+&\mathcal{F}_{4}(\epsilon )\left(\epsilon -\frac{1}{3 \epsilon }-\frac{2}{3}\right)\Big]\nonumber\\
&+&\mathcal{F}_5(\epsilon )\frac{2}{3} \text{} \left(-6 \epsilon ^2-\frac{3}{\epsilon ^2}+43 \epsilon +\frac{27}{\epsilon }-61\right)\nonumber\\
&-&\mathcal{F}_7(\epsilon )\frac{4}{9} \text{} (\epsilon -1)\nonumber\\
&+&\mathcal{F}_9 (\epsilon )\text{} \frac{3 \epsilon ^2-2 \epsilon -1}{18 \epsilon +3}\nonumber\\
+&\Big[&\mathcal{F}_1(\epsilon )2 \text{} \frac{1}{\epsilon ^5 (2 \epsilon -1)}\nonumber\\
&&\times\left(72 \epsilon ^9-1332 \epsilon ^8-3014 \epsilon ^7+22217 \epsilon ^6-39799 \epsilon ^5\right.\nonumber\\
&&\left.+34159 \epsilon ^4-15710 \epsilon ^3+3841 \epsilon ^2-454 \epsilon +20\right)\nonumber\\
&+&\mathcal{F}_{4}(\epsilon )\left(\epsilon -\frac{1}{3 \epsilon }-\frac{2}{3}\right)\nonumber\Big]\Bigg\}\nonumber
\eea
\bea
&=&\frac{2^2}{3^6}\frac{1}{2!}\frac{1}{8N_c(N_c^2-1)}(4\pi\alpha_S)^3\Phi_4^S(\epsilon) C_F  c_H^2(C_A-2C_F)\nonumber\\
\times\Bigg\{&&\frac{277020}{\epsilon ^5}-\frac{3661524}{\epsilon ^4}+\frac{1}{\epsilon ^3}\big(-189540 \zeta _2+17139195\big)\nonumber\\
&+&\frac{1}{\epsilon ^2}\big(-947700 \zeta _3+2489778 \zeta _2-36505620\big)+\frac{1}{\epsilon }\big(-2850390 \zeta _4+12171870 \zeta _3\nonumber\\
&-&11487906 \zeta _2+36053523\big)-5846580 \zeta _5-554040 \zeta _2 \zeta _3+35305713 \zeta _4-54062316 \zeta _3\nonumber\\
&+&23579532 \zeta _2-12809922+\epsilon\big(-12501135 \zeta _6-1006020 \zeta _3^2+66995586 \zeta _5\nonumber\\
&+&6861348 \zeta _2 \zeta _3-146716434 \zeta _4+103965822 \zeta _3-20700180 \zeta _2-893224\big)\nonumber\\
&+&C_A C_F\Bigg[\frac{1370520}{\epsilon ^5}-\frac{17969364}{\epsilon ^4}+\frac{1}{\epsilon ^3}\big(-962280 \zeta _2+83256498\big)+\frac{1}{\epsilon ^2}\big(-5277960 \zeta _3\nonumber\\
&+&12711816 \zeta _2-175936887\big)+\frac{1}{\epsilon }\big(-17787600 \zeta _4+68755392 \zeta _3-59276286 \zeta _2\nonumber\\
&+&174582603\big)-45927000 \zeta _5-2391120 \zeta _2 \zeta _3+227399400 \zeta _4-313487658 \zeta _3\nonumber\\
&+&124360218 \zeta _2-65975079+\epsilon\big(-124170570 \zeta _6-5248800 \zeta _3^2+572721840 \zeta _5\nonumber\\
&+&29492424 \zeta _2 \zeta _3-1004568156 \zeta _4+635422050 \zeta _3-116229744 \zeta _2-466460\big)\Bigg] \nonumber \\
&+&6C_A^2C_F^2\Bigg[\frac{291600}{\epsilon ^5}-\frac{3785940}{\epsilon ^4}+\frac{1}{\epsilon ^3}\big(-194400 \zeta _2+17329869\big)+\frac{1}{\epsilon ^2}\big(-1108080 \zeta _3\nonumber\\
&+&2577420 \zeta _2-36309654\big)+\frac{1}{\epsilon }\big(-3883140 \zeta _4+14537556 \zeta _3-12100158 \zeta _2+36217701\big)\nonumber\\
&-&10711440 \zeta _5-427680 \zeta _2 \zeta _3+50265198 \zeta _4-67142142 \zeta _3+25733718 \zeta _2-14421072\nonumber\\
&+&\epsilon\big(-30383100 \zeta _6-1001160 \zeta _3^2+136655748 \zeta _5+5256576 \zeta _2 \zeta _3-227203596 \zeta _4\nonumber\\
&+&139522482 \zeta _3-24943128 \zeta _2+612040\big)\Bigg]+\ord(\eps^2)\Bigg\} \nonumber.
\eea

\newpage
\bea
\sigma^{S(1)}_{g\,q\rightarrow H+q\,q\,\bar{q}}&=&\sigma^{S(1)}_{g\,\bar{q}\rightarrow H+q\,\bar{q}\,\bar{q}}=\frac{1}{2!}\frac{1}{8(1-\epsilon)N_c(N_c^2-1)}(4\pi\alpha_S)^3\Phi_4^S(\epsilon) C_F c_H^2(C_A-2 C_F)\nonumber\\
\times\Bigg\{C_A^3 &\Big[-&\mathcal{F}_1(\epsilon )8 \text{} \frac{36 \epsilon ^7-1908 \epsilon ^6+7349 \epsilon ^5-11636 \epsilon ^4+9364 \epsilon ^3-3952 \epsilon ^2+807 \epsilon -60}{\epsilon ^4 (2 \epsilon -3)}\nonumber\\
&-&\mathcal{F}_5(\epsilon )\frac{8}{3} \text{} \frac{6 \epsilon ^3-13 \epsilon ^2+8 \epsilon -1}{\epsilon  (2 \epsilon -3)}\Big]\\
+C_A^2&\Big[-&\mathcal{F}_1(\epsilon )\frac{2}{3} \text{} \frac{1}{\epsilon ^5 (2 \epsilon -1)}\nonumber\\
&&\times\left(216 \epsilon ^9-1188 \epsilon ^8+6834 \epsilon ^7-17631 \epsilon ^6+22878 \epsilon ^5\right.\nonumber\\
&&\left.-17064 \epsilon ^4+7853 \epsilon ^3-2212 \epsilon ^2+334 \epsilon -20\right)\nonumber\\
&+&\mathcal{F}_2(\epsilon )\frac{2}{9} \text{} \frac{216 \epsilon ^5-306 \epsilon ^4+165 \epsilon ^3-110 \epsilon ^2+39 \epsilon -4}{\epsilon ^2 (2 \epsilon -1)}\nonumber\\
&+&\mathcal{F}_5(\epsilon )\frac{2}{3} \text{} \left(12 \epsilon ^2+\frac{2}{\epsilon ^2}-20 \epsilon -\frac{15}{\epsilon }+21\right)\nonumber\\
&-&\mathcal{F}_8(\epsilon )\frac{4}{9} \text{} (\epsilon -1)\nonumber\\
&+&\mathcal{F}_{10}(\epsilon )\left(\epsilon -\frac{1}{3 \epsilon }-\frac{2}{3}\right)\nonumber\Big]\\
+C_A&\Big[&\mathcal{F}_1(\epsilon )8 \text{} \frac{36 \epsilon ^7-288 \epsilon ^6+869 \epsilon ^5-1331 \epsilon ^4+1129 \epsilon ^3-527 \epsilon ^2+122 \epsilon -10}{\epsilon ^4 (2 \epsilon -3)}\Big]\nonumber\\
+&\Big[&\mathcal{F}_1(\epsilon )2 \text{} \frac{1}{\epsilon ^5 (2 \epsilon -1)}\nonumber\\
&&\times\left(72 \epsilon ^9-396 \epsilon ^8-1178 \epsilon ^7+7947 \epsilon ^6-15222 \epsilon ^5\right.\nonumber\\
&&\left.+14472 \epsilon ^4-7593 \epsilon ^3+2212 \epsilon ^2-334 \epsilon +20\right)\nonumber\\
&+&\mathcal{F}_2(\epsilon )2 \text{} \frac{18 \epsilon ^3-27 \epsilon ^2+10 \epsilon -1}{\epsilon  (2 \epsilon -1)}\nonumber\\
&+&\mathcal{F}_{10}(\epsilon )\left(\epsilon -\frac{1}{3 \epsilon }-\frac{2}{3}\right)\Big]\Bigg\}\nonumber
\label{eq:4quarks}
\eea

\newpage
\bea
&=&\frac{2}{3^6}\frac{1}{2!}\frac{1}{8N_c(N_c^2-1)}(4\pi\alpha_S)^3\Phi_4^S(\epsilon) C_F c_H^2\nonumber\\
\times\Bigg\{&&\frac{1}{\epsilon }\big(14580 \zeta _4-87480 \zeta _3\big)+174960 \zeta _5-58320 \zeta _2 \zeta _3\nonumber\\
&-&943326 \zeta _4+965196 \zeta _3+\epsilon\big(767880 \zeta _6-612360 \zeta _3^2-6770952 \zeta _5+973944 \zeta _2 \zeta _3\nonumber\\
&+&9188316 \zeta _4-3175524 \zeta _3\big)\nonumber\\
&+&4C_A\Bigg[-\frac{12150}{\epsilon ^4}+\frac{146205}{\epsilon ^3}+\frac{1}{\epsilon ^2}\big(9720 \zeta _2-580500\big)+\frac{1}{\epsilon }\big(77760 \zeta _3-116964 \zeta _2+936135\big)\nonumber\\
&+&374220 \zeta _4-935712 \zeta _3+464400 \zeta _2-556890+\epsilon\big(1389960 \zeta _5+58320 \zeta _2 \zeta _3-4503114 \zeta _4\nonumber\\
&+&3715200 \zeta _3-748908 \zeta _2+22400\big)\Bigg]\nonumber\\
&+&2C_AC_F\Bigg[\frac{1}{\epsilon ^2}\big(29160 \zeta _3-43740 \zeta _2+47385\big)+\frac{1}{\epsilon }\big(255150 \zeta _4-793152 \zeta _3+511758 \zeta _2\nonumber\\
&-&330237\big)+1020600 \zeta _5+145800 \zeta _2 \zeta _3-5551335 \zeta _4+6515802 \zeta _3-1928934 \zeta _2+309825\nonumber\\
&+&\epsilon\big(4740930 \zeta _6+218700 \zeta _3^2-22424040 \zeta _5-2172420 \zeta _2 \zeta _3+40779531 \zeta _4-21139542 \zeta _3\nonumber\\
&+&2834352 \zeta _2-174960\big)\Bigg]\nonumber\\
&+&24 C_A^2C_F\Bigg[-\frac{4860}{\epsilon ^4}+\frac{57267}{\epsilon ^3}+\frac{1}{\epsilon ^2}\big(3240 \zeta _2-221427\big)+\frac{1}{\epsilon }\big(25920 \zeta _3-38988 \zeta _2+351261\big)\nonumber\\
&+&124740 \zeta _4-311904 \zeta _3+154800 \zeta _2-209463+\epsilon\big(463320 \zeta _5+19440 \zeta _2 \zeta _3-1501038 \zeta _4\nonumber\\
&+&1238400 \zeta _3-249636 \zeta _2+11990\big)\Bigg]+\ord(\eps^2)\Bigg\}\nonumber.
\eea
\newpage
\bea
\sigma^{S(1)}_{g\,q_1\rightarrow H+q_1\,q_2\,\bar{q_2}}&=&\sigma^{S(1)}_{g\,\bar{q_1}\rightarrow H+q_2\,\bar{q_2}\,\bar{q_1}}=\frac{1}{8(1-\epsilon)N_c(N_c^2-1)}(4\pi\alpha_S)^3\Phi_4^S(\epsilon) C_F c_H^2(N_f-1)\\
\times\Bigg\{C_A^2&\Big[-&\mathcal{F}_1(\epsilon )4 \text{} \frac{36 \epsilon ^7-1908 \epsilon ^6+7349 \epsilon ^5-11636 \epsilon ^4+9364 \epsilon ^3-3952 \epsilon ^2+807 \epsilon -60}{\epsilon ^4 (2 \epsilon -3)}\nonumber\\
&-&\mathcal{F}_5(\epsilon )\frac{4}{3} \text{} \frac{6 \epsilon ^3-13 \epsilon ^2+8 \epsilon -1}{\epsilon  (2 \epsilon -3)}\Big]\nonumber\\
+&\Big[&\mathcal{F}_1(\epsilon )4 \text{} \frac{36 \epsilon ^7-288 \epsilon ^6+869 \epsilon ^5-1331 \epsilon ^4+1129 \epsilon ^3-527 \epsilon ^2+122 \epsilon -10}{\epsilon ^4 (2 \epsilon -3)}\Big]\Bigg\}\nonumber\\
&=& \frac{2^2}{3^7}\frac{1}{8N_c(N_c^2-1)}(4\pi\alpha_S)^3\Phi_4^S(\epsilon) C_F c_H^2(N_f-1)\nonumber\\
\times\Bigg\{&-&\frac{36450}{\epsilon ^4}+\frac{438615}{\epsilon ^3}+\frac{1}{\epsilon ^2}\big(29160 \zeta _2-1741500\big)\nonumber\\
&+&\frac{1}{\epsilon }\big(233280 \zeta _3-350892 \zeta _2+2808405\big)+1122660 \zeta _4-2807136 \zeta _3+1393200 \zeta _2\nonumber\\
&-&1670670+\epsilon\big(4169880 \zeta _5+174960 \zeta _2 \zeta _3-13509342 \zeta _4+11145600 \zeta _3\nonumber\\
&-&2246724 \zeta _2+67200\big)+\epsilon ^2\big(15578730 \zeta _6+699840 \zeta _3^2\nonumber\\
&-&50177556 \zeta _5-2105352 \zeta _2 \zeta _3+53638200 \zeta _4-17973792 \zeta _3+1336536 \zeta _2+44800\big)\nonumber\\
&+&6C_AC_F\Bigg[-\frac{14580}{\epsilon ^4}+\frac{171801}{\epsilon ^3}+\frac{1}{\epsilon ^2}\big(9720 \zeta _2-664281\big)+\frac{1}{\epsilon }\big(77760 \zeta _3-116964 \zeta _2\nonumber\\
&+&1053783\big)+374220 \zeta _4-935712 \zeta _3+464400 \zeta _2-628389+\epsilon\big(1389960 \zeta _5\nonumber\\
&+&58320 \zeta _2 \zeta _3-4503114 \zeta _4+3715200 \zeta _3-748908 \zeta _2+35970\big)\nonumber\\
&+&\epsilon ^2\big(5192910 \zeta _6+233280 \zeta _3^2-16725852 \zeta _5-701784 \zeta _2 \zeta _3\nonumber\\
&+&17879400 \zeta _4-5991264 \zeta _3+445512 \zeta _2+15232\big)\Bigg]+\ord(\eps^2)\Bigg\}\nonumber.
\eea


\newpage

\section{Analytic computation of the master integrals}
\label{sec:masters_computation}
In this section we present some details on how to evaluate analytically all the master integrals $\cF_i$ defined in Section~\ref{sec:masters_definition}. The analytic result for the soft phase space volume can easily be obtained from the expression for the phase space volume in general kinematics and will not be discussed here (see Appendix~\ref{sec:psv} for the derivation).

In general, our strategy is to follow the steps outlined in Section~\ref{sec:PS_soft}: we use the `energies and angles' parametrization to obtain a representation for the integral where the energies and angles appear in a factorized form. This may require the introduction of MB integrations in order to factorize sums of invariants in a denominator. We then integrate out the energies and angles to obtain a multifold MB representation for each master integral. Whenever we are able to do so, we evaluate the remaining MB integrals to all orders in $\eps$ in terms of hypergeometric functions that can easily be expanded into a Laurent series $\eps$ using the {\tt HypExp} package~\cite{Huber:2005yg}. In those cases where we did no manage to perform the MB integral in closed form for finite values of $\eps$, we only compute the Laurent expansion of the integral around $\eps=0$, e.g., by resolving singularities in $\eps$ and summing up harmonic sums or by converting the MB integral to a parametric integral which can be computed more easily. Details on how to perform these steps for the different master integrals will be given in the rest of this section.


\subsection{The master integral ${\mathcal{F}_3}$}
\label{sec:F3}
In this section we compute the master integral $\cF_3$ defined  by
\beq
\label{eq:integral_F3_again}
\Phi_4^S(\eps)\,\cF_3(\eps)=\int \frac{d\Phi_4^S}{s_{14}s_{23}s_{34}}\,.
\eeq
The integrand only involves two-particle invariants, so following the discussion in Section~\ref{sec:PS_soft} we can immediately insert the energies and angles parametrization and integrate out the energies in terms of $\Gamma$ functions. The remaining angular integrations can easily be carried out using the formulas given in Section~\ref{sec:PS_soft}. Note that in the present case all the angular integrals can be performed in closed form, so there is no need to introduce MB representations for the angular integrals. We obtain
\beq\bsp
\cF_3(\eps)&\,=2^{6 \epsilon -7} \pi ^{3 \epsilon -3}\frac{ \Gamma (6-6 \epsilon ) \Gamma (2-2 \epsilon ) \Gamma (1-2 \epsilon )^2}{\eps^2\, \Gamma (2-6 \epsilon ) \Gamma (1-\epsilon )^3}\int\frac{d\Omega_3^{(D-1)}\,d\Omega_4^{(D-1)}\,d\Omega_5^{(D-1)}}{(\beta_1\cdot\beta_4)\,(\beta_2\cdot\beta_3)\,(\beta_3\cdot\beta_4)}\\
&\,=2^{6 \epsilon -7} \pi ^{3 \epsilon -3}\frac{ \Gamma (6-6 \epsilon ) \Gamma (2-2 \epsilon ) \Gamma (1-2 \epsilon )^2}{ \eps^2\,\Gamma (2-6 \epsilon ) \Gamma (1-\epsilon )^3}\,\Omega_{3-2\eps}\,\int\frac{d\Omega_3^{(D-1)}}{(\beta_2\cdot\beta_3)}\,\Omega_{D-1}^{(1,1)}(\beta_1\cdot\beta_3)\\
&\,=-2^{2\epsilon -1}\,\frac{ \Gamma (6-6 \epsilon ) \Gamma (1-2 \epsilon ) }{\epsilon ^3\, \Gamma (2-6 \epsilon ) \Gamma (1-\epsilon )^2 }\int_{-1}^1\frac{d\cos\theta_3}{(1+\cos \theta _3)^{1+\eps}(1-\cos\theta_3)^{\eps}}\\
&\,\qquad\times {_2F_1}\left(1,1;1-\epsilon ;\frac{1+\cos\theta _3}{2} \right)\,.
\esp\eeq
The remaining integral can be brought into a more standard form by the change of variable $\cos\theta_3=2y-1$,
\beq\bsp
\int_{-1}^1&\frac{d\cos\theta_3}{(1+\cos \theta _3)^{1+\eps}(1-\cos\theta_3)^{\eps}}
\, _2F_1\left(1,1;1-\epsilon ;\frac{1+\cos\theta _3}{2} \right)\\
&\,\quad = 2^{-2\eps}\int_0^1dy \,y^{-1-\eps}\,(1-y)^{-\eps}\,_2F_1(1,1;1-\epsilon ;y)\,.
\esp\eeq
The integral over $y$ is now easily performed using the recursive definition of the hypergeometric function
\beq\bsp
{_{p+1}F_p}&(a_1,\ldots,a_{p+1};b_1,\ldots,b_{p};z) = \frac{\Gamma(b_p)}{\Gamma(a_{p+1})\Gamma(b_p-a_{p+1})}\\
&\times\int_0^1dt\,t^{a_{p+1}-1}\,(1-t)^{b_p-a_{p+1}-1}\,{_{p}F_{p-1}}(a_1,\ldots,a_{p};b_1,\ldots,b_{p-1};zt)\,.
\esp\eeq
We immediately get
\beq
\cF_3(\eps)=\frac{\Gamma (6-6 \epsilon )}{2 \epsilon ^4 \Gamma (2-6 \epsilon )}\,_3F_2(1,1,-\epsilon ;1-2 \epsilon ,1-\epsilon ;1)\,.
\eeq
The $_3F_2$ function can be expanded to the desired order in $\eps$ using the {\tt HypExp} package~\cite{Huber:2005yg}, and we arrive immediately at the Laurent series of eq.~\eqref{eq:F3_result}.


\subsection{The master integral $\mathcal{F}_2$}
The integrand of the master integral $\cF_2$ involves a sum of two-particle invariants in the denominator. We replace the sum by a product to the price of introducing a MB integration via eq.~\eqref{eq:mb_int},
\beq\bsp
\Phi_4^S(\eps)\,\cF_2(\eps) =\int \frac{d\Phi_4^S}{(s_{13}+s_{15})s_{34}}= \mbint\frac{dz_1}{2\pi i}\Gamma \left(-z_1\right) \Gamma \left(z_1+1\right)\int\frac{d\Phi_4^S}{s_{13}^{z_1+1} s_{15}^{-z_1}s_{34}}\,.
\esp\eeq
The phase space integral is now in the form~\eqref{eq:toy_integral}, and so we can introduce the energies and angles parametrization and integrate out all the energy and the angular variables. This results in the following two-fold integral representation for $\cF_2$, which is of mixed MB and Euler-type,
\beq\bsp
\cF_2(\eps)&\,= \frac{ \Gamma (6-6 \epsilon ) \Gamma (1-2 \epsilon )}{ \eps\,\Gamma (3-6 \epsilon ) \Gamma (1-\epsilon )^4}
\, 
\mbint\frac{dz_1}{2\pi i}\Gamma \left(-z_1\right) \Gamma \left(z_1+1\right) \Gamma \left(-\epsilon -z_1\right) \\
&\qquad\qquad\times\Gamma \left(1-\epsilon +z_1\right)
\int_{0}^1dy\,y^{-\eps}\,(1-y)^{-\eps}\, _2F_1\left(1,z_1+1;1-\epsilon ;y\right)
\,.
\esp\eeq
The Euler integral over $y$ could immediately be performed in terms of a $_3F_2$ function, but after that we still need to integrate over the MB parameter $z_1$. We therefore prefer not to perform the integration over $y$, but we rather insert the MB representation for the hypergeometric function in the integrand
\beq\bsp\label{eq:pFq_MB}
{_{p}F_q}&(a_1,\ldots,a_p;b_1,\ldots,b_q;x)\\
&\, = \mbint\frac{dz}{2\pi i}(-x)^{z}\,\Gamma(-z)\,\left[\prod_{i=1}^p\frac{\Gamma(a_i+z)}{\Gamma(a_i)}\right]\,\left[\prod_{i=1}^q\frac{\Gamma(b_i)}{\Gamma(b_i+z)}\right]\,.
\esp\eeq
The integral over $y$ evaluates to a Beta function, and we are left with the following two-dimensional MB integral
\beq\bsp
\cF_2(\eps) &\,=\frac{\Gamma (6-6 \epsilon ) \Gamma (1-2 \epsilon )}{ \eps\,\Gamma (3-6 \epsilon ) \Gamma (1-\epsilon )^2 }\mbint\frac{dz_1dz_2}{(2\pi i)^2}\,(-1)^{z_2}\, \Gamma \left(-z_1\right) \Gamma \left(-z_2\right) \\
&\,\times\frac{\Gamma \left(z_2+1\right) \Gamma \left(z_1+z_2+1\right) \Gamma \left(-\epsilon -z_1\right) \Gamma \left(1-\epsilon +z_1\right)}{\Gamma \left(2-2 \epsilon +z_2\right)}\,.
\esp\eeq
The integral over $z_1$ is easily performed using Barnes' first lemma, and the remaining one-fold MB integral can immediately be recognized as a $_3F_2$ function (see eq.~\eqref{eq:pFq_MB}). We finally obtain the following result for the master integral $\cF_2$, in agreement with eq.~\eqref{eq:F2_result},
\beq\bsp
\cF_2(\eps)&\,=\frac{\Gamma (6-6 \epsilon ) \Gamma (1-2 \epsilon )^2}{\epsilon\,  \Gamma (3-6 \epsilon ) \Gamma (2-2 \epsilon )^2}\, _3F_2(1,1,1-\epsilon ;2-2 \epsilon ,2-2 \epsilon ;1)\,.
\esp\eeq


\subsection{The master integral $\mathcal{F}_7$}

The integrand of $\cF_7$ only contains two-particle invariants,
\beq
\Phi_4^S(\eps)\,\cF_7(\eps)=\int \frac{d\Phi_4^S}{s_{15}s_{24}s_{34}s_{35}}\,.
\eeq 
We can therefore immediately integrate out the energy and the angular variables. We obtain
\beq\bsp
\cF_7(\eps) &\,= 3\frac{\Gamma (6-6 \epsilon ) \Gamma (1-2 \epsilon ) }{\eps^4\, \Gamma (1-\epsilon )^2 \Gamma (1-6 \epsilon )}
\int_0^1dy\,y^{-\epsilon }\,(1-y)^{-\eps}  \\
&\,\qquad\times {_2F_1}(1,1;1-\epsilon ;1-y) \, _2F_1(1,1;1-\epsilon ;y)\,.
\esp\eeq
In order to perform the integral over $y$, we introduce an MB representation for each $_2F_1$ function in the integrand and perform the $y$-integration. This leaves us with the following two-fold MB representation for $\cF_7$,
\beq\bsp
\cF_7(\eps) &\,= 3\frac{\Gamma (6-6 \epsilon ) \Gamma (1-2 \epsilon ) }{\eps^4\, \Gamma (1-\epsilon )^2 \Gamma (1-6 \epsilon )}\\
&\,\times
\mbint\frac{dz_1dz_2}{(2\pi i)^2}\,
(-1)^{z_1+z_2}
\frac{ \Gamma \left(-z_1\right) \Gamma \left(z_1+1\right)^2 \Gamma \left(-z_2\right) \Gamma \left(z_2+1\right)^2 }{ \Gamma \left(-2 \epsilon +z_1+z_2+2\right)}\,.
\esp\eeq
To proceed, we notice that one of the two integrations evaluates to a ${_2F_1}$, which can be reduced to $\Gamma$ functions using Gauss' identity,
\beq\bsp
\mbint\frac{dz_1}{2\pi i}\,
(-1)^{z_1}
\frac{ \Gamma \left(-z_1\right) \Gamma \left(z_1+1\right)^2 }{ \Gamma \left(-2 \epsilon +z_1+z_2+2\right)}&\, = \frac{1}{\Gamma(2-2\eps+z_2)}\,_2F_1(1,1;2-2\eps+z_2;1)\\
&\,=\frac{\Gamma(-2\eps+z_2)}{\Gamma(1-2\eps+z_2)^2}\,.
\esp\eeq
Inserting this result into the two-fold MB integral, we immediately see that the remaining MB integral evaluates to a $_3F_2$ function, and we get
\beq\bsp
\cF_7(\eps) &\,= -\frac{3}{2}\,\frac{\Gamma (6-6 \epsilon) }{ \eps^5\, \Gamma (1-6 \epsilon )}\, _3F_2(1,1,-2 \epsilon ;1-2 \epsilon ,1-2 \epsilon ;1)\,.
\esp\eeq


\subsection{The master integral $\mathcal{F}_{4}$}
The integral $\cF_{4}$ is defined by
\beq
\Phi_4^S(\eps)\,\cF_{4}(\eps)=\int \frac{d\Phi_4^S}{s_{13}s_{15}s_{34}s_{45}}\,.
\eeq
The integrand only contains two-particle invariants and we can immediately integrate out the energy and angular variables in the usual way. We obtain a one-fold MB representation,
\beq\bsp
\cF_{4}(\epsilon) &= \frac{\Gamma (6-6 \epsilon ) \Gamma (-2 \epsilon )}{\epsilon ^4 \Gamma (-6 \epsilon ) \Gamma (-\epsilon )^2} \mbint\frac{dz_1}{2\pi i}\Gamma \left(-z_1\right)\\
&\times\frac{\Gamma \left(z_1+1\right)^2 \Gamma \left(z_1-2 \epsilon \right) \Gamma \left(-z_1-\epsilon -1\right) \Gamma \left(z_1-\epsilon +1\right)}{\Gamma \left(z_1-2 \epsilon +1\right)^2}.
\esp\eeq
Closing the integration contour to the right and summing up residues at $z_1=n$ and $z_1=-1-\eps+n$, $n\in\mathbb{N}^{\times}$, we immediately see that $\cF_{4}$ can be expressed as a combination of hypergeometric functions,
\beq\bsp
\cF_{4}(\eps)&\,=-\frac{3\, \Gamma (6-6 \epsilon ) \Gamma (1-2 \epsilon ) }{2 \epsilon ^4 \Gamma (1-6 \epsilon )}\\
&\,\times\Bigg[\frac{3\, \Gamma (1-2 \epsilon ) \Gamma (\epsilon +1)}{(1+3 \epsilon) \Gamma (1-3 \epsilon )}\,{_3F_2}(-3 \epsilon -1,-2 \epsilon ,-\epsilon ;-3 \epsilon ,-3 \epsilon ;1)\\
&\,+\frac{1}{(1+\epsilon ) \Gamma (1-2 \epsilon )}\,{_4F_3}(1,1,1-\epsilon ,-2 \epsilon ;1-2 \epsilon ,1-2 \epsilon ,2+\epsilon;1)\Bigg]\,.
\esp\eeq


\subsection{The master integral $\mathcal{F}_6$}
The integral $\cF_6$ contains two sums in the denominator, which we can replace by products to the price of introducing two MB integrations,
\beq\bsp
\Phi&_4^S(\eps)\,\cF_6(\eps)=\int \frac{d\Phi_4^S}{(s_{13}+s_{14})(s_{14}+s_{15})s_{23}s_{34}}\\
&= \mbint\frac{dz_1dz_2}{(2\pi i)^2}\Gamma \left(-z_1\right) \Gamma \left(z_1+1\right) \Gamma \left(-z_2\right) \Gamma \left(z_2+1\right) \int\frac{d\Phi_4^S}{ s_{13}^{-z_1} s_{14}^{z_1+z_2+2} s_{15}^{-z_2}s_{23} s_{34}}\,.
\esp\eeq
We then proceed in the by now familiar way and integrate out the energies and the angles, and we arrive at the following two-fold MB representation for $\cF_6$,
\beq\bsp
\cF_6(\eps) &\,= \frac{\Gamma (6-6 \epsilon ) }{\eps\,
\Gamma (1-6 \epsilon ) \Gamma (1-\epsilon )^2}\,\mbint\frac{dz_1dz_2}{(2\pi i)^2}\,\Gamma \left(-z_1\right) \Gamma \left(z_1+1\right) \Gamma \left(-z_2\right)\Gamma \left(z_2+1\right) \\
&\,\qquad\times
\frac{ \Gamma \left(-\epsilon +z_1-z_2\right) \Gamma \left(z_2-\epsilon \right) \Gamma \left(-2 \epsilon -z_1+z_2\right) \Gamma \left(-\epsilon -z_1+z_2\right)}{\Gamma \left(-\epsilon +z_2+1\right) \Gamma \left(-2 \epsilon -z_1+z_2+1\right)}\,.
\esp\eeq
Unlike in the previous cases, we were not able to reduce this integral for generic $\eps$ to simple hypergeometric functions. We therefore only compute the Laurent expansion of the integral. We proceed in the standard way: we apply the packages {\tt MB}~\cite{Czakon:2005rk}, {\tt MBresolve}~\cite{Smirnov:2009up} and {\tt barnesroutines}~\cite{barnesroutines} to resolve singularities in $\eps$ and to expand the resulting integrals under the integration sign and apply Barnes' lemmas in an automated way. The resulting MB integrals are at most two-fold, and all of them can easily be done by closing the contours to the right and summing up residues in terms of nested harmonic sums defined recursively  by~\cite{Vermaseren:1998uu}
\beq
S_i(n) = \sum_{k=1}^n\frac{1}{k^i} {\rm~~and~~}S_{i\vec \jmath}(n) = \sum_{k=1}^n\frac{S_{\vec \jmath}(k)}{k^i}\,.
\eeq
Note that in the limit $n\to \infty$ harmonic sums immediately reduce to combinations of multiple zeta values.
The result for $\cF_6$ reads
\beq\bsp
\cF_6(\eps) &\,= \frac{10}{\eps^{5}} - \frac{137}{\eps^{4}} + \frac{1}{\eps^{3}} \Big( 40\,\zeta_2+675 \Big) + \frac{1}{\eps^{2}} \Big( 320\,\zeta_3-548\,\zeta_2-1530 \Big)\\
&\, + \frac{1}{\eps} \Big( 1500\,\zeta_4-4384\,\zeta_3+2700\,\zeta_2+1620 \Big) + 5160\,\zeta_5+320\,\zeta_2\,\zeta_3-20550\,\zeta_4\\
&\,+21600\,\zeta_3-6120\,\zeta_2-648 + \eps \Big( 18340\,\zeta_6+1280\,\zeta_3^2-70692\,\zeta_5-4384\,\zeta_2\,\zeta_3\\
&\,+101250\,\zeta_4-48960\,\zeta_3+6480\,\zeta_2 \Big)+\ord(\eps^2)\,.
\esp\eeq


\subsection{The master integral $\mathcal{F}_{10}$}
The integrand of $\cF_{10}$ involves two sums in the denominator, so we start by introducing two MB representations,
\beq\bsp
\Phi&_4^S(\eps)\,\cF_{10}(\eps) = \int \frac{d\Phi_4^S}{(s_{23}+s_{24})(s_{24}+s_{25})s_{34}s_{45}}\\
&=\mbint\frac{dz_1dz_2}{(2\pi i)^2}
\Gamma \left(-z_1\right) \Gamma \left(z_1+1\right) \Gamma \left(-z_2\right) \Gamma \left(z_2+1\right)\int\frac{d\Phi_4^S}{s_{23}^{-z_1} s_{24}^{2+z_1+z_2} s_{25}^{-z_2}s_{34} s_{45}}\,.
\esp\eeq
Integrating over the angles of the particles three and four yields two hypergeometric functions, and we introduce an MB representation for each of them. Performing the integration over the last angle, we obtain a four-fold MB representation for $\cF_{10}$. Two integrations can immediately be perfumed using Barnes' lemmas, and we obtain
\beq\bsp
&\cF_{10}(\eps)=
6\frac{\Gamma (6-6 \epsilon )}{\Gamma (1-\epsilon )^3\Gamma(1-6\eps)}
\mbint\frac{dz_2dz_3}{(2\pi i)^2}\Gamma \left(-z_2\right)\Gamma \left(-z_3\right) \Gamma \left(z_2+1\right)  \Gamma \left(z_3+1\right) \\
&\times\frac{ \Gamma \left(-2 \epsilon -z_2-1\right) \Gamma \left(-\epsilon +z_2+1\right) \Gamma \left(-2 \epsilon -z_3-1\right) \Gamma \left(-\epsilon -z_2-z_3-1\right)\Gamma \left(z_3-\epsilon \right) }{ \Gamma \left(-2 \epsilon -z_2\right) \Gamma \left(-2 \epsilon -z_3\right)}\,.
\esp\eeq
After resolving the singularities in $\eps$ and expanding under the integration sign, all the twofold integrals can be reduced to onefold integrals using Barnes' lemmas and their corollaries. The remaining onefold integrals are trivial to compute by closing the contour and summing up residues. We find,
\beq\bsp
\cF_{10}(\eps)&\,=-\frac{120}{\epsilon ^4}+\frac{2004}{\epsilon ^3}-\frac{14112}{\epsilon ^2}+\frac{1}{\epsilon }\Big(240\, \zeta _3+60696\Big)+1980\, \zeta _4-4008\, \zeta _3 \\
&\,-201528 
+\epsilon\Big(6960\, \zeta _5+1680\, \zeta _2\, \zeta _3-33066\, \zeta _4+28224\, \zeta _3+612360\Big)\\
&\,+\eps^2\Big(32700\, \zeta _6+6840\, \zeta _3^2-116232\, \zeta _5
-28056\, \zeta _2\, \zeta _3+232848\, \zeta _4-121392\, \zeta _3\\
&\,-1837080\Big)+\ord(\eps^3)\,.
\esp\eeq


\subsection{The master integral $\mathcal{F}_5$}
We start by replacing the sums in the denominator of the integrand of $\cF_5$ by three MB integrals,
\beq\bsp
\Phi_4^S(\eps)\,\cF_5(\eps)&\,=\int \frac{d\Phi_4^S}{(s_{14}+s_{15})s_{23}s_{345}}\\
&\,=\mbint\frac{dz_1dz_2dz_3}{(2\pi i)^3}\Gamma(-z_1)\Gamma(-z_2)\Gamma(-z_3)\Gamma(1+z_1+z_2)\Gamma(1+z_3)\\
&\,\times\int\frac{d\Phi_4^S}{s_{14}^{1+z_1}s_{15}^{-z_1}s_{23}s_{34}^{1+z_1+z_2}s_{45}^{-z1}s_{35}^{-z2}}\,.
\esp\eeq
We insert the energies and angles parametrization for the final-state particles and introduce MB integrations for the angular integrals. Note that the first angular integral necessarily involves three massless propagators and can thus not be done in closed form as a ${_2F_1}$, so we insert the three-fold MB representation~\eqref{eq:Omega_i_j_k} for it. We then arrive at a representation for $\cF_5$ as a six-fold MB integral convoluted with two angular integrations. It turns out that after performing the change of variables $z_3\to z_3+z_5$ and $z_6\to z_6+z_1+z_2$, the integrals over $z_1$, $z_2$ and $z_5$ can be done in closed form using Barnes' first lemma. 
The remaining two angular integrations can easily be performed in terms of hypergeometric functions, and all but one MB integration can be performed using Barnes' lemmas.
We thus arrive at a one-fold MB representation for $\cF_5$,
\beq\bsp
\cF_5(\eps)&\,=-\frac{ \Gamma (6-6 \epsilon ) \Gamma (1-3 \epsilon ) }{\eps\,\Gamma (2-6 \epsilon ) \Gamma (2-2 \epsilon ) \Gamma (1-\epsilon )}\\
&\,\times\mbint\frac{dz_1}{2\pi i}\frac{\Gamma \left(-z_1\right) \Gamma \left(z_1+1\right)^2\Gamma \left(-z_1-2 \epsilon \right) \Gamma \left(z_1-\epsilon +1\right)}{ \Gamma \left(z_1-3 \epsilon +2\right)}\,.
\esp\eeq
One might be tempted to sum up the residues at $z_1=n$ and $z_1=-2\eps+n$, $n\in\mathbb{N}^\times$, for finite values of $\eps$ to obtain an expression for $\cF_5$ as a combination of two hypergeometric functions at 1 valid to all orders in $\eps$. The two hypergeometric functions are however separately divergent (even for finite values of $\eps$) and only their sum is finite. We therefore only compute a Laurent series for $\cF_{5}$.
Resolving singularities in $\eps$ and summing up residues in terms of harmonic sums we obtain,
\beq\bsp
\cF_5(\eps) &\,= -\frac{120}{\eps} \,\zeta_2 -960\,\zeta_3+684\,\zeta_2 + \eps \Big( -4620\,\zeta_4+5472\,\zeta_3-1188\,\zeta_2 \Big)\\
&\, + \eps^{2} \Big( -17160\,\zeta_5-720\,\zeta_2\,\zeta_3+26334\,\zeta_4-9504\,\zeta_3+648\,\zeta_2 \Big)\\
&\, + \eps^{3} \Big( -64110\,\zeta_6-2880\,\zeta_3^2+97812\,\zeta_5+4104\,\zeta_2\,\zeta_3-45738\,\zeta_4+5184\,\zeta_3 \Big)\\
&\,+\ord(\eps^4)\,.
\esp\eeq

\subsection{The master integral $\mathcal{F}_8$}
The master integral $\cF_8$ is defined by
\beq
\Phi_4^S(\eps)\,\cF_8(\eps) = \int \frac{d\Phi_4^S}{(s_{13}+s_{15})(s_{23}+s_{24})s_{34}s_{35}}\,.
\eeq
We start by introducing two MB integrations in order to remove the sums in the denominator of the integral $\cF_8$. The energies are integrated out in terms of $\Gamma$ functions, and the angular integrations over the particles four and five are performed using eq.~\eqref{eq:Omega_j_k}. At this stage we have three integrations left to do: the two MB integrations and the integral over $\cos\theta_3=2y-1$
\beq\bsp\label{eq:F81}
\cF_8(\eps) &\,= -\frac{6\Gamma (6-6 \epsilon ) }{\eps\,\Gamma (1-\epsilon )^4 \Gamma (1-6 \epsilon )}\,\mbint\frac{dz_1dz_2}{(2\pi i)^2}\int_0^1dy\,y^{z_2-\eps}\,
(1-y)^{z_1-\epsilon } \,\Gamma \left(-z_1\right) \\
&\,\qquad\times \Gamma \left(z_1+1\right)\Gamma \left(-z_2\right) \Gamma \left(z_2+1\right)  \Gamma \left(-\epsilon -z_1\right) \Gamma \left(-\epsilon -z_2\right) \Gamma \left(-2 \epsilon +z_1+z_2\right) \\
&\,\qquad\times {_2F_1}\left(1,z_1+1;1-\epsilon ;y\right) \, _2F_1\left(1,z_2+1;1-\epsilon ;1-y\right)\,.
\esp\eeq
In order to proceed, we first apply the identity
\beq\label{eq:Pfaff}
_2F_1(a,b;c;x) = (1-x)^{-b}\,_2F_1\left(c-a,b;c;\frac{x}{x-1}\right)\,,
\eeq
and then insert an MB representation for each 
hypergeometric function in the integrand of eq.~\eqref{eq:F81}.
The reason to apply eq.~\eqref{eq:Pfaff} before inserting the MB integrations comes from the fact that in this way one of the four MB integrations can be performed using Barnes' first lemma.
We then arrive at the following three-fold MB representation for $\cF_8$,
\beq\bsp
\cF_8(\eps) &\,= -\frac{6\eps\,\Gamma (6-6 \epsilon )}{\Gamma (1-\epsilon )^4 \Gamma (1-6 \epsilon )}\,
\mbint\frac{dz_2dz_3dz_4}{(2\pi i)^3}\,\Gamma \left(-z_2\right) \Gamma \left(-z_3\right)\Gamma \left(-z_4\right)\\
&\,\times\Gamma \left(z_3+1\right) \Gamma \left(z_2-2 \epsilon \right) \Gamma \left(-z_2-z_4\right)  \Gamma \left(z_2+z_4+1\right) \Gamma \left(-\epsilon -z_3\right)\Gamma \left(z_3-\epsilon \right)\\
&\,\times\frac{  \Gamma \left(-2 \epsilon +z_2-z_3\right)  \Gamma \left(-\epsilon -z_4\right) \Gamma \left(z_4-\epsilon \right)}{\Gamma \left(-2 \epsilon +z_2+1\right) \Gamma \left(-2 \epsilon -z_3-z_4\right)}\,.
\esp\eeq
In the rest of this section we show how we can compute a Laurent expansion for this integral. We proceed in the standard way and resolve singularities in $\eps$. At the end of this procedure, we have a collection of MB integrals of dimensionality at most three with integration contours that are straight vertical lines. These integrals can then be safely expanded in $\eps$ under the integration sign.
In the following we discuss the computation of the two and three-fold integrals.

\paragraph{Three-fold MB integrals.}
There is one three-fold integral contributing to $\cF_8$,
\beq\bsp
\cF_{8,3}=&\,\mbint\frac{dz_2dz_3dz_4}{(2\pi i)^3}\Gamma \left(-z_2\right) \Gamma \left(z_2\right)\Gamma \left(-z_4\right)^2\Gamma \left(z_4\right)\Gamma \left(-z_2-z_4\right) \Gamma \left(-z_3-z_4\right)\\
&\,\qquad\times\frac{ \Gamma \left(1-z_3-z_4\right)   \Gamma \left(z_2+z_4+1\right) \Gamma \left(z_3+z_4\right)^2 \Gamma \left(z_2+z_3+z_4\right)}{\Gamma \left(z_2+1\right) \Gamma \left(z_3\right)}\,,
\esp\eeq
where we omit all $\Gamma$ function prefactors and
where the integration contours are straight vertical lines defined by
\beq
\mathrm{Re}(z_3) = 0.28 {\rm~~and~~} \mathrm{Re}(z_4) = 0.97 {\rm~~and~~}\mathrm{Re}(z_5) = -0.36\,.
\eeq
We start by closing the $z_3$ contour to the right and take residues,
\beq\bsp
\mbint&\frac{dz_2dz_4}{(2\pi i)^2}\,\frac{\Gamma \left(-z_2\right) \Gamma \left(z_2\right) \Gamma \left(-z_2-z_4\right) \Gamma \left(-z_4\right)^2 \Gamma \left(z_4\right) \Gamma \left(z_2+z_4+1\right)}{\Gamma \left(z_2+1\right)}\\
&\,\times\sum_{n=1}^\infty\Bigg\{
\frac{ \Gamma \left(n+z_2\right) }{n\,  \Gamma \left(n-z_4\right)}\psi\left(n+z_2\right)
-\frac{ \Gamma \left(n+z_2\right) }{n\,  \Gamma \left(n-z_4\right)}\psi\left(n-z_4\right)
-\frac{ \Gamma \left(n+z_2\right) }{n^2\,  \Gamma \left(n-z_4\right)}\Bigg\}\,.
\esp\eeq
The first two sums can be performed in terms of $\Gamma$ functions and their derivatives. We illustrate this on the first term (the second term is similar),
\beq\bsp
\sum_{n=1}^\infty&\,\frac{\Gamma \left(n+z_2\right) }{n\,  \Gamma \left(n-z_4\right)}\psi\left(n+z_2\right)
 = \lim_{\eta\to0}\frac{\partial}{\partial\eta}\sum_{n=1}^\infty\frac{\Gamma \left(n+z_2+\eta\right) }{n\,  \Gamma \left(n-z_4\right)}\\
&\, =\lim_{\eta\to0}\frac{\partial}{\partial\eta}\Bigg\{
 \frac{\Gamma \left(\eta +z_2\right)}{\Gamma \left(-z_4\right)} \,\left[\psi\left(-z_4\right)-\psi\left(-\eta -z_2-z_4\right)\right]\Bigg\}\\
 &\,=-\frac{\Gamma \left(z_2\right)}{\Gamma \left(-z_4\right)} \Big[\psi\left(z_2+1\right) \psi\left(-z_2-z_4\right)-\frac{1}{z_2}\psi\left(-z_2-z_4\right)-z_2 \psi\left(z_2+1\right) \psi\left(-z_4\right)\\
 &\,+\frac{1}{z_2}\psi\left(-z_4\right)-\psi'\left(-z_2-z_4\right)\Big]\,.
\esp \eeq
In this way, the first two terms can effectively be reduced to the computation of two-fold integrals, and we will therefore not discuss them any further in this section.

The third term can also be summed up in closed form. However, unlike the first two terms, the sum cannot be expressed in terms of $\Gamma$ functions and their derivatives alone, but the sum evaluates to a $_4F_3$ function. We therefore arrive at the following single three-fold MB integral,
\beq\bsp
\cF_{8,3} &\,= -\mbint\frac{dz_2dz_4}{(2\pi i)^2}
\frac{\Gamma \left(-z_2\right) \Gamma \left(z_2\right) \Gamma \left(-z_2-z_4\right) \Gamma \left(-z_4\right)^2 \Gamma \left(z_4\right)  }{\Gamma \left(1-z_4\right)}
\\&\,\qquad\times \,\Gamma \left(z_2+z_4+1\right)\,{_4F_3}\left(1,1,1,z_2+1;2,2,1-z_4;1\right)\,.
\esp\eeq
Although it looks like we have managed to reduce the three-fold integral to a two-fold integral, it is still secretly three-fold, except that we have `hidden' one integration inside the ${_4F_3}$ function. The advantage of this representation is that we can change the representation for the ${_4F_3}$ function in the integrand. More precisely, we perform the change of variables $z_4\to-z_4-z_2$ and chose the contours to be straight vertical lines given by
\beq
\mathrm{Re}(z_2) = \frac{1}{3}{\rm~~and~~}\mathrm{Re}(z_4) = \frac{1}{5}\,.
\eeq
We then insert an Euler integral representation for the ${_4F_3}$ function as an integral over a ${_3F_2}$. The ${_3F_2}$ function turns out to be reducible,
\beq
_3F_2\left(1,1,1;2,2;t\right) = \frac{\mathrm{Li}_2(t)}{t}\,,
\eeq
and so we finally arrive at
\beq\bsp
\cF_{8,3} &\,= -\mbint\frac{dz_2dz_4}{(2\pi i)^2}\int_0^1dt\,t^{z_2-1}\,(1-t)^{z_4-1}\,\text{Li}_2(t)\\
&\,\times
\frac{   \Gamma \left(-z_2\right) \Gamma \left(z_2\right) \Gamma \left(1-z_4\right) \Gamma \left(-z_2-z_4\right) \Gamma \left(z_2+z_4\right)^2}{\Gamma \left(z_2+1\right)}\,,
\esp\eeq
Next, we would like to exchange the MB and the Euler integration and sum up the residues of the poles of the $\Gamma$ functions in the integrand. However, we are only allowed to do so if the Euler integration does not produce any new poles whose residues need to be taken into account. It is easy to see that in our case the Euler integral converges whenever $\mathrm{Re}(z_2)$ and $\mathrm{Re}(z_4)$ are positive. We can thus close both contours to the right and exchange the Euler and MB integrations and then sum up the residues coming from the $\Gamma$ functions.

We start by taking residues in $z_2$ and then the residues in $z_4$. There are several cases to be considered separately:
\begin{enumerate}
\item The poles at $z_2=n_2\in\mathbb{N}^\times$ give rise to poles in $z_4$ at $n_4\in\mathbb{N}^\times$. We hence obtain double sums of the form
\beq\label{eq:binom_sum}
\sum_{n_2,n_4=1}^\infty\binom{n_2+n_4}{n_2}\,\frac{S_{\vec{\imath}_1}(n_2)}{n_2^{j_1}}\,\frac{S_{\vec{\imath}_2}(n_4)}{n_4^{j_2}}\,\frac{S_{\vec{\imath}_3}(n_2+n_4)}{(n_2+n_4)^{j_3}}\,t^{n_2}\,(1-t)^{n_4}\,.
\eeq
Sums of this type can be performed using {\tt XSummer}~\cite{Moch:2001zr,Moch:2005uc}, and give rise to complicated multiple polylogarithms whose arguments are rational functions of $t$ and $(1-t)$. Using symbols (see Appendix~\ref{app:parametric_integrals}), all these complicated multiple polylogarithms can be reduced to harmonic polylogarithms with indices 0 and 1 in $t$.
\item Taking the residues at the poles at $z_2=-z_4+n_2$, $n_2\in\mathbb{N}^\times$ gives rise to the expression
\beq\label{eq:F8_triple_residue}
-\int_0^1dt\,\text{Li}_2(t)\,\sum_{n_2=1}^\infty\,\mbint\frac{dz_4}{2\pi i}
\frac{z_4^2\, n_2!\,   \Gamma \left(-z_4\right)^2 \Gamma \left(z_4\right) }{n_2^2 \left(n_2-z_4\right)^2 \Gamma \left(n_2-z_4\right)}\,t^{n_2-z_4-1}\,(1-t)^{z_4-1}\,.
\eeq
Next we want to close the $z_4$ contour and sum up the corresponding residues. From the previous discussion, we are forced to close the contour to the right, and the summand obviously only has poles at $z_4=n_4\in\mathbb{N}^\times$ inside the integration contour. There is however a subtlety, and we cannot just sum up the residues. Indeed, it is easy to see that
\begin{itemize}
\item for $n_4<n_2$, there are double poles. Shifting the summation variable $n_2 \to n_2+n_4$, these residues give rise to double sums similar to eq.~\eqref{eq:binom_sum} and can again be performed using {\tt XSummer}. We obtain complicated multiple polylogarithms with rational functions of $t$ as argument. Using symbols, they can again be simplified to harmonic polylogarithms with indices 0 and 1 in $t$. 
\item for $n_4=n_2$, there is a simple pole. This gives rise to a simple sum which is trivial to perform in terms of harmonic polylogarithms in $t$.
\item for $n_4>n_2$, there is a simple pole. Shifting the summation variable $n_4\to n_2+n_4$ this gives rise to a single double sum $S(1-t,1-1/t)$, with
\beq
S(x,y) = -\sum_{n_2,n_4=0}^\infty\frac{n_2!\,n_4!}{(n_2+n_4+1)!}\,\frac{x^{n_2}}{n_2+1}\,\frac{y^{n_4}}{n_4+1}\,.
\eeq
This sum is not of the type~\eqref{eq:binom_sum}, and we therefore need a different way to sum up the series. This procedure will be discussed in the rest of this section.
\end{itemize}
\end{enumerate}
One way to sum the series $S(x,y)$ is to recognize that it is related to an integral over an Appell $F_3$ function. More precisely we can write
\beq\bsp
S(x,y)&\, = -\frac{1}{xy}\sum_{n_2,n_4=0}^\infty\frac{n_2!\,n_4!}{(n_2+n_4+1)!}\,\frac{x^{n_2+1}}{n_2+1}\,\frac{y^{n_4+1}}{n_4+1}\\
&\,=-\frac{1}{xy}\int_0^xd\xi\int_0^yd\chi\sum_{n_2,n_4=0}^\infty\frac{(n_2!)^2\,(n_4!)^2}{(n_2+n_4+1)!}\,\frac{\xi^{n_2}}{n_2!}\,\frac{\chi^{n_4}}{n_4!}\\
&\,=-\frac{1}{xy}\int_0^xd\xi\int_0^yd\chi\,F_3(1,1,1,1;2;\xi,\chi)\,.
\esp\eeq
The particular Appell $F_3$ function we obtain turns out to be reducible,
\beq
F_3(\alpha,\gamma-\alpha,\beta,\gamma-\beta;\gamma;\xi,\chi) = (1-\chi)^{\alpha+\beta-\gamma}\,{_2F_1}(\alpha,\beta;\gamma;\xi+\chi-\xi\chi)\,,
\eeq
and so we obtain a simple two-fold integral representation for $S$,
\beq\bsp
S(x,y)&\, = -\frac{1}{xy}\int_0^xd\xi\int_0^yd\chi\,_2F_1(1,1;2;\xi+\chi-\xi\chi)\\
&\,=\frac{1}{xy}\int_0^xd\xi\int_0^yd\chi\,\frac{\log(1-\xi)+\log(1-\chi)}{\xi+\chi-\xi\chi}\,.
\esp\eeq
This integral is trivial to perform in terms of multiple polylogarithms with rational functions in $x$ and $y$ as arguments (see Appendix~\ref{app:parametric_integrals}). 
It then follows that $S(1-t,1-1/t)$ can be expressed in terms of multiple polylogarithms with rational functions in $t$ as arguments. 
Using symbols, we arrive at the following simple expression,
\beq\bsp
S\left(1-t,1-\frac{1}{t}\right) = \frac{t}{(1-t)^2}\Big[&-4 \text{Li}_3(t)+2 \text{Li}_2(t) \log t\\
&+\frac{1}{6}\log ^3t+\frac{\pi^2}{3} \log  t+4 \zeta_3\Big]\,.
\esp\eeq
Putting everything together, we arrive at a representation for $\cF_{8,3}$ as a one-fold integral over harmonic polylogarithms. This integral is trivial to perform, and we obtain
\beq
\cF_{8,3} = \frac{23 \pi ^6}{22680}-2 \zeta_3^2\,.
\eeq

\paragraph{Two-fold MB integrals.}
There is only one two-fold MB integral contributing to $\cF_8$ that cannot be reduced to simpler integrals by Barnes' lemmas and their corollaries. This integral reads
\beq\bsp\label{eq:F_8_2}
\cF_{8,2}(\eps) &\,= \mbint\frac{dz_3dz_4}{(2\pi i)^2}
\frac{\Gamma \left(-z_3\right)^3 \Gamma \left(z_3\right) \Gamma \left(z_3+1\right) \Gamma \left(-z_4\right)^3 \Gamma \left(z_4\right) \Gamma \left(z_4+1\right) }{2 \epsilon  \Gamma \left(-z_3-z_4\right)}\\
&\,\qquad\times\Big[3 \epsilon  \psi\left(-z_3\right)+\epsilon  \psi\left(z_3\right)-2 \epsilon  \psi\left(-z_3-z_4\right)+\epsilon  \psi\left(-z_4\right)+\epsilon  \psi\left(z_4\right)-1\Big]\,,
\esp\eeq
where the integration contours are straight vertical lines defined by
\beq\label{eq:F_8_2_contour}
\mathrm{Re}(z_3)=-0.64 {\rm~~and~~}\mathrm{Re}(z_4) = -0.22\,.
\eeq
The integral could in principle be done by closing contours to the left and summing up residues. This leads to double sums of the form~\eqref{eq:binom_sum} -- thousands of them due to the presence of multiple poles --  but without any parametric dependence. 
We therefore took a different route, which we present in the following.

We start by noting that the \emph{integrand} of eq.~\eqref{eq:F_8_2} agrees, up to higher order terms of $\ord(\eps)$, with the function
\beq\label{eq:F_8_2_integrand}
-\frac{\Gamma \left(z_3+1\right) \Gamma \left(-z_4\right)^2 \Gamma \
\left(z_4+1\right) \Gamma \left(-\epsilon -z_3\right)^3 \Gamma \
\left(z_3-\epsilon \right) \Gamma \left(-\epsilon -z_4\right) \Gamma \
\left(z_4-\epsilon \right)}{2 \epsilon  \Gamma \left(-2 \epsilon \
-z_3-z_4\right)}\,.
\eeq
It would however be wrong to conclude that then necessarily the \emph{integrals}, seen as a Laurent expansion in $\eps$, are also equal to the same accuracy, because the Laurent expansion of eq.~\eqref{eq:F_8_2_integrand} around $\eps=0$ might require to shift the integration contours to avoid pinch singularities in the limit $\eps\to0$. It is however easy to see that, for the contours given in eq.~\eqref{eq:F_8_2_contour}, no new pinch singularity is created for $\eps\to0$ in eq.~\eqref{eq:F_8_2_integrand}. We therefore consider from now on $\eps$ to be infinitesimal but finite: it is large enough to separate poles at, e.g., $z_2=0$ from $z_2=\eps$, but small enough to ensure that no poles change their nature, i.e., poles that were left (right) of the contour~\eqref{eq:F_8_2_contour} in eq.~\eqref{eq:F_8_2} remain left (right) of the contour in eq.~\eqref{eq:F_8_2_integrand}. If we chose $\eps$ in this way, we conclude that
\beq
\cF_{8,2}(\eps) = \widetilde{\cF}_{8,2}(\eps)+\ord(\eps)\,,
\eeq
where $\widetilde{\cF}_{8,2}$ is given by eq.~\eqref{eq:F_8_2_integrand} integrated over the straight vertical lines defined by eq.~\eqref{eq:F_8_2_contour}.
Our aim will be to find an $\eps$ expansion for $\widetilde{\cF}_{8,2}$.

We perform the change of variables $(z_3,z_4)\to (-1-z_3,-z_4)$, and we close the $z_3$ contour to the right and take residues. There are two towers of poles we need to take into account:
\beq
z_3 = n_3\in\mathbb{N} {\rm~~and~~} z_3 = -1-\eps+n_3, \,n_3\in\mathbb{N}^\times\,.
\eeq
Two comments are in order:
\begin{enumerate}
\item At this stage our assumption that $\eps$ be infinitesimal but finite is vital, because otherwise the two towers of poles would glue together and thus give rise to double poles.
\item The second tower of poles runs over the set $\{-\eps,-\eps+1,\ldots\}$. For technical reasons that will become clear below, it is easier to explicitly take into account the residue at $z_3=-1-\eps$, and to compensate for this by adding it back,
\beq
\widetilde{\cF}_{8,2}(\eps) = \cR_{8,2}(\eps) + F_{8,2}(\eps)\,,
\eeq
where $F_{8,2}$ is the integral obtained from $\widetilde{\cF}_{8,2}$ by deforming the $z_3$ contour such that the pole at $z_3=-1-\eps$ is now to the right of the contour. The residue at $z_3=-1-\eps$ can be computed in closed form and gives rise to
\beq\bsp
\cR_{8,2}(\eps)&\,=-\frac{\Gamma (1-\epsilon )^2 \Gamma (1+\epsilon ) \Gamma (1-2 \epsilon )^3 }{16\eps^5\,(1+\eps)\,  \Gamma (1-3 \epsilon )}\,{_3F_2}(1,1,1-\epsilon ;1-3 \epsilon ,\epsilon +2;1)\\
&\,+\frac{\Gamma (1-2 \epsilon )^4 \Gamma (1-\epsilon )^2 \Gamma (1+\epsilon )^2}{8 \epsilon ^6\, \Gamma (1-4 \epsilon )}-\frac{3 \Gamma (1-2 \epsilon )^3 \Gamma (1-\epsilon )^3 \Gamma (1+\epsilon )^2}{16 \epsilon ^6\, \Gamma (1-3 \epsilon )}\,.
\esp\eeq
\end{enumerate}
Next we compute $F_{8,2}$ by closing the $z_3$ contour to the right and summing up residues. The resulting sums can easily be performed in terms of hypergeometric functions, and we get
\beq\bsp
F_{8,2}(\eps) &\,= -\mbint\frac{dz_4}{2\pi i}
\frac{\Gamma \left(1-z_4\right) \Gamma \left(z_4\right)^2 \Gamma \left(-\epsilon -z_4\right) \Gamma \left(z_4-\epsilon \right) }{2 \epsilon  \Gamma \left(z_4-3 \epsilon \right) \Gamma \left(-2 \epsilon +z_4+1\right)}\\
&\,\times\Big[\Gamma (-\epsilon -1) \Gamma (1-\epsilon )^3 \Gamma \left(z_4-3 \epsilon \right) \, _3F_2\left(1-\epsilon ,1-\epsilon ,1-\epsilon ;\epsilon +2,-2 \epsilon +z_4+1;1\right)\\
&\,\qquad+\Gamma (-2 \epsilon )^3 \Gamma (\epsilon +1) \Gamma \left(-2 \epsilon +z_4+1\right) \, _3F_2\left(-2 \epsilon ,-2 \epsilon ,-2 \epsilon ;-\epsilon ,z_4-3 \epsilon ;1\right)\Big]\,.
\esp\eeq
In order to proceed, we insert an Euler integral representation for each of the ${_3F_2}$ functions. It is then easy to see that the Euler integrals are convergent for $\mathrm{Re}(z_4)>0$ and $\eps$ infinitesimal but finite, and so we can exchange the Euler and MB integrations provided that we close the integration contour to the right. The important point is that the ${_2F_1}$ functions appearing inside the Euler integrals are independent of $z_4$, and so they can be pulled out of the MB integral. Summing up the residues in $z_4$, we then arrive at the following integral representation for $F_{8,2}$.
\begin{eqnarray}
F_{8,2}(\eps) &=& \frac{\Gamma(1-\eps)}{2\eps^3}\int_0^1dt\,t^{-\eps}(1-t)^{-\eps}\\
\nonumber&\times&\Bigg\{(1-t)^{-\epsilon }
 \frac{\Gamma (1-\epsilon )^4 \Gamma (1+\epsilon)}{\epsilon ^2 (1+\epsilon )(1-t)}\,{_2F_1}(1-\epsilon ,1-\epsilon ;\epsilon +2;t)\\
\nonumber&&\phantom{\Bigg\{}+t^{- \epsilon -1}\,(1-t)^{-1-\epsilon }\,\frac{\Gamma (1-2 \epsilon )^2 \Gamma (1-\epsilon ) \Gamma
(1+\epsilon )^2}{4 \epsilon^3}\, _2F_1(-2 \epsilon ,-2 \epsilon ;-\epsilon ;t)\\
\nonumber&&\phantom{\Bigg\{}- (1-t)^{-1-\epsilon } \frac{\Gamma (1-2 \epsilon )^2 \Gamma (1-\epsilon ) \Gamma
(1+\epsilon )^2}{4 \epsilon ^3\,t}\,_2F_1(-2 \epsilon ,-2 \epsilon ;-\epsilon ;t)\\
\nonumber&&\phantom{\Bigg\{}- 
\frac{\Gamma (1-2 \epsilon )^2 \Gamma (1+\epsilon )}{4 \epsilon (1+\epsilon )\,t}\, _2F_1(-2 \epsilon ,-2 \epsilon ;-\epsilon ;t) \,{_2F_1}(\epsilon +1,\epsilon +1;\epsilon +2;1-t)\\
\nonumber&&\phantom{\Bigg\{}-t^\eps\,(1-t)^{-\epsilon }\,\frac{\Gamma (1-\epsilon )^4 \Gamma (1+\epsilon)}{\epsilon ^2 (1+\epsilon )\,(1-t)} \, _2F_1(1-\epsilon ,1-\epsilon ;\epsilon +2;t)\\
\nonumber&&\phantom{\Bigg\{}-t^{\eps}
\frac{\Gamma (1-\epsilon )^3}{ (1+\epsilon)^2}\, _2F_1(1-\epsilon ,1-\epsilon ;\epsilon +2;t) \, {_2F_1}(\epsilon +1,\epsilon +1;\epsilon +2;1-t)
\Bigg\}\,.
\end{eqnarray}
The terms involving a single ${_2F_1}$ function in the integrand immediately evaluate to ${_3F_2}$ functions, which can be expanded in $\eps$ using {\tt HypExp}. In addition, the fourth term can be done in closed form as follows: We insert an MB representation for each ${_2F_1}$ and perform the Euler integration as a Beta function. One of the two remaining MB integrals evaluates to a ${_2F_1}$ evaluated at 1, which reduces to $\Gamma$ functions through Gauss' identity. The remaining one-fold MB integral then immediately evaluates to a ${_4F_3}$ function, which can be expanded in $\eps$ using {\tt HypExp}. We were not able to find a closed form for the last remaining Euler integral. We therefore insert an Euler integration for each ${_2F_1}$, and we obtain the expression,
\beq
-\frac{ \Gamma (1-\epsilon )^3 \Gamma (1+\epsilon )}{2 \epsilon^3\,  \Gamma (1+2 \epsilon )}\,\cI(\eps)\,,
\eeq
where $\cI$ denotes the integral
\beq
\cI(\eps)=\int_0^1dt\,du\,dv\,(1-t)^{-\epsilon } (1-u)^{2 \epsilon } u^{-\epsilon } v^{\epsilon } (1-t u)^{\epsilon -1} (1-(1-t) v)^{-\epsilon -1}\,.
\eeq
It is easy to see that $\cI$ is finite as $\eps\to0$, and so we can expand in $\eps$ under the integration sign and perform the integration over $t$, $u$ and $v$ recursively using the techniques of Appendix~\ref{app:parametric_integrals}. This is a trivial exercise that leads to
\beq
\cI(\eps) =2 \zeta_3-\frac{7 \pi ^4  }{180}\epsilon
+\left(25 \zeta_5-\frac{\pi ^2 }{2}\zeta_3\right) \epsilon ^2+
 \left(-6 \zeta_3^2-\frac{809 \pi ^6}{22680}\right) \epsilon ^3+\ord(\eps^4)\,.
 \eeq
 We have thus obtained the $\eps$ expansion of $F_{8,2}$, and thus of $\cF_{8,2}$. We find,
 \beq
 \cF_{8,2}(\eps) = \left(\frac{\pi ^2}{6} \zeta_3-\frac{9 }{2}\zeta_5\right)\frac{1}{\epsilon }-18 \gamma_E  \zeta_5+4 \zeta_3^2+\frac{2}{3} \gamma_E  \pi ^2 \zeta_3-\frac{817 \pi ^6}{45360}+\ord(\eps)\,,
 \eeq
where $\gamma_E=\Gamma'(1)$ denotes the Euler-Mascheroni constant.

\paragraph{The result for $\mathcal{F}_8$.}
We have now computed all the two and three-fold integrals contributing to $\cF_8$. The remaining one-fold integrals are trivial to compute and we obtain
\beq\bsp
\cF_8(\eps) &\,= -\frac{60}{\eps^{5}} + \frac{822}{\eps^{4}} + \frac{1}{\eps^{3}} \Big( 240\,\zeta_2-4050 \Big) + \frac{1}{\eps^{2}} \Big( 2400\,\zeta_3-3288\,\zeta_2+9180 \Big)\\
&\, + \frac{1}{\eps} \Big( 13320\,\zeta_4-32880\,\zeta_3+16200\,\zeta_2-9720 \Big) + 51840\,\zeta_5+3360\,\zeta_2\,\zeta_3\\
&\,-182484\,\zeta_4+162000\,\zeta_3-36720\,\zeta_2+3888 + \eps \Big( 207600\,\zeta_6+11760\,\zeta_3^2\\
&\,-710208\,\zeta_5-46032\,\zeta_2\,\zeta_3+899100\,\zeta_4-367200\,\zeta_3+38880\,\zeta_2 \Big)+\ord(\eps^2)\,.
\esp\eeq



\subsection{The master integral $\mathcal{F}_9$}
\label{sec:F9}
In this section we describe the computation of the most complicated master integral, the integral
\beq
\Phi_4^S(\eps)\,\cF_9(\eps)=\int \frac{d\Phi_4^S}{s_{15}(s_{14}+s_{15})s_{23}s_{34}s_{345}}\,.
\eeq
We start in the usual way and derive an MB representation for $\cF_9$ by inserting the energies and angles parametrisation and inserting MB integrations for the angular integrals. After applying Barnes' lemmas and their corollaries several times, we arrive at the following MB representation for $\cF_9$,
\beq
\cF_9(\eps) = \cF_{9,1}(\eps) + \cF_{9,2}(\eps)\,,
\eeq
where
\beq\bsp\label{eq:F91}
\cF_{9,1} &\,= \frac{3 (1+6 \epsilon) \Gamma (6-6 \epsilon ) \Gamma (1-4 \epsilon
)}{2 \epsilon  (1+4 \epsilon) \Gamma (1-6 \epsilon ) \Gamma(1-\epsilon )^3}
\mbint\frac{dz_1dz_2dz_3dz_4}{(2\pi i)^4}\, \Gamma \left(-z_2\right)\Gamma \left(-z_3\right)\\
&\,\times\Gamma \left(1-z_1-z_2\right) \Gamma \left(1-z_1-z_3\right)  \Gamma \left(z_1+z_2+z_3\right)\Gamma \left(-z_1-z_2-z_3-z_4\right)\\
&\,\times\Gamma \left(z_1+z_2+z_4+1\right) \Gamma \left(z_1+z_3+z_4+1\right)\Gamma \left(-z_1-z_2-2 \epsilon \right)\\
&\,\times\frac{      \Gamma \left(z_1-\epsilon -1\right) \Gamma \left(-z_1-z_4-\epsilon -1\right) \Gamma \left(z_4-\epsilon +1\right)}{\Gamma \left(1-z_2\right) \Gamma \left(1-z_3\right)  \Gamma \left(-z_1-z_2-4 \epsilon \right) \Gamma \left(z_4-2 \epsilon +1\right)}\,,
\esp\eeq
\beq\bsp
\label{eq:F92}
\cF_{9,2} &\,= -\frac{3 (1+6 \epsilon) \Gamma (6-6 \epsilon ) \Gamma (1-4 \epsilon
)}{2 \epsilon  (1+4 \epsilon) \Gamma (1-6 \epsilon ) \Gamma(1-\epsilon )^3}
\mbint\frac{dz_1dz_2dz_3dz_4}{(2\pi i)^4}\,\Gamma \left(1-z_1\right) \Gamma \left(-z_2\right)\\
&\,\times\Gamma \left(-z_3\right)\Gamma \left(-z_1-z_3+1\right)  \Gamma \left(z_1+z_2+z_3\right) \Gamma \left(-z_1-z_2-z_3-z_4\right) \\
&\,\times\Gamma \left(z_1+z_2+z_4+1\right)\Gamma \left(z_1+z_3+z_4+1\right)  \Gamma \left(-z_1-z_2-2 \epsilon \right) \\
&\,\times\frac{  \Gamma \left(z_1-\epsilon -1\right) \Gamma \left(-z_1-z_4-\epsilon -1\right) \Gamma \left(z_4-\epsilon +1\right)}{\Gamma \left(1-z_2\right) \Gamma \left(1-z_3\right)  \Gamma \left(-z_1-4 \epsilon \right) \Gamma \left(z_4-2 \epsilon +1\right)}\,.
\esp\eeq
We can resolve singularities for $\cF_{9,1}$ and $\cF_{9,2}$ and expand in $\eps$ up to and including $\ord(\eps)$. The result is a collection of high dimensional MB integrals which, after closing the contour and taking residues, result in multi-fold harmonic sums. While we were able to perform all the harmonic sums in terms of zeta values for all MB integrals up to $\ord(\eps^0)$, at $\ord(\eps)$ new polygamma functions appear in the integrand which make the combinatorics of the sums rather intricate. We therefore chose a different method to evaluate the integral $\cF_9$, which we describe in the rest of this section.

We start by noting that the $\cF_9$ is finite in $D=6$ dimensions. This can easily be checked by replacing $\eps$ by $\eps-1$ in the MB representations~\eqref{eq:F91} and~\eqref{eq:F92} and resolving singularities. Our goal is to find a parametric integral representation for $\cF_9$ in $D=6-2\eps$ dimensions and to expand under the integration and perform the parametric integrations recursively. The result in $D=6-2\eps$ can then be related to the (divergent) result in $D=4-2\eps$ using the dimensional recurrence relation for $\cF_9$ of Section~\ref{sec:masters_definition}.

 It is easy to derive a parametric representation for $\cF_9$ using the technique described in Appendix~\ref{app:MB_to_param}. We find
\beq
\cF_9(D=6-2\eps) = \frac{\Gamma (12-6 \epsilon ) \Gamma (3-3 \epsilon ) \Gamma (1-\epsilon )}{\Gamma (5-6 \epsilon ) \Gamma (2-\epsilon )^4}\Big[\cI_{9,1}(\eps) + \cI_{9,2}(\eps)\Big]\,,
\eeq
with
\begin{eqnarray}
\nonumber
\cI_{9,1}(\eps) &=& -\int_0^\infty dt_1\,dt_2\,\int_0^1dx_1\,dx_2\,dx_3\,t_1^{2-4 \epsilon } \left(1+t_1\right)^{\epsilon -1} t_2^{1-2 \epsilon } \\
\label{eq:I91}&\times& x_1^{-\epsilon }\left(1-x_1\right)^{2-4 \epsilon }  x_2^{1-3 \epsilon }\left(1-x_2\right)^{-\epsilon }  x_3^{-\epsilon } \left(1+t_2 x_3\right)^{1-3 \epsilon } \left(1+t_2 x_2 x_3\right)^{\epsilon } \\
\nonumber&\times&\left(t_1 t_2^2 x_1 x_2 x_3+t_2^2 x_2 x_3+t_1 t_2 x_1 x_2+t_1 t_2 x_3+t_2 x_2 x_3+t_2+t_1+1\right)^{3 \epsilon -3}\,,\\
\phantom{A}\nonumber&&\\
\nonumber
\cI_{9,2}(\eps) &=& \int_0^\infty dt_1\,dt_2\,\int_0^1dx_1\,dx_2\,dx_3\,
t_1^{2-4 \epsilon } \left(1+t_1\right)^{\epsilon -1} t_2^{1-2 \epsilon } \\
\label{eq:I92}&\times&x_1^{1-\epsilon }\left(1-x_1\right)^{2-4 \epsilon }x_2^{1-3 \epsilon }  \left(1-x_2\right)^{-\epsilon }  x_3^{-\epsilon } \left(1+t_2 x_3\right)^{1-3 \epsilon } \left(1+t_2 x_2 x_3\right)^{\epsilon }\\
\nonumber&\times&\left(t_1 t_2^2 x_1 x_2 x_3+t_2^2 x_1 x_2 x_3+t_2 x_1+t_1 t_2 x_1 x_2+t_1 t_2 x_3+t_2 x_1 x_2 x_3+t_1+x_1\right)^{3 \epsilon -3},
\end{eqnarray}
Several comments are in order about the parametric integrals we just defined. First, one can easily check that both $\cI_{9,1}$ and $\cI_{9,2}$ are individually finite as $\eps\to0$. Second, at first glance our goal to integrate out the integration variables one-by-one seems rather hopeless due to the appearance of the huge polynomial factor. However, as we will see shortly, there is a sufficient condition that allows one to test whether a parametric integral can be performed in terms of multiple polylogarithms, and this criterion is fulfilled for the integrands of $\cI_{9,1}$ and $\cI_{9,2}$. We very briefly summarize this criterion in the following, and we refer to ref.~\cite{Brown:2008um} or to Appendix~\ref{app:parametric_integrals} for more details. In order to understand the criterion, it is important to first understand multiple polylogarithms and their integration.

Multiple polylogarithms are generalizations of the ordinary logarithm and the \emph{classical} polylogarithms,
\beq\label{eq:classical_polylog}
\ln z = \int_1^z{d t\over t} {\rm~~and~~} \mathrm{Li}_n(z) = \int_0^z{d t\over t}\,\mathrm{Li}_{n-1}(t)\,,
\eeq
with $\mathrm{Li}_1(z) = -\ln(1-z)$. Multiple polylogarithms are defined recursively via the iterated integral~\cite{Goncharov:1998, Goncharov:2001}
 \beq\label{eq:Mult_PolyLog_def}
 G(a_1,\ldots,a_n;z)=\,\int_0^z\,{d t\over t-a_1}\,G(a_2,\ldots,a_n;t)\,,\\
\eeq
with $G(z)=1$ and where $a_i, z\in \mathbb{C}$. In the special case where all the $a_i$'s are zero, we define, using the obvious vector notation $\vec a_n=(\underbrace{a,\dots,a}_{n})$,
\beq
G(\vec 0_n;z) = {1\over n!}\,\ln^n z\,.
\eeq
In the special case where $a_i\in\{-1,0,1\}$, multiple polylogarithms are related to harmonic polylogarithms,
\beq
H(a_1,\ldots,a_n;z) = (-1)^p\,G(a_1,\ldots,a_n;z)\,,
\eeq
where $p$ denotes the number of elements in $(a_1,\ldots,a_n)$ equal to $+1$. 

Suppose now that we are given some integral over some rational function. We would like to be able to perform the integrations one-by-one in terms of multiple polylogarithms. Unfortunately, such a procedure is not possible for a generic integral. In ref.~\cite{Brown:2008um} a sufficient condition was derived that allows one to test whether such a recursive integration is possible. For a detailed description of the criterion we refer to ref.~\cite{Brown:2008um}, or to Appendix~\ref{app:parametric_integrals}, where (a version of) the criterion is reviewed. In a nutshell, as can easily be seen from the definition of multiple polylogarithms, a sufficient condition to perform the parametric integrations recursively using multiple polylogarithms is that at each integration step we can find an integration variable in which all the denominators appearing in the integrand are linear. More formally, let $S^{(i)}$ be the set of all polynomial factor in the integrand of $\cI_{9,i}$. Next, we define the sets $S^{(i)}_{(x,y,\ldots)}$, where $x,y,\ldots$ are integration variables, as the sets of irreducible non-monomial polynomial factors in the integrand after integration over $(x,y,\ldots)$ in this order. For more details how to construct these sets we refer to ref.~\cite{Brown:2008um} and to Appendix~\ref{app:parametric_integrals}. Here it suffices to say that the sets $S^{(i)}_{(x,y,\ldots)}$ can be constructed solely from the knowledge of the sets $S^{(i)}$. A sufficient criterion for a recursive integration in terms of multiple polylogarithms is that there is an ordering of the integration variables such that the sets $S^{(i)}_{(x,y,\ldots)}$ corresponding to this ordering contain at least one integration variable in which all the polynomials of the set are linear. 
Miraculously, this condition is satisfied for both $\cI_{9,1}$ and $\cI_{9,2}$. We have
\begin{eqnarray}
S^{(1)} &=& \{1+t_1,1-x_1,1-x_2,1+t_2 x_3,1+t_2 x_2 x_3,\\
\nonumber&&\phantom{\{}t_1 t_2^2 x_1 x_2 x_3+t_2^2 x_2 x_3+t_1 t_2 x_1 x_2+t_1 t_2 x_3+t_2 x_2 x_3+t_2+t_1+1\}\,,\\
S^{(2)} &=&\{1+t_1,1-x_1,1-x_2,1+t_2 x_3,1+t_2 x_2 x_3,\\
\nonumber&&\phantom{\{}t_1 t_2^2 x_1 x_2 x_3+t_2^2 x_1 x_2 x_3+t_2 x_1+t_1 t_2 x_1 x_2+t_1 t_2 x_3+t_2 x_1 x_2 x_3+t_1+x_1\}\,\\
S^{(1)}_{(t_1)}&=&\{1-x_1,1-x_2,1+t_2 x_1 x_2,1+t_2 x_3,1+t_2 x_2 x_3,\\
\nonumber&&\phantom{\{}-t_2 x_3 x_2+t_2 x_1 x_3 x_2+x_1 x_2-x_3 x_2+x_3-1\}\,,\\
S^{(2)}_{(t_1)}&=&\{1-x_1,1-x_2,1+t_2 x_1 x_2,1+t_2 x_3,1+t_2 x_2 x_3,\\
\nonumber&&\phantom{\{}t_2 x_1-t_2 x_2 x_1+t_2 x_2 x_3 x_1-t_2 x_3+x_1-1\}\,,\\
S^{(1)}_{(t_1,x_1)}&=&\{1-x_2,1+t_2 x_2,1-x_3,1+t_2 x_3,1+t_2 x_2 x_3,\\
\nonumber&&\phantom{\{}t_2 x_2 x_3+x_2 x_3-x_3+1\}\,,\\
S^{(2)}_{(t_1,x_1)}&=&\{1-x_2,1+t_2 x_2,1-x_3,1+t_2 x_3,1+t_2 x_2 x_3,\\
\nonumber&&\phantom{\{}-t_2 x_2+t_2 x_2 x_3+t_2+1\}\,,\\
S^{(1)}_{(t_1,x_1,x_2)}&=&\{1-x_3,1+t_2 x_3,2 t_2 x_3-t_2+x_3\}\,,\\
S^{(2)}_{(t_1,x_1,x_2)}&=&\{2+t_2-x_3,1-x_3,1+t_2 x_3\}\,,\\
S^{(1)}_{(t_1,x_1,x_2,x_3)}&=&\{1+t_2,1+2t_2\}\,,\\
S^{(2)}_{(t_1,x_1,x_2,x_3)}&=&\{1+t_2,2+t_2\}\,.
\end{eqnarray}
We see that if we perform the integration in the order $(t_1,x_1,x_2,x_3,t_2)$ then at each step all the polynomials are linear in the next integration variable. 
The actual integration can be carried out in an algorithmic way. The procedure is however rather lengthy, so we do not discuss it here in detail, but we refer to Appendix~\ref{app:parametric_integrals} for a detailed description of the integration algorithm. The result is
\beq\bsp
\cF&_{9}(D=6-2\eps) = 1663200 \zeta_3-554400 \pi ^2+3326400\\
&\,+120\eps\Bigg( +1309 \pi ^4-244203 \zeta_3+2861 \pi ^2+135294\Bigg)\\
&\,-2\eps^2\Bigg(-25779600 \zeta_5-970200 \pi ^2 \zeta_3+838657 \pi ^4-8149392 \zeta_3-201756 \pi ^2\\
&\,-31378284\Bigg)+\frac{4}{15} \eps^3\Bigg(960575 \pi ^6+180873000 \zeta_3^2-1978358850 \zeta_5\\
&\,-33612075 \pi ^2 \zeta_3+56663280 \zeta_3+3240501 \pi ^4+6836130 \pi ^2+810381510\Bigg)\\
&\,+\ord(\eps^4)\,.
\esp\eeq
Using the dimensional recurrence relations for $\cF_9$ derived in Section~\ref{sec:masters_definition} we then finally find the value of $\cF_9$ in $D=4-2\eps$ dimensions,
\beq\bsp
\cF_9(\eps) &\,= \frac{160}{\eps^{5}}  - \frac{1712}{\eps^{4}} + \frac{1}{\eps^{3}} \Big( -120\,\zeta_2+2784 \Big) + \frac{1}{\eps^{2}} \Big( -120\,\zeta_3+1284\,\zeta_2+31968 \Big) \\
&\,+ \frac{1}{\eps} \Big( 2520\,\zeta_4+1284\,\zeta_3-2088\,\zeta_2-216864 \Big) + 15720\,\zeta_5+1920\,\zeta_2\,\zeta_3\\
&\,-26964\,\zeta_4-2088\,\zeta_3-23976\,\zeta_2+795744 + \eps \Big( 82520\,\zeta_6+9600\,\zeta_3^2\\
&\,-168204\,\zeta_5-20544\,\zeta_2\,\zeta_3+43848\,\zeta_4-23976\,\zeta_3+162648\,\zeta_2-2449440 \Big)\\
&\,+\ord(\eps^2)\,.
\esp\eeq

\section{Conclusions and outlook}
\label{sec:conclusion}

In this article, we presented a method for the expansion of phase-space integrals in kinematic 
parameters. We applied our method to perform  a threshold expansion of triple-real radiation 
partonic cross-sections for Higgs production at hadron colliders. Our method reduces the 
phase-space integrals in the coefficients of the expansion to a small number of {\it soft} master 
integrals. These master integrals emerge not only in Higgs production processes but in all processes 
which are topologically equivalent, such as Drell-Yan production. We have computed the master integrals 
using a combination of techniques for Mellin-Barnes integration and the integration of real parametric integrals. 
We presented explicit expressions for the first two coefficients in the threshold expansion of all triple-real radiation 
partonic cross-sections which are necessary for the inclusive Higgs cross-section at \n3lo, both in terms of {\it soft} 
master integrals and as a Laurent series expansion in the dimensional regulator $\epsilon$.

We can apply our method to compute further coefficients in the  threshold expansion. This requires  the reduction of higher rank 
cut loop integrals in their powers of propagators and irreducible numerators. It also requires the evaluation of some new master integrals 
originating from diagrams which did not contribute to the first two terms of the threshold expansion. We believe that it is feasible to 
compute in this way a sufficient number of terms in the threshold expansion for phenomenology purposes.  
A second possibility is to apply the reverse-unitarity method to compute the same cross-sections without resorting to a threshold expansion. 
The  work that we have presented here will be particularly important for solving the unexpanded master integrals.  With reverse unitarity, one can 
derive differential equations for the master integrals in arbitrary kinematics. For their solution, we can use the soft limits presented here 
as boundary conditions. Subleading terms in the expansion can serve as a check of the solutions or as an aid to guess them.

A complete computation of the \n3lo coefficient requires also virtual contributions integrated over the phase space of up to two  
real partons in the final state.  The corresponding one-loop, two-loop and three-loop amplitudes are already known in the literature
as a Laurent expansion in $\epsilon$~\cite{Badger:2006us,Berger:2006sh,Badger:2007si,Glover:2008ffa,Dixon:2009uk,Badger:2009vh,Badger:2009hw,Gehrmann:2011aa,Baikov:2009bg,Lee:2010cga,Gehrmann:2010ue}. Integrating them over phase-space requires in addition their universal single-real and 
double-real infrared limits at higher orders in $\epsilon$.  A threshold expansion of mixed real and virtual corrections requires an extension of 
our method.  
While the threshold expansion for real emission matrix elements can be performed by uniformly rescaling all final-state parton momenta, $p_i\rightarrow \bar{z} p_i$, and expanding in the small parameter $\bar{z}$,  for real-virtual matrix-elements the scaling of the loop momentum cannot be determined without further consideration. To determine the correct expansion at threshold we consider the strategy of regions~\cite{Beneke:1997zp,smirnovBook}. We applied this method to all 2-loop real-virtual master integrals contributing to Higgs production at NNLO and could reproduce the soft limits of the integrals. In the following we will illustrate this technique on the example of the integral
\vspace{-.5cm}
\beq
\mathcal{I}=\YD{27} =\int d\Phi_2 \frac{1}{(p_3+p_{2})^2} \int d^Dk \frac{1}{k^2(k+p_1)^2(k-p_{3})^2(k+p_{12})^2}.
\vspace{.5cm}
\eeq
We introduce the light-cone coordinates, e.g. for $q^{\mu}$,
\beq
q_{\pm}= \frac{1}{\sqrt{2}}\left(q^0 \pm q^1\right),
\eeq
and $q_{\perp}^{\mu}$ denotes the remaining $(D-2)$ Euclidian transverse components.
Furthermore we choose our reference frame such that
\beq\bsp
p_{1,+} &= \frac{1}{\sqrt{2}}, \ \ p_{1,-} = 0, \ \ p_{1,\perp} = 0, \\
p_{2,+} &= 0, \ \ p_{2,-} = \frac{1}{\sqrt{2}}, \ \ p_{1,\perp} = 0.
\esp\eeq
We make the soft region manifest by redefining the loop momentum and the final state parton momentum in terms of the soft-expansion parameter $\bar{z}$
 \beq
 k_-\to\bar{z}\,k_-,\hspace{.5cm}k_\perp\to\sqrt{\bar{z}}\,k_\perp, \hspace{.5cm} p_3\to\bar{z}\,p_3.
\eeq
After expanding in $\bar{z}$ we find the leading soft integral to be

\bea
\mathcal{I}^S&=&\bar{z}^{-1-3 \epsilon }\int d\Phi_2^S \frac{1}{ p_{3+}} \int d^Dk \frac{1}{k^2(k^2+k_+)(k^2+k_+p_{3-})(k_++1)}\nonumber.
\eea
All external kinematic scales are factored out of the integrand and the integral can be easily computed using Schwinger parameters. This yields
\bea
\mathcal{I}^S&=&-3\ 2^{2 \epsilon -3} \pi ^{-1+\epsilon}  \bar{z}^{-1-3 \epsilon } \frac{\Gamma (1-\epsilon )^2 \Gamma (\epsilon   +1)}{\epsilon^3\Gamma (1-3 \epsilon )},
\eea
in agreement with the result in ref.~\cite{Anastasiou:2012kq}. In full kinematics other regions, which are suppressed in the soft limit, contribute to the integral.

Finally, the computation of the collinear counterterms for the partonic cross-section 
has been  performed in refs.~\cite{Hoschele:2012xc,BuehlerPDF}.  Alternatively, this task can easily be achieved numerically following the procedure of ref.~\cite{Buehler:2012cu} 
and using the results for the partonic cross-sections at NNLO through $\ord(\eps)$ of refs.~\cite{Anastasiou:2012kq,Pak:2011hs}.

We believe that the prospects for a complete computation of the \n3lo coefficient for Higgs production and other processes are excellent, 
and we are looking forward to completing this task in future publications.


\section*{Acknowledgements}
The authors are grateful to F.~Herzog for stimulating discussions.
This work was supported by the ERC grant ``IterQCD'' and the Swiss National Science Foundation 
project $200021\_143781$.

\appendix
\section{Parametrisation of the $2\rightarrow H + (N-2)$ phase space}
\label{app:PSparam}
In this appendix we derive the parametrisation of the $2\rightarrow H + (N-2)$ phase space in terms of Lorentz invariant quantities.
The usual $D$-dimensional phase space measure is given in eq.~\eqref{eq:psm} and  we rescale the momenta as defined in eq.~\eqref{eq:mom_scaling} and find for $s=1$,
\beq\bsp
&d\Phi_{N-1}(p_H,p_3,\dots,p_N;z;D) \\&= \bar{z}^{(D-2)(N-2)}\frac{d^Dp_H}{(2\pi)^{D-1}}\delta_+\left(p_H^2-M^2\right)\prod_{i=3}^N\frac{d^Dp_i}{(2\pi)^{D-1}}\delta_+\left(p_i^2\right)(2\pi)^D\delta^{(D)}\left(p_{12}-p_H-p_{3\dots N}\right)\,,
\esp\eeq
for two incoming partons with momenta $p_1$ and $p_2$ producing a Higgs boson with momentum $p_H$ and $n$ additional partons with momenta $p_i$. 
We parametrise the final-state momenta as $p_i = \left(p_{i,0},p_{i,z},p_{i,\perp}\right)$, so that $p_{i,\perp}$ is the $(D-2)$-dimensional transverse component of the momentum. Then we can rewrite the integration over the momenta of the final-state partons as,
\beq
d^Dp_i\delta_+\left(p_i^2\right) = dp_{i,0}dp_{i,z}d^{D-2}p_{i,\perp}\Theta\left(p_{i,0}\right)\delta\left(p_{i,0}^2-p_{i,z}^2-p_{i,\perp}^2\right)\,,
\eeq
Next we choose a frame such that
\beq
p_1 = \frac{1}{2}\left(1,1,0,\dots,0\right) \ \ \text{and} \ \ p_2 = \frac{1}{2}\left(1,-1,0,\dots,0\right)\,.
\eeq
In this frame we can write the $t$-channel invariants ($i \geq 3$) as
\beq\bsp
s_{1i}&= \left(p_1-p_i\right)^2=-2p_1 p_i = -\left(p_{i,0}-p_{i,z}\right)\,,\\
s_{2i}&= \left(p_2-p_i\right)^2=-2p_2 p_i = -\left(p_{i,0}+p_{i,z}\right)\,,\\
\esp\eeq
or equivalenty
\beq
p_{i,0} = -\frac{1}{2}\left(s_{1i}+s_{2i}\right) \ \ \text{and} \ \ p_{i,z} = -\frac{1}{2}\left(s_{2i}-s_{1i}\right)\,,
\eeq
such that
\beq
p_{i,\perp}^2 = s_{1i}s_{2i}\,.
\eeq
The $s$-channel invariants ($i,j \geq 3$) can be written as
\beq
\label{eq:sci}
s_{i,j} =(s_{1i}s_{2j}+s_{1j}s_{2i})-2p_{i,\perp}\cdot p_{j,\perp}\,.
\eeq
This yields
\beq
d^Dp_i\delta_+\left(p_i^2\right) = \frac{1}{2}d^{D-2}p_{i,\perp}ds_{1i}ds_{2i}\Theta\left(-s_{1i}\right)\Theta\left(-s_{2i}\right)\delta\left(s_{1i}s_{2i}-p_{i,\perp}^2\right)\,.
\eeq
After eliminating the integration over the Higgs momentum using momentum conservation the measure becomes
\beq\bsp
d\Phi_{N-1}(D) &= \bar{z}^{(D-2)(N-2)-1}\prod_{i=3}^N\frac{d^{D-2}p_{i,\perp}}{(2\pi)^{D-1}}ds_{1i}ds_{2i}\Theta(-s_{1i})\Theta(-s_{2i})\delta(s_{1i}s_{2i}-p_{i,\perp}^2)\\&\times\frac{\pi}{2^{N-3}}\delta_+
\left(\left(p_{3\dots N}-p_{12}\right)^2-z p_{3\dots N}^2\right)\,.
\esp\eeq
It is well known that rotationally invariant integrals of the form
\beq
I = \int \prod_i d^Dp_i f(p_i\cdot p_j)
\eeq
can be rewritten by making the integration over the scalar products manifest:
\beq
I = \frac{1}{2^N}\int\prod_{i=1}^N\Omega_{D+1-i}\prod_{j=1}^i d(p_i\cdot p_j) \left[\textbf{Gram}(p_1,\dots,p_N)\right]^{\frac{D-N-1}{2}}\Theta\left[\textbf{Gram}(p_1,\dots,p_N)\right]f(p_i\cdot p_j)\,.
\eeq
Here $\textbf{Gram}(p_1,\dots,p_N)$ denotes the Gram determinant of the vectors $p_i$, which is defined as 
\beq
\textbf{Gram}(p_1,\dots,p_N)=\det(p_i\cdot p_j)_{1\leq i,j\leq N}.
\eeq
We observe that the measure as well as the matrix element are invariant under rotations in the transverse space.
Therefore, we can use this relation to rewrite the $D-2$ dimensional transverse integration 
\beq\bsp
\prod_{i=3}^N d^{D-2}p_{i,\perp}&=2^{2-N}\prod_{i=3}^N\Omega_{D+1-i}\prod_{j=3}^i d(p_{i,\perp}\cdot p_{j,\perp})\\
&\times\left[\textbf{Gram}(p_{3,\perp},\dots,p_{N,\perp})\right]^{\frac{D-N-1}{2}}\Theta\left[\textbf{Gram}(p_{3,\perp},\dots,p_{N,\perp})\right]\,.
\esp\eeq
Using the definition \eqref{eq:sci} of the $s$-channel invariants, we can rewrite the integration over the scalar products as
\beq
\prod_{i=3}^N\prod_{j=3}^i d(p_{i,\perp}\cdot p_{j,\perp}) = (-2)^{\frac{(N-2)(N-3)}{2}}\left(\prod_{i=3}^N dp_{i,\perp}^2\right)\left(\prod_{3\leq i<j\leq N}ds_{ij}\right)\,.
\eeq
The integration over the norm of the transverse components can be performed using the on-shell $\delta$-functions, so that we find
\beq\bsp
\prod_{i=3}^N d^Dp_i\delta_+(p_i^2)&=(-1)^{\frac{(N-2)(N-3)}{2}}2^{6-3N-\frac{(N-2)(N-3)}{2}}\\
&\times\left(\prod_{i=3}^N\Omega_{D+1-i}\right)\left(\prod_{\substack{1\leq i,j \leq N\\i\neq j,(i,j)\neq(1,2)}}\!\!\!\!ds_{ij}\right)G_N(\{s_{ij}\})^{\frac{D-N-1}{2}}\Theta[G_N(\{s_{ij}\})]\,,
\esp\eeq
where we have written the Gram determinant as 
\beq
G_N(\{s_{ij}\}) = \det(s_{1i}s_{2j}+s_{1j}s_{2i}- s_{ij})_{3\leq i,j \leq N}\,.
\eeq
Thus we finally arrive at the phase space measure 
\beq\bsp
\label{eq:psme}
d\Phi_{N-1}(D)&= \mathcal{N}_{N-2}(D)\bar{z}^{(N-2)(D-2)-1}\left(\prod_{\substack{1\leq i,j \leq N\\i\neq j,(i,j)\neq(1,2)}}\!\!\!\!ds_{ij}\right)\\
&\times\delta\left(1-\sum_{i=3}^N(s_{1i}+s_{2i})+(1-z)\sum_{i=3}^{N}\sum_{j=3}^{i-1}s_{ij}\right)G_N(\{s_{ij}\})^{\frac{D-N-1}{2}}\Theta[G_N(\{s_{ij}\})]\,,
\esp\eeq
with
\beq
\mathcal{N}_{N-2}(D) = (-1)^{\frac{(N-2)(N-3)}{2}}2^{-(N-2)\frac{D}{2}}(2\pi)^{(N-1)-(N-2)D}\prod_{i=3}^{N}\Omega_{D-i+1}\,,
\eeq
as in eq.~\eqref{eq:normN} and
\beq
\Omega_D = \frac{2 \pi^{\frac{D}{2}}}{\Gamma\left(\frac{D}{2}\right)}\,.
\eeq
\section{Derivation of the phase space volume	}
\label{sec:psv}
In this appendix we derive an expression for the $2\rightarrow H + (N-2)$ phase space volume that is valid for all orders in $z$ and $\eps$.
We start by recalling the general phase space factorisation:
\beq
\label{eq:psfac}
d\Phi_{k+1}(m^2,s) = \int_{m^2}^s \frac{d\mu^2}{2\pi} d\Phi_{l+1}(\mu^2,s)d\Phi_{k-l+1}(m^2,\mu^2)
\eeq
If we assume that we know the phase space volume at N$^k$LO, i.e. for $2\rightarrow H + k$,
\beq
\Phi_{k+1}(m^2,s) = \int d\Phi_{k+1}(m^2,s),
\eeq
this relation allows us to rewrite the phase space volume at N$^{k+l}$LO as
\beq
\Phi_{k+l+1}(m^2,s) = \int \int_{m^2}^{s}\frac{d\mu^2}{2\pi}d\Phi_{l+1}(\mu^2,s)d\Phi_{k+1}(m^2,\mu^2).
\eeq
We are specifically interested in rewriting the phase space volume at N$^{k+1}$LO as a convolution of the N$^{k}$LO phase space with a two particle phase space. Specialising to $l=1$, we can use this formula to inductively derive the phase space volume for arbitrary orders.
We have,
\beq\bsp
\Phi_{k+2}(m^2,s)& = \int \int_{m^2}^{s}\frac{d\mu^2}{2\pi}d\Phi_2(\mu^2,s)d\Phi_{k+1}(m^2,\mu^2) \\
&=\int_{m^2}^{s}\frac{d\mu^2}{2\pi}d\Phi_2(\mu^2,s)\Phi_{k+1}(m^2,\mu^2).
\esp\eeq
From eq.~\eqref{eq:psme} we obtain an explicit parametrisation of the two particle phase space and we find
\beq\bsp
\Phi_{k+2}(m^2,s)&=\frac{1}{4}(2\pi)^{-2+2\eps}s^{-\eps}\Omega_{2-2\eps}\int_{m^2}^{s}\frac{d\mu^2}{2\pi}\int_{0}^{\infty}dx_{1,k+3}dx_{2,k+3}(x_{1,k+3}x_{2,k+3})^{-\eps} \\
&\times \delta(1-x_{1,k+3}-x_{2,k+3})\left(1-\frac{\mu^2}{s}\right)^{1-2\eps}\Phi_{k+1}(m^2,\mu^2).
\esp\eeq
We can perform the integral over $x_{1,k+3}$ and $x_{2,k+3}$ in terms of beta functions and make the transformation $\mu^2 = s x$, $m^2=s z$ to obtain 
\beq\label{eq:PS_rec}
\Phi_{k+2}(z,s) = (4\pi)^{-2+\eps}s^{1-\eps}\frac{\Gamma(1-\eps)}{\Gamma(2-2\eps)}\int_{z}^{1}dx\,(1-x)^{1-2\eps}\Phi_{k+1}(z s, x s).
\eeq

In the following we use eq.~\eqref{eq:PS_rec} to prove inductively the following result:
\beq\bsp
\label{eq:psao}
\Phi_{n+1}(m^2,s)&=\frac{1}{2}(4\pi)^{1-2n+n\eps}\frac{\Gamma(1-\eps)^n}{\Gamma(2n(1-\eps))}s^{n-1-n\eps}(1-z)^{2n-1-2n\eps} \\
&\times{}_2F_1\left((n-1)(1-\eps),n(1-\eps),2n(1-\eps);1-z\right),
\esp\eeq
First, eq.~\eqref{eq:psao} correctly describes the phase space volume for $n=1$ and $n=2$. In order to derive $\Phi_{n+2}$ iteratively from $\Phi_{n+1}$ we use eq.~\eqref{eq:psao} and find
\beq\bsp
\Phi_{n+2}(z,s) &= \frac{1}{2} (4\pi)^{1-2(n+1)+(n+1)\eps}\frac{\Gamma(1-\eps)^{n+1}}{\Gamma(2n(1-\eps))\Gamma(2-2\eps)}s^{n-(n+1)\eps} \\
&\times \int_{z}^{1}dx(1-x)^{1-2\eps}x^{n-1-n\eps}\left(1-\frac{z}{x}\right)^{2n-1-2n\eps}\\
&\times{}_2F_1\left((n-1)(1-\eps),n(1-\eps),2n(1-\eps);1-\frac{z}{x}\right).
\esp\eeq
To solve the integral, we make the transformation $x = 1-(1-z)y$ and find
\beq\bsp
\Phi_{n+2}(z,s)&= \mathcal{C}\int_0^1dy\ y^{1-2\eps}(1-y)^{2n-1-2n\eps}(1-(1-z)y)^{-n+n\eps}\\
&\times{}_2F_1\left((n-1)(1-\eps),n(1-\eps),2n(1-\eps);(1-z)\frac{1-y}{1-(1-z)y}\right),
\esp\eeq
where we have factored out 
\beq
\mathcal{C} = \frac{1}{2}(4\pi)^{1-2(n+1)+(n+1)\eps}\frac{\Gamma(1-\eps)^{n+1}}{\Gamma(2n(1-\eps))\Gamma(2-2\eps)}s^{n-(n+1)\eps}(1-z)^{2(n+1)-1-2(n+1)\eps}.
\eeq
Next, we introduce a Mellin-Barnes representation for ${}_2F_1$ and exchange the MB integration with the parametric integration. This allows us to perform the integration over $y$ in terms of another ${}_2F_1$. Then we find
\beq\bsp
\Phi_{n+2}(z,s)&=\mathcal{C}\frac{\Gamma(2n(1-\eps))\Gamma(2-2\eps)}{\Gamma((n-1)(1-\eps))\Gamma(n(1-\eps))}\int_{-i\infty}^{+i\infty}\frac{d\xi}{2\pi i}(z-1)^{\xi}\Gamma(-\xi)\frac{\Gamma((n-1)(1-\eps)+\xi)}{\Gamma(2(n+1)(1-\eps)+\xi)} \\
&\times \Gamma(n(1-\eps)+\xi){}_2F_1\left(n(1-\eps)+\xi,2-2\eps,2(n+1)(1-\eps)+\xi;1-z\right)\,.
\esp\eeq
Introduce another MB integral for the ${}_2F_1$, we arrive at
\beq\bsp
\Phi_{n+2}(z,s)&=\mathcal{C}\frac{\Gamma(2n(1-\eps))}{\Gamma((n-1)(1-\eps))\Gamma(n(1-\eps))}\int_{-i\infty}^{+i\infty}\frac{d\xi d\chi}{(2\pi i)^2}(z-1)^{\xi+\chi}\Gamma(-\xi)\Gamma(-\chi)\\
&\times \frac{\Gamma((n-1)(1-\eps)+\xi)\Gamma(n(1-\eps)+\xi+\chi)\Gamma(2-2\eps+\chi)}{\Gamma(2(n+1)(1-\eps)+\xi+\chi)}.
\esp\eeq
In the next step we perform the change of variables $\xi\rightarrow\xi+\chi$, such that
\beq\bsp
\Phi_{n+2}(z,s)&=\mathcal{C}\frac{\Gamma(2n(1-\eps))}{\Gamma((n-1)(1-\eps))\Gamma(n(1-\eps))}\int_{-i\infty}^{+i\infty}\frac{d\xi d\chi}{(2\pi i)^2} (z-1)^{\xi}\Gamma(\chi-\xi)\Gamma(-\xi)\\
&\times \frac{\Gamma((n-1)(1-\eps)+\xi-\chi)\Gamma(n(1-\eps)+\xi)\Gamma(2-2\eps+\xi)}{\Gamma(2(n+1)(1-\eps)+\xi)}\,.
\esp\eeq
The integral over $\chi$ can now be performed using Barnes' first lemma, and we find
\beq\bsp
\Phi_{n+2}(z,s)&=\mathcal{C}\frac{\Gamma(2n(1-\eps))}{\Gamma((n-1)(1-\eps))\Gamma(n(1-\eps))}\int_{-i\infty}^{+i\infty}\frac{d\xi}{2\pi i}\Gamma(-\xi)(z-1)^{\xi}\\
&\times\frac{\Gamma(n(1-\eps)+\xi)\Gamma((n+1)(1-\eps)+\xi)}{\Gamma(2(n+1)(1-\eps)+\xi)}.
\esp\eeq
The integral over $\xi$ is just the MB representation of a ${}_2F_1$ so that we finally find
\beq\bsp
\Phi_{n+2}(z,s)&=\frac{1}{2}(4\pi)^{1-2(n+1)+(n+1)\eps}s^{n-(n+1)\eps}(1-z)^{2(n+1)-1-2(n+1)\eps}\frac{\Gamma(1-\eps)^{n+1}}{\Gamma(2(n+1)(1-\eps))} \\
&\times {}_2F_1\left(n(1-\eps),(n+1)(1-\eps),2(n+1)(1-\eps);1-z\right).
\esp\eeq
We have therefore inductively shown the validity of \eqref{eq:psao} for all $n$.
\section{From Mellin-Barnes integrals to parametric integrals}
\label{app:MB_to_param}
In this appendix we describe how one can derive an Euler-type integral from a Mellin-Barnes integral with a balanced integrand. 
Roughly speaking, a MB integral is said to be \emph{balanced} if for each integration variable $z_i$ the number of $\Gamma$ functions of the form $\Gamma(\ldots-z_i)$ is equal to the number of $\Gamma$ functions of the form  $\Gamma(\ldots+z_i)$. More precisely, the integral
\beq
\mbint\frac{dz_i}{2\pi i}\prod_{k_1=1}^{n_+}\Gamma(a_{k_1}+z_i)^{\alpha_{k_1}}\prod_{k_2=1}^{n_-}\Gamma(b_{k_2}-z_i)^{\beta_{k_2}}\,, \quad \alpha_{k_i},\beta_{k_i}\in\mathbb{Z}\,,
\eeq
is said to be \emph{balanced} if $\sum_{k_1}^{n_+}\alpha_{k_1} = \sum_{k_2}^{n_-}\beta_{k_2}$. We assume in the following that the contours are straight vertical lines such that the real parts of the arguments of all the $\Gamma$ functions are positive\footnote{Note that in dimensional regularisation we might need to require $\eps$ to be finite for such a contour to exist.}. In that case we can always derive an Euler-type integral representation for the MB integral. We start by noting that if an integral is balanced, then we can always express its integrand as a product of Beta functions,
\beq\label{eq:Beta_def}
B(x,y) = \int_0^\infty dt\,t^{x-1}\,(1+t)^{-x-y}\,.
\eeq
The integral~\eqref{eq:Beta_def} is convergent whenever $\mathrm{Re}(x), \mathrm{Re}(y)>0$. It is easy to convince oneself that this condition is satisfied whenever the real parts of all arguments of the $\Gamma$ functions are positive in the original MB integral. We can therefore replace each Beta function by its integral representation~\eqref{eq:Beta_def} in the \emph{integrand} of the MB integral and, because all the integral are convergent, we can exchange the MB integrations and the integrations coming from the Beta functions. This leaves us with an integral of the form
\beq\label{eq:MBtoparam}
\int_0^\infty \left(\prod_{n=1}^Ndt_i\right) R_{0}({\bf t})\,R_{\eps}({\bf t})^\eps\,\mbint\prod_{m=1}^M\frac{dz_i}{2\pi i} R_i({\bf t})^{z_i}\,,
\eeq
where ${\bf t} = (t_1,\ldots,t_N)$ and the $R_k$ are ratios of products of the $t_i$ and $1+t_i$. Next, we would like to perform the MB integrations. This can be done using the formula
\beq\label{eq:fourier}
\int_{z_0-i\infty}^{z_0+i\infty}\frac{dz}{2\pi i}a^z = \delta(1-a)\,,\qquad a>0\,.
\eeq
Indeed, parametrizing the contour as $z=z_0+it$, we obtain
\beq
\int_{z_0-i\infty}^{z_0+i\infty}\frac{dz}{2\pi i}a^z = a^{z_0}\int_{-\infty}^{+\infty}\frac{dt}{2\pi}e^{it\ln a} = a^{z_0}\delta(\ln a) = \delta(1-a)\,.
\eeq
Equation~\eqref{eq:MBtoparam} can thus be written in the form
\beq
\int_0^\infty \left(\prod_{n=1}^Ndt_i\right) R_{0}({\bf t})\,R_{\eps}({\bf t})^\eps\,\prod_{m=1}^M\delta\left(1- R_i({\bf t})\right)\,.
\eeq
We can solve the $\delta$ constraints, and the result is the desired parametric integral.

In the following we discuss two very simple examples that illustrate the above procedure. We start by discussing Barnes' first lemma, i.e., we consider the integral
\beq
\cI = \mbint\frac{dz}{2\pi i}\,\Gamma(a+z)\,\Gamma(b+z)\,\Gamma(c-z)\,\Gamma(d-z)\,.
\eeq
We assume that the integration contour and $a$, $b$, $c$ and $d$ are such that the real parts of all $\Gamma$ functions are positive. We rewrite the integrand in terms of Beta functions,
\begin{eqnarray}
\cI&=&\Gamma(a+c)\,\Gamma(b+d)\,\mbint\frac{dz}{2\pi i}\,B(a+z,c-z)\,B(b+z,d-z)\\
\nonumber&=&\Gamma(a+c)\,\Gamma(b+d)\,\mbint\frac{dz}{2\pi i}\int_0^\infty dt_1\,dt_2\,t_1^{a+z-1}\,(1+t_1)^{-a-c}\,t_2^{b+z-1}\,(1+t_2)^{-b-d}\\
\nonumber&=&\Gamma(a+c)\,\Gamma(b+d)\,\int_0^\infty dt_1\,dt_2\,t_1^{a-1}\,(1+t_1)^{-a-c}\,t_2^{b-1}\,(1+t_2)^{-b-d}\,\delta(1-t_1t_2)\,,
\end{eqnarray}
 where the last step follows from eq.~\eqref{eq:fourier}. Solving the $\delta$-function constraint leads to a one-fold integral that can immediately be recognized as a Beta function, and we recover the usual form of Barnes' first lemma,
 \beq\bsp
 \cI&\,=\Gamma(a+c)\,\Gamma(b+d)\,\int_0^\infty dt_1\,t_1^{a+d-1}\,(1+t_1)^{-a-b-c-d}\\
 &\, = \frac{\Gamma(a+c)\,\Gamma(a+d)\,\Gamma(b+c)\,\Gamma(b+d)}{\Gamma(a+b+c+d)}\,.
 \esp\eeq
 
 The second example we are going to discuss is Gauss' hypergeometric function. We consider the integral
 \beq
 \mathcal{J} = \mbint\frac{dz}{2\pi i}\Gamma(-z)\,\frac{\Gamma(a+z)\Gamma(b+z)}{\Gamma(c+z)}\,x^z\,.
 \eeq
 We again assume that all conditions for convergence are satisfied. Rewriting the integrand in terms of Beta functions, we obtain
 \beq\bsp
 \mathcal{J} &\,= \frac{\Gamma(b)}{\Gamma(c-a)}\mbint\frac{dz}{2\pi i}B(b+z,-z)\,B(a+z,c-a)\,x^z\\
 &\, = \frac{\Gamma(b)}{\Gamma(c-a)}\int_0^\infty dt_1\,dt_2\,t_1^{a-1}\,(1+t_1)^{-c}\,t_2^{b-1}\,(1+t_2)^{-b}\,\delta\left(1-\frac{x\,t_1\,t_2}{1+t_1}\right)\,.
 \esp\eeq
 Solving the $\delta$-function constraint with respect to $t_2$ and performing the change of variables $t_1\to \xi/(1-\xi)$, we immediately arrive at the usual integral representation for the $_2F_1$ function,
 \beq\bsp
 \mathcal{J} &\,=  \frac{\Gamma(b)}{\Gamma(c-a)}\int_0^\infty dt_1\,t_1^{a-1}\,(1+t_1)^{b-c}\,(1+t_1+x\,t_1)^{-b}\\
 &\,=\frac{\Gamma(b)}{\Gamma(c-a)}\int_0^1 d\xi\,\xi^{a-1}\,(1-\xi)^{c-a-1}\,(1+x\,\xi)^{-b}\\
 &\,=\frac{\Gamma(a)\,\Gamma(b)}{\Gamma(c)}\,{_2F_1}(a,b;c;-x)\,.
 \esp\eeq
 
\def \sha{{\,\amalg\hskip -3.6pt\amalg\,}}
\newcommand{\cS}[0]{\mathcal{S}}

\section{Symbolic integration of parametric integrals}
\label{app:parametric_integrals}
In this section we describe an algorithmic approach to compute certain classes of parametric integrals. 
We stress that this approach is not genuinely new, but has already been successfully applied, in some variant or another, to the computation of Feynman integrals, e.g.,~\cite{Brown:2008um,DelDuca:2009ac,DelDuca:2010zg,Ablinger:2012qm,Chavez:2012kn,Bogner:2012dn}. More precisely, consider an integral of the form
\beq\label{eq:toy_int}
\cI(\{y_j\};\eps)=\int_0^\infty\left(\prod_{i=1}^ndx_i\right)\,\prod_{k=1}^m P_k(\{x_i\};\{y_j\})^{a_k+\eps\,b_k}\,,
\eeq
where $a_k$ and $b_k$ are integers and the $P_k(\{x_i\};\{y_j\})$ are polynomials with integer coefficients, which we assume irreducible over $\mathbb{Z}$, i.e., they cannot be factorized into a product of non-constant polynomials of lower degree. We assume that the integration range is $[0,\infty]$ for each integration variable $x_i$. While many statements remain true for generic integration boundaries, in certain cases the algorithm we are going to describe breaks down, due to the appearance of boundary contributions. We can however always map a generic integration region $x_i\in[a,b]$ to $y_i\in[0,\infty]$ by the change of variable
\beq
y_i = \frac{x_i-b}{x_i-a}\,.
\eeq
We furthermore assume that the integral is convergent for $\eps=0$, and so we can expand in $\eps$ under the integration sign. Note that in the simplest divergent cases where the singularities in the integrand factorize we can reduce the problem to a convergent integral by subtracting the divergencies. 

Our goal is to compute the coefficients of the Taylor expansion of $\cI(\{y_j\};\eps)$ by integrating out the integration variables recursively one-by-one in terms of multiple polylogarithms, eq.~\eqref{eq:Mult_PolyLog_def}. Obviously, not every integral can be performed in this way. In the rest of this appendix we describe a sufficient condition, first obtained in ref.~\cite{Brown:2008um}, to be satisfied by the integrand of $\cI(\{y_j\};\eps)$ so that it can be integrated in terms of multiple polylogarithms. We then review an algorithm for integrating these classes of integrals, and illustrate the procedure explicitly on a simple example.


\subsection{Denominator reduction}
In this section we review (a variant of) the sufficient condition of ref.~\cite{Brown:2008um} to determine whether an integral can be performed recursively in terms of multiple polylogarithms. The main idea is that we need to determine an ordering of the integration variables such that at each step during the integration all the denominators are linear in the next integration variable. We start by defining the set $S$ of all the polynomials that are not monomials and that appear inside the integrand of eq.~\eqref{eq:toy_int},
\beq
S=\Big\{P_k(\{x_i\};\{y_j\})\Big\}\,.
\eeq
To start the integration, we have to assume that there is one integration variable, say $x_a$ such that all the element of $S$ are linear in $x_a$. In that case we may write
\beq
P_k(\{x_i\};\{y_j\}) = Q_k^a(\{x_i\};\{y_j\})\,x_a + R_k^a(\{x_i\};\{y_j\})\,,
\eeq
where $Q_k^a(\{x_i\};\{y_j\})$ and $R_k^a(\{x_i\};\{y_j\})$ are polynomials that are independent of $x_a$. Note that, after expansion in $\eps$, the integrand may also contain logarithms of the $P_k(\{x_i\};\{y_j\})$, which can be rewritten in terms of multiple polylogarithms,
\beq\bsp\label{eq:LogToG}
\log P_k&\,= \log\Big(Q_k^a\,x_a + R_k^a\Big) = \log R_k^a + \log\left(1 + \frac{Q_k^a}{R_k^a}x_a\right) \\
&\,= \log R_k^a + G\left(-\frac{R_k^a}{Q_k^a};x_a\right)\,,
\esp\eeq
where for clarity we suppressed the arguments of the polynomials. Furthermore, we can use the shuffle algebra of multiple polylogarithms to replace every product of multiple polylogarithms by a sum,
  \beq\bsp\label{eq:G_shuffle}
  G(a_1,\ldots,a_{n_1};z) \, G(a_{n_1+1},\ldots,a_{n_1+n_2};z) &\,=\sum_{\sigma\in\Sigma(n_1, n_2)}\,G(a_{\sigma(1)},\ldots,a_{\sigma(n_1+n_2)};z),\\
      \esp\eeq
where $\Sigma(n_1,n_2)$ denotes the set of all shuffles of $n_1+n_2$ elements, \emph{i.e.}, the subset of the symmetric group $S_{n_1+n_2}$ defined by
\beq\bsp\label{eq:Sigma_def}
&\Sigma(n_1,n_2) \\
&=\{\sigma\in S_{n_1+n_2} |\, \sigma^{-1}(1)<\ldots<\sigma^{-1}({n_1}) {\rm~~and~~} \sigma^{-1}(n_1+1)<\ldots<\sigma^{-1}(n_1+{n_2})\}\,.
\esp\eeq 
Thus we can assume without loss generality that the integration over $x_a$ takes the form
\beq
\int_0^\infty\frac{dx_a}{(Q_1^ax_a+R_1^a)^{-a_1}\ldots(Q_m^ax_a+R_m^a)^{-a_m}}\,G(\vec a;x_a)\,.
\eeq
Partial fractioning the factors in the denominator, e.g., 
\beq\label{eq:partial_fractioning}
\frac{1}{(Q_k^ax_a+R_k^a)(Q_l^ax_a+R_l^a)} = \frac{1}{Q_k^aR_l^a - Q_l^aR_k^a}\left(\frac{1}{x_a +R_k^a/Q_k^a}  - \frac{1}{x_a +R_l^a/Q_l^a}\right)\,,
\eeq
we obtain a sum of integrals that can be reduced to the recursive definition of multiple polylogarithms, eq.~\eqref{eq:Mult_PolyLog_def}. We can thus easily compute a primitive with respect to $x_a$, and then take the limits $x_a\to0$ and $x_a\to \infty$. We will address the issue of limits in the next subsection.

We would like to iterate this procedure and integrate over the next variable. This is however only possible if inside the new integrand we can still find an integration variable in which all polynomials are linear. The polynomials appearing inside the integrand are $Q_k^a$ and $R_k^a$, which have been introduced through eq.~\eqref{eq:LogToG} and~\eqref{eq:partial_fractioning}, as well as the combinations $Q_k^aR_l^a - Q_l^aR_k^a$ introduced by partial fractioning. This last polynomial however need not necessarily be linear, even if the $Q_k^a$ and $R_k^a$ are. In order to proceed, it is therefore mandatory that all the $Q_k^aR_l^a - Q_l^aR_k^a$ factor into polynomials that are linear in a certain variable.

In ref.~\cite{Brown:2008um} a criterion was given that allows one to determine a priori whether the above procedure terminates, i.e., whether there is an ordering of the integration variables such that all the denominators stay linear at each integration step. We start by defining the set 
$S_{(x_a)}$ as the set of irreducible factors that appear inside the polynomials $Q_k^a$, $R_k^a$ and $Q_k^aR_l^a - Q_l^aR_k^a$. Then, if we can find an integration variable $x_b$ such that all the elements of $S_{(x_a)}$ are linear in $x_b$, we can restart the above procedure and integrate over $x_b$. If we iterate this procedure and are able to construct a sequence of sets of polynomials 
\beq\label{eq:set_sequence}
S_{(x_a)}, S_{(x_a, x_b)}, S_{(x_a, x_b,x_c)},\ldots
\eeq 
such that in each set all the polynomials are linear in at least one integration variable, then we have found an ordering of the integration variables such that we can recursively integrate out all the integration variables in terms of multiple polylogarithms. We stress that this condition is sufficient, but not necessary: even if we fail to find a suitable sequence~\eqref{eq:set_sequence}, the integral might still be expressible in terms of multiple polylogarithms (e.g., after a suitable change of variables). In addition, note that for the last integration step it is not necessary for the polynomials to be linear: we can then factor the polynomials into linear factors, whose roots involve algebraic expressions of the parameters $\{y_j\}$.

\subsection{Symbolic integration}
We have now a criterion to determine if a given integral of the type~\eqref{eq:toy_int} can be integrated recursively in terms of multiple polylogarithms, and we have already explained how to perform the first integration step. We assume from now on that we have found a sequence~\eqref{eq:set_sequence} and the associated ordering of the integration variables, and that we have performed the integration over $x_a$ as described in the previous subsection. 

In order to continue and perform the next integration, we need to address two issues
\begin{enumerate}
\item We need to take the limits $x_a\to0$ and $x_a\to\infty$ of the primitive with respect to $x_a$.
\item If we want to integrate next over $x_b$, we have to face the problem that $x_b$ might be `hidden' inside the arguments of multiple polylogarithms of the form
\beq\label{eq:G_bad}
G\left(\ldots,-\frac{R_k^a}{Q_k^a},\ldots\right)\,,
\eeq
i.e., $x_b$ appears inside the polynomials $Q_k^a$ and $R_k^a$. In order to compute a primitive with respect to $x_b$, we need to rewrite all multiple polylogarithms of the form~\eqref{eq:G_bad} as $G(\vec a;x_b)$, where $\vec a$ is independent of $x_b$. 
\end{enumerate}

The limit $x_a\to0$ can easily by taken by using the fact that
\beq\label{eq:lim0}
\lim_{x_a\to0}G(\vec a_1;x_a) = 0\,, \qquad\vec a_1\neq\vec 0\,.
\eeq
While eq.~\eqref{eq:lim0} is sufficient to compute the value at $x_a=0$ in many circumstances, it can happen that the primitive has spurious poles and / or logarithmic singularities at $x_a=0$. In the presence of poles we need to expand the multiple polylogarithms around $x_a=0$. This task can easily be achieved using the series representation of multiple polylogarithms. Indeed, multiple polylogarithms can equally well be represented as multiple nested sums,
\beq
\textrm{Li}_{m_k,\ldots,m_1}(z_k,\ldots,z_1) = \sum_{n_1>\ldots>n_k>0}{z_1^{n_1}\over n_1^{m_1}}\ldots{z_k^{n_k}\over n_k^{m_k}}
=\sum_{n=1}^\infty{z_1^{n}\over n^{m_1}}\,Z_{m_2,\ldots,m_k}(n-1;z_2,\ldots,z_k)\,,
\eeq
where $Z_{m_2,\ldots,m_k}(n-1;z_2,\ldots,z_k)$ denote the $Z$-sums defined in ref.~\cite{Moch:2001zr}.
The integral representation and the series representation are related by
\beq
G(\underbrace{0,\ldots,0}_{m_1-1},a_1, \ldots, \underbrace{0,\ldots,0}_{m_k-1}, a_k;x)
= (-1)^k\,
\textrm{Li}_{m_k,\ldots,m_1}\left(\frac{a_{k-1}}{a_k},\ldots,\frac{x}{a_1}\right)\,,\qquad a_i\neq0\,.
\eeq
If $a_k=0$, we can use the shuffle algebra to extract pure logarithms in $x_a$, e.g., if $b\neq 0$,
\beq
G(b,0;x_a) = G(0;x_a)G(b;x_a)-G(0,b;x_a)\,.
\eeq
We can thus easily obtain the expansion around $x_a=0$ in terms of $Z$-sums to arbitrary order.

Next, we discuss how to take the limit $x_a\to\infty$. If we define $\bar{x}_a=1/x_a$, then the problem can be reduced to taking the limit $\bar{x}_a\to0$,
\beq\label{eq:liminfty}
\lim_{x_a\to\infty}G(\vec a;x_a) = \lim_{\bar{x}_a\to0}G\left(\vec a;\frac{1}{\bar{x}_a}\right)\,.
\eeq
We would like to use the inversion relations for the multiple polylogarithms in the right-hand side to express them as linear combinations of multiple polylogarithms of the form $G\left(\ldots;{\bar{x}_a}\right)$, for which the limit $\bar{x}_a\to0$ is trivial. Unfortunately, the corresponding functional equations are often unknown. Note at this point that the problem of finding the inversion relations is formally equivalent to the second problem we have not yet addressed, namely how to bring the primitive with respect to $x_a$ into a form where $x_b$ only enters through multiple polylogarithms of the form $G(\vec a;x_b)$: in both cases we are looking for functional equations that bring a certain multiple polylogarithm into a `canonical form' where a certain variable only appears as the explicit upper integration boundary of the multiple polylogarithms. 
In the rest of this section we argue that, under certain conditions which we will later show to be equivalent to the criterion derived in the previous subsection, plus the hypothesis that the integration ranges are $[0,\infty]$, we can always bring multiple polylogarithms into the `canonical form'
\beq\label{eq:canon}
\sum_{i} c_i \,G(\vec a_i;x)\,,
\eeq
for some variable $x$ such that $a_i$ is independent of $x$ and the coefficients $c_i$ involve only multiple polylogarithms that are independent of $x$. Note that this result is similar to the result obtained in ref.~\cite{Brown:2008um}. 
In the following we give a constructive algorithm that allows us to derive the canonical form~\eqref{eq:canon}.

In order to achieve a rewriting of our multiple polylogarithms in canonical form, we need to derive the corresponding functional equations.
The natural language to discuss functional equations among multiple polylogarithms are symbols~\cite{Goncharov-simple-Grassmannian,symbolsC,symbolsB,Goncharov:2010jf,Duhr:2011zq} and the Hopf algebra of multiple polylogarithms~\cite{Goncharov:2001}. We start by giving a concise review of symbols.

One possible way to define the symbol of a multiple polylogarithm is to consider its total differential~\cite{Goncharov:2001},
\beq\label{eq:MPL_diff_eq}
dG(a_{n-1},\ldots,a_{1};a_n) = \sum_{i=1}^{n-1}G(a_{n-1},\ldots,{a}_{i-1},{a}_{i+1},\ldots,a_{1};a_n)\, d\ln\left({a_i-a_{i+1}\over a_i-a_{i-1}}\right)\,,
\eeq
and to define the symbol recursively by~\cite{Goncharov:2010jf}
\beq\label{eq:GSVV_def}
\cS(G(a_{n-1},\ldots,a_{1};a_n)) = \sum_{i=1}^{n-1}\cS(G(a_{n-1},\ldots,{a}_{i-1},{a}_{i+1},\ldots,a_{1};a_n))\otimes\left({a_i-a_{i+1}\over a_i-a_{i-1}}\right)\,.
\eeq
As an example, the symbols of the classical polylogarithms and the ordinary logarithms are given by
\beq
\cS(\mathrm{Li}_{n}(z)) = -(1-z)\otimes\underbrace{z\otimes\ldots\otimes z}_{(n-1)\textrm{ times}} {\rm~~and~~} \cS\left(\frac{1}{n!}\ln^nz\right) = \underbrace{z\otimes\ldots\otimes z}_{n\textrm{ times}}\,.
\eeq
In addition the symbol satisfies the following identities,
\beq\bsp
\ldots\otimes (a\cdot b)\otimes\ldots &\,= \ldots\otimes a\otimes\ldots + \ldots\otimes b\otimes\ldots\,,\\
 \ldots\otimes(\pm1)\otimes\ldots &\,= 0\,,\\
 \cS\left(G(\vec a;x)\,G(\vec b;y)\right) &\,=  \cS(G(\vec a;x))\sha \cS(G(\vec b;y))\,,
\esp\eeq
where $\sha$ denotes the shuffle product on tensors.

We can make the following observation about the symbol of a multiple polylogarithm: if the $a_i$ are independent of $x$, then the symbol of $G(a_1,\ldots,a_n;x)$ contains exactly one term hat contains $x$ in all its entries, and this term can be chosen to be of the form
\beq\label{eq:topterm}
\cS(G(a_1,\ldots,a_n;x)) = (a_n-x)\otimes\ldots\otimes(a_1-x) + \ldots\,.
\eeq
In order to proof this statement, we focus on the term in the total differential of $G(a_1,\ldots,a_n;x)$ proportional to $dx$,
\beq\bsp
dG(a_1,\ldots,a_n;x) &\,= G(a_2,\ldots,a_n;x)\,\frac{dx}{x-a_1}+\ldots\\
&\,= G(a_2,\ldots,a_n;x)\,d\log(a_1-x)+\ldots\,,
\esp\eeq
where the last step follows from
\beq
d\log(a_1-x) = \frac{dx-da_1}{x-a_1}\,,
\eeq
and where the dots indicate terms in the total differential that are independent of $dx$. The statement then follows recursively.
Note that this statement is independent of whether the $a_i$ are zero or not. 

Assume now that we are given an integrable symbol $T$ (which will correspond later to the symbol of the multiple polylogarithm we want to bring into canonical form) which is of uniform weight $w$ and has rational coefficients. If $T$ does not satisfy this last condition, we deal separately with the contributions of different weight and / or different rational function prefactor. Let us suppose that the entries are drawn from a set $S$. Without loss of generality, we may assume $S$ to consist of irreducible polynomials in some variables $x_i$, $1\le i\le n$, and for simplicity we assume for now that the polynomials are linear in \emph{all} the $x_i$ (we will see in the next subsection that the correct restriction is that $S$ satisfies the reduction criterion of the previous subsection). Furthermore, we assume that we have fixed an ordering on the variables, which we will take in the following as $(x_1,\ldots,x_n)$. 
For each variable $x_i$, we define a linear map $\phi_{x_i}$ which acts on elementary tensors $s$ by
\beq
\phi_{x_i}(s) = \left\{\begin{array}{ll}
G\left(-\frac{b_1}{a_1},\ldots,-\frac{b_w}{a_w};x_i\right)\,,&\textrm{if } s=(a_wx_i+b_w)\otimes\ldots\otimes(a_1x_i+b_1)\,,\\
0\,,&\textrm{otherwise}\,.
\end{array}\right.
\eeq
Morally speaking, the map $\phi_{x_i}$ assigns to $T$ the combination of multiple polylogarithms which will give the same terms that have $x_i$ in all entries through eq.~\eqref{eq:topterm}.

Using the maps $\phi_{x_i}$, we can now formulate an algorithm that assigns to $T$ a multiple polylogarithm in the canonical form associated to the ordering of the variables $(x_1,\ldots,x_n)$. We start by defining a new symbol by subtracting off the contribution from $\phi_{x_1}(T)$,
\beq
T_1 = T - \cS[\phi_{x_1}(T)]\,.
\eeq
By construction, each term in the symbol $T_1$ has at least one entry that is independent of $x_1$. Next, concentrate on the terms of the form
\beq
\sum_{(i_1,\ldots,i_w)}c_{i_1,\ldots,i_w}\,b_{i_1}\otimes(a_{i_2}x_1+b_{i_2})\otimes\ldots\otimes(a_{i_w}x_1+b_{i_w}))\,,
\eeq
where by hypothesis the $b_{i_k}$ are independent of $x_1$. As we have subtracted of the contribution from $\phi_{x_1}(T)$, these terms cannot come from a multiple polylogarithm of the form $G(\ldots;x_1)$ of weight $w$, but it can only arise from the product 
\beq
\log b_{i_1}\,G\left(-\frac{b_{i_w}}{a_{i_w}},\ldots,-\frac{b_{i_2}}{a_{i_2}};x_1\right) \to \phi_{x_2}(b_{i_1})\,\phi_{x_1}\Big((a_{i_2}x_1+b_{i_2})\otimes\ldots\otimes(a_{i_w}x_1+b_{i_w}))\Big)\,.
\eeq
It is easy to convince oneself that the difference
\beq
T_2 = T_1 - \sum_{(i_1,\ldots,i_w)}c_{i_1,\ldots,i_w}\,\cS\left[\phi_{x_2}(b_{i_1})\,\phi_{x_1}\Big((a_{i_2}x_1+b_{i_2})\otimes\ldots\otimes(a_{i_w}x_1+b_{i_w}))\Big)\right]
\eeq
contains only terms for which at most $(w-2)$ entries depend on $x_1$. We can now go on an recursively subtract contributions with different multiplicities of $x_1$. Assume for example that we have subtracted all contributions where $x_1$ appears in more than $(w-r)$ entries, and that the resulting symbol is $T_r$. We can then concentrate on the terms
\beq
\sum_{(i_1,\ldots,i_w)}c_{i_1,\ldots,i_w}\,b_{i_1}\otimes\ldots\otimes b_{i_r}\otimes(a_{i_{r+1}}x_1+b_{i_{r+1}})\otimes\ldots\otimes(a_{i_w}x_1+b_{i_w}))\,.
\eeq
It is easy to convince oneself that in the difference
\beq\bsp
T_{r+1} &\,=T_r - \\
&\sum_{(i_1,\ldots,i_w)}c_{i_1,\ldots,i_w}\,\cS\left[\phi_{x_2}\Big(b_{i_1}\otimes\ldots\otimes b_{i_r}\Big)\,\phi_{x_1}\Big((a_{i_{r+1}}x_1+b_{i_{r+1}})\otimes\ldots\otimes(a_{i_w}x_1+b_{i_w}))\Big)\right]
\esp\eeq
$x_1$ appear in at most $(w-r-1)$ entries. We continue this procedure until we reach $T_w$, which is independent of $x_1$, and we restart the algorithm with $T_w$ and $\phi_{x_2}$. We then repeat this procedure until we have exhausted all the integration variables, and the algorithm stops.
The result of this algorithm is, by construction, a function of the form
\beq
\sum_{(i_1,\ldots,i_n)}c_{i_1,\ldots,i_n}\,G(\vec a_{i_n};x_n)\ldots G(\vec a_{i_1};x_1)\,,
\eeq
whose symbol is $T$ and
such that the $\vec a_{i_k}$ are sequences of rational functions that are in the variables $x_k$, $k>i_k$, i.e., the sought-for canonical form for $T$.

Using this algorithm we can easily take the limit $x_a\to\infty$, by applying it to the right-hand side of eq.~\eqref{eq:liminfty} with the ordering $(\bar{x}_a,x_b,\ldots)$. We can also use it to rewrite the integral over $x_a$ such that we can easily integrate over the next integration variable $x_b$. However, the result of the algorithm will at this stage still deliver the wrong answer, as we have so far only constructed a function in canonical form whose \emph{symbol} matches the symbol of a given function. To illustrate this, assume we run the algorithm on the function in the right-hand side of eq.~\eqref{eq:liminfty} with the ordering $(\bar{x}_a,x_b,\ldots)$. The result is a function $\mathcal{G}(\bar{x}_a,x_b,\ldots)$ in canonical form such that
\beq\label{eq:same_symbol}
\cS\left[G\left(\vec a;\frac{1}{\bar{x}_a}\right) - \mathcal{G}(\bar{x}_a,x_b,\ldots)\right] = 0\,.
\eeq
It would however be wrong to conclude that $G\left(\vec a;\frac{1}{\bar{x}_a}\right)$ and $\mathcal{G}(\bar{x}_a,x_b,\ldots)$ are equal, because the symbol maps to 0 all terms proportional to multiple zeta values. In ref.~\cite{Duhr:2012fh} an algorithm was described that allows to reconstruct these zeta-valued terms using the full Hopf algebra structure of multiple polylogarithms, augmented by some ideas by Brown~\cite{Brown:2011ik}. In the following we only give a very brief account on how to reconstruct the zeta-valued terms, and we refer to ref.~\cite{Duhr:2012fh} for a detailed description of the algorithm.

In the following we assume for simplicity that the function $G\left(\vec a;\frac{1}{\bar{x}_a}\right) - \mathcal{G}(\bar{x}_a,x_b,\ldots)$ is real\footnote{There is no obstacle to consider complex-valued functions. Imaginary parts can be extracted in exactly the same way using $\Delta_{1,\ldots,1}$.}, so we do not need to worry about imaginary parts proportional to $i\pi$. Next we act with $\Delta_{2,1,\ldots,1}$ on the difference, where $\Delta_{2,1,\ldots,1}$ is the component of the iterated coproduct where the first component has weight two and all other components have weight one. Without loss of generality we can write
\beq\bsp\label{eq:delta2111}
\Delta_{2,1,\ldots,1}&\left[G\left(\vec a;\frac{1}{\bar{x}_a}\right) - \mathcal{G}(\bar{x}_a,x_b,\ldots)\right]\\
&\,=
\sum_{(i_1,\ldots,i_{w-1})}c_{i_1,\ldots,i_{w-1}}\,A_{i_1}\otimes\log a_{i_2}\otimes\ldots\otimes\log a_{i_{w-1}}\,,
\esp\eeq
where the $a_{i_k}$ are irreducible polynomials and $A_{i_1}$ is a combination of multiple polylogarithms of weight two. Without loss of generality we can assume that we have collected all term that have the same `tail' $\log a_{i_2}\otimes\ldots\otimes\log a_{i_{w-1}}$, i.e., we can assume 
\beq 
\log a_{i_2}\otimes\ldots\otimes\log a_{i_{w-1}} \neq \log a_{j_2}\otimes\ldots\otimes\log a_{j_{w-1}} \textrm{ if } (i_2,\ldots,i_{w-1})\neq(j_2,\ldots,j_{w-1})\,.
\eeq
As we know that the symbol of the function vanishes, eq.~\eqref{eq:same_symbol}, we necessarily conclude that $A_{i_1}$ is proportional to $\zeta_2$, i.e., $A_{i_1} = k_{i_1}\zeta_2$, for some rational number $k_{i_1}$. This rational number can easily be determined by evaluating $A_{i_1}$ numerically at a single point using any of the standard libraries to evaluate multiple polylogarithms~\cite{Gehrmann:2001pz,Gehrmann:2001jv,Maitre:2005uu,Maitre:2007kp,Vollinga:2004sn,Bauer:2000cp,Buehler:2011ev}, and running for example the PSLQ algorithm~\cite{pslq}. Equation~\eqref{eq:delta2111} then takes the form
\beq\bsp
\Delta_{2,1,\ldots,1}&\left[G\left(\vec a;\frac{1}{\bar{x}_a}\right) - \mathcal{G}(\bar{x}_a,x_b,\ldots)\right]\\
&\,=
\sum_{(i_1,\ldots,i_{w-1})}c_{i_1,\ldots,i_{w-1}}\,k_{i_1}\,\zeta_2\otimes\log a_{i_2}\otimes\ldots\otimes\log a_{i_{w-1}}\,,
\esp\eeq
Next we drop $\zeta_2$, i.e., we only keep the tail of each elementary tensor. If we also drop the $\log$ signs, we obtain a symbol associated with the terms proportional to $\zeta_2$,
\beq\label{eq:symb_2111}
\Delta_{2,1,\ldots,1}\left[G\left(\vec a;\frac{1}{\bar{x}_a}\right) - \mathcal{G}(\bar{x}_a,x_b,\ldots)\right] \to 
\sum_{(i_1,\ldots,i_{w-1})}c_{i_1,\ldots,i_{w-1}}\,k_{i_1}\,a_{i_2}\otimes\ldots\otimes a_{i_{w-1}}\,.
\eeq
Running the algorithm described at the beginning of this section on the symbol in eq.~\eqref{eq:symb_2111} we obtain a function $\mathcal{G}_2(\bar{x}_a,x_b,\ldots)$ of weight $w-2$ in canonical form such that
\beq
\Delta_{2,1,\ldots,1}\left[G\left(\vec a;\frac{1}{\bar{x}_a}\right) - \mathcal{G}(\bar{x}_a,x_b,\ldots)-\zeta_2\,\mathcal{G}_2(\bar{x}_a,x_b,\ldots)\right]=0\,.
\eeq
We have in this way determined all the contributions proportional to $\zeta_2$, and the result is by construction in canonical form. We then repeat exactly the same exercise by acting with $\Delta_{3,1,\ldots,1}$ to determine the terms proportional to $\zeta_3$, and we continue in this way until we have exhausted all possibilities and the algorithms stops. As a result we obtain the expression of $G\left(\vec a;\frac{1}{\bar{x}_a}\right)$ in canonical form at \emph{function level}.

This terminates the algorithm to bring multiple polylogarithms into the canonical form corresponding to a certain ordering of the variables. This operation is sufficient to take the limit $x_a\to\infty$, and to bring the integral into a form where the next integration can easily be performed.
We have however not yet shown that this algorithm will always terminate on the class of integrals we consider. In particular, when formulating the algorithm, we assumed that all entries that appear in the symbol are linear in all variables. In the next subsection we show that this condition can be relaxed, and that the algorithm always terminates if the integrand satisfies the reduction criterion of the previous subsection and if the integration range is $[0,\infty]$.

\subsection{Symbolic integration and denominator reduction}
In this subsection we show that the algorithm we just described always terminates for integrals of the type~\eqref{eq:toy_int} such that the set $S$ of polynomials satisfies the reduction criterion. From now on we assume that we have found an ordering of the integration variables, which we take as $(x_1,\ldots,x_n)$ and we can find a sequence
\beq
S,S_{(x_1)},S_{(x_1,x_2)},S_{(x_1,x_2,x_3)},\ldots
\eeq
such that all the elements of $S_{(x_1,\ldots x_k)}$ are linear in $x_{k+1}$. We start by showing that under these hypotheses and after having integrated out $(x_1,\ldots,x_{k-1})$,
\begin{enumerate}
\item the symbol of the primitive with respect to $x_k$ has all its entries drawn from the set $\overline{S}_{(x_1,\ldots x_{k-1})}=\{x_k,\ldots,x_n\} \cup S_{(x_1,\ldots x_{k-1})}$.
\item the symbol of the function after integration over $x_k$, i.e., after taking the limits $x_k\to0,\infty$ of the primitive, has all its entries drawn from $\widetilde{S}_{(x_1,\ldots x_{k})}=\{x_{k+1},\ldots,x_n\} \cup S_{(x_1,\ldots x_{k})}$.
\end{enumerate}
We start by proving the first statement, the second then immediately follows by taking the appropriate limits. As in the following we constantly switch between polynomials of the sets we just defined, and polynomials as entries of symbols, we introduce the notation $[P]$ to refer to $P$ as an entry of a symbol. Note that we then have the identities
\beq\bsp
[P\,Q] &\,= [P]+[Q]\,,\\ 
[P/Q] &\,= [P]-[Q]\,,\\ 
[\pm1] &\,= 0\,,\\ 
\textrm{etc.}
\esp\eeq

We proceed by iteration in the number $k$ of variables we have already integrated out. We start by analyzing what happens after the first integration. It is obvious from~eqs.~\eqref{eq:LogToG} and~\eqref{eq:partial_fractioning} that after the first integration the primitive only involves multiple polylogarithms $G(a_1,\ldots,a_w;x_1)$ with
\beq
a_i\in\left\{0,-R_k^1/Q_k^1\right\}\,.
\eeq
It is then easy to see (e.g., from the polygon approach to the symbol of ref.~\cite{Duhr:2011zq}) that the symbol can only have the following entries:
\beq\bsp
&[x_k], [R_k^1]\,, \\
&\left[-R_k^1/Q_k^1\right] = [R_k^1] - [Q_l^1]\,,\\
&\left[1+\frac{x_1\,Q_k^1}{R_k^1}\right] = [R_k^1+Q_k^1\,x_1] - [R_k^1] = [P_k] - [R_k^1]\,,\\
&\left[1+\frac{Q_k^1\,R_l^1}{R_k^1\,Q_l^1}\right] = [Q_k^1\,R_l^1-R_k^1\,Q_l^1] - [R_k^1] - [Q_l^1]\,.
\esp\eeq
The polynomials that appear inside the symbol are precisely those that appear in $S$ and $S_{(x_1)}$, which finishes the proof of the first statement for the first integration.

Next consider taking the limits $x_1\to0$ and $x_1\to\infty$. By definition, $R_k^1$ and $Q_k^1$ are independent of the limit, so we only need to consider the limits of $[x_1]$ and $[P_k]$. $[x_1]$ will give rise to logarithmic singularities in the limit, and these terms must cancel if the integral is convergent. For $[P_k]$ we have,
\beq\bsp
\lim_{x_1\to0} [P_k] &\,= [R_k^1]\,,\\
\lim_{x_1\to\infty}[P_k]&\,=\lim_{\bar{x}_1\to0}\{[\bar{x}_1\,R_k^1+Q_k^1]-[\bar{x}_1]\} = [Q_k^1] + \lim_{\bar{x}_1\to0}[x_1]\,.
\esp\eeq
The logarithmic singularity in the last line must again cancel for convergent integrals, and so we see that the only polynomials that appear in the limit are those in $S_{(x_1)}$. This finishes the proof of the second statement for the first integration. We stress that it is important that the integration region is $[0,\infty]$, because otherwise we have to take into account effects coming from the integration boundaries.

Let us now suppose that the two statements are true for the first $r-1$ integrations. The set $S_{(x_1,\ldots,x_{r-1})}$ then consists of polynomials of the form $\tilde{P}_k = \tilde{Q}_k^r\,x_r+\tilde{R}_k^r$. Let us now compute the primitive with respect to $x_r$. We start by running the algorithm to bring the multiple polylogarithms in the integrand into canonical form. As this involves the application of the map $\phi_{x_r}$, we obtain multiple polylogarithms of the form $G(a_1,\ldots,a_w;x_r)$ with 
\beq\label{eq:ai_xr}
a_i\in\left\{0,-\tilde{R}_k^r/\tilde{Q}_k^r\right\}\,.
\eeq
If we compute the primitive, we integrate over linear functions in $x_r$ from the set $S_{(x_1,\ldots,x_{r-1})}$, and it is easy to see that this does not change eq.~\eqref{eq:ai_xr}.
Using exactly the same argument as for the first integration, we see that these multiple polylogarithms only contribute terms to the symbols whose entries are drawn from $\overline{S}_{(x_1,\ldots x_{r-1})}$. In addition we have multiple polylogarithms that are independent of $x_r$, whose symbols involve those elements of $\widetilde{S}_{(x_1,\ldots x_{r-1})}$ that are independent of $x_r$. As these functions do not change if we take the primitive with respect to $x_r$, we see that they do not alter the conclusion. So the symbol of the primitive with respect to $x_r$ must have all its entries drawn from $\overline{S}_{(x_1,\ldots x_{r-1})}$. Taking the limits $x_r\to0$ and $x_r\to\infty$ just like for the first integration then finishes the proof.

Having proved the two statements, we can show that our algorithm always terminates for the class of integral we consider. More precisely, we have to show that our algorithm can always produce the canonical form for the next integration step. This is done by applying the map $\phi_{x_r}$, which requires all the entries in the symbol to be either independent of $x_r$ or linear in $x_r$. By construction, this condition is always fulfilled for the elements of $\widetilde{S}_{(x_1,\ldots x_{r-1})}$. It is easy to check that the same argument shows that the map $\phi_{x_{r+1}}$, which is called recursively by $\phi_{x_r}$ is well-defined, and so we can always find a canonical form for the integrand.

\subsection{A toy example}
In this section we discuss an example that illustrates how the algorithm described in the previous subsection can be used to compute parametric integrals. In particular, we used this algorithm to compute the parametric integrals~\eqref{eq:I91} and~\eqref{eq:I92}. While these integrals are a straightforward application of the algorithm, intermediate expression are rather long, so we prefer to use a simpler integral where we can explicitly show all the steps. The integral we are going to consider is
\beq\bsp
\cI(\eps) &\,= \int_0^\infty dx_1dx_2dx_3\,x_1^{\epsilon } \left(1+x_1\right)^{3 \epsilon -2} x_2^{-\epsilon } \left(1+x_2\right)^{-4 \epsilon -2} x_3^{2 \epsilon } \left(1+x_3\right)^{-\epsilon -1} \\
&\,\times\left(1+x_2+x_3+x_1 x_3\right)^{-2 \epsilon -1}\,.
\esp\eeq
The integral is finite as $\eps\to0$, and we want to compute the first few terms in the Taylor expansion
\beq
\cI(\eps) = \cI_0+\cI_1\,\eps+\cI_2\,\eps^2+\cI_3\,\eps^3+\ord(\eps^4)\,.
\eeq
The first coefficient is trivial to compute,
\beq
\cI_0 = \frac{\pi^2}{9}-\frac{2}{3}\,.
\eeq
We will now illustrate our algorithm in detail on the coefficient $\cI_1$. 
If we integrate out the integration variables in the order $(x_1,x_2,x_3)$, we obtain
\beq\bsp
S&\,=\{1+x_1,1+x_2,1+x_3,1+x_2+x_3+x_1 x_3\}\,,\\
S_{(x_1)} &\,=\{1+x_2,1+x_3,1+x_2+x_3\}\,,\\
S_{(x_1,x_2)}&\,=\{1+x_3\}\,.
\esp\eeq
We see that all the sets are linear in all integration variables, so we perform the integrations one after the other.

The coefficient $\cI_1$ is given by the integral
\beq\bsp
\cI_1&\,=\int_0^\infty \frac{dx_1dx_2dx_3}{\left(1+x_1\right)^2 \left(1+x_2\right)^2 \left(1+x_3\right) \left(1+x_2+x_3+x_1 x_3\right)}\\
&\,\times\Bigg[3 G\left(-1;x_1\right)-4 G\left(-1;x_2\right)-3 G\left(-1;x_3\right)+G\left(0;x_1\right)-G\left(0;x_2\right)\\
&\,+2 G\left(0;x_3\right)-2 G\left(-x_3-1;x_2\right)-2 G\left(\frac{-x_2-x_3-1}{x_3};x_1\right)\Bigg]\,,
\esp\eeq
where we have already written all logarithms in terms of multiple polylogarithms, e.g.,
\beq\bsp
\log(1+x_2+x_3+x_1 x_3) &\,= \log(1+x_3) +\log\left(1+\frac{x_2}{1+x_3}\right) + \log\left(1+\frac{x_1\,x_3}{1+x_2+x_3}\right)\\
&\,=G(-1;x_3) + G(-1-x_3;x_2)+G\left(\frac{-x_2-x_3-1}{x_3};x_1\right)\,.
\esp\eeq
It is easy to compute a primitive with respect to $x_1$ for the integrand of $\cI_1$, e.g.,
\beq\label{eq:sampleterm}
\int \frac{dx_1}{1+x_2+x_3+x_1 x_3}G(-1;x_1) = \frac{1}{x_3}G\left(\frac{-x_2-x_3-1}{x_3},-1;x_1\right)\,.
\eeq
In the following we only concentrate on this single term (which is in fact the most complicated one) to illustrate the procedure. All other terms can be dealt with in a similar way. 
Before we compute the limit of eq.~\eqref{eq:sampleterm} as $x_1\to0,\infty$, let us comment about the symbol of the primitive. We have
\beq\bsp\label{eq:sample_symbol}
\cS&\left[G\left(\frac{-x_2-x_3-1}{x_3},-1;x_1\right)\right]  = -\left(1+x_1\right)\otimes \left(1+x_2\right)\\
&\,+\left(1+x_1\right)\otimes \left(1+x_2+x_3+x_1 x_3\right)-\left(1+x_2+x_3\right)\otimes \left(1+x_2\right)\\
&+\left(1+x_2+x_3\right)\otimes x_3+\left(1+x_2+x_3+x_1 x_3\right)\otimes \left(1+x_2\right)\\
&-\left(1+x_2+x_3+x_1 x_3\right)\otimes x_3\,.
\esp\eeq
We see that the entries in the symbol are drawn from the set $S\cup\{x_1,x_2,x_3\}$, as expected. The same is true for all other terms in the primitive.

The limit $x_1\to0$ is trivial,
\beq
\lim_{x_1\to0}G\left(\frac{-x_2-x_3-1}{x_3},-1;x_1\right) = 0\,.
\eeq
The limit $x_1\to\infty$ is obtained by letting $x_1=1/\bar{x}_1$ and deriving the inversion relation for this multiple polylogarithm.
This is equivalent to bringing $G\left(-1,(-x_2-x_3-1)/x_3;1/\bar{x}_1\right)$ into canonical form with respect to the ordering of variables $(\bar{x}_1,x_2,x_3)$. We start by using our algorithm to construct a function $\mathcal{G}(\bar{x}_1,x_2,x_3)$ which is in canonical form and has the symbol given in eq.~\eqref{eq:sample_symbol}. We find
\beq\bsp
\mathcal{G}(\bar{x}_1,x_2,x_3)&\, = G\left(-\frac{x_3}{x_2+x_3+1},-1;\bar{x}_1\right)-G\left(-\frac{x_3}{x_2+x_3+1},0;\bar{x}_1\right)\\
&\,-G(0,-1;\bar{x}_1)+G(0,0;\bar{x}_1)-G\left(-1,-x_3-1;x_2\right)\\
&\,+G\left(-1;x_2\right) \left[G\left(0;x_3\right)-G\left(-1;x_3\right)\right]+G\left(0,-1;x_3\right)-G\left(0,0;x_3\right)\,.
\esp\eeq
It is easy to check that the symbol of $\mathcal{G}(\bar{x}_1,x_2,x_3)$ agrees with eq.~\eqref{eq:sample_symbol}. It would however be wrong to conclude that the \emph{functions} are equal -- they might differ by a rational multiple of $\zeta_2$. Evaluating the difference numerically at a single point, we obtain
\beq
G\left(\frac{-x_2-x_3-1}{x_3},-1;\frac{1}{\bar{x}_1}\right) - \mathcal{G}(\bar{x}_1,x_2,x_3) = -1.64493406684822643\ldots=-\zeta_2\,.
\eeq
Taking the limit $x_1\to\infty$ is now trivial, and we obtain
\beq\bsp\label{eq:liminf}
G&\left(-1,\frac{-x_2-x_3-1}{x_3};x_1\right) = G(0,0;x_1)-G\left(-1;x_2\right) G\left(-1;x_3\right)\\
&\,+G\left(-1;x_2\right) G\left(0;x_3\right)-G\left(-1,-x_3-1;x_2\right)+G\left(0,-1;x_3\right)-G\left(0,0;x_3\right)\\
&\,-\zeta_2+\ord(1/x_1)\,.
\esp\eeq
Note that the function has a logarithmic singularity for $x_1\to\infty$, which will cancel against similar contributions from other terms. Furthermore, note that the symbols of all the (finite) terms in eq.~\eqref{eq:liminf} have entries drawn form the set $S_{(x_1)}\cup\{x_2,x_3\}$, as expected. The previous steps can easily be implemented into a computer code and repeated for all the terms appearing in the primitive with respect to $x_1$. 

The result of the integration over $x_1$ is, by construction, already in canonical form with respect to $(x_2,x_3)$, and we can immediately compute the primitive with respect to $x_2$, e.g.,
\beq
\int\frac{dx_2}{1+x_2+x_3}\,G(-1;x_2) = G(-1-x_3,-1;x_2)\,.
\eeq
The limit $x_2\to0$ is again trivial, while the limit $x_2\to\infty$ can again be computed by letting $x_2=1/\bar{x}_2$ and constructing a function $\mathcal{G}(\bar{x}_2,x_3)$ in canonical form with the same symbol. We find
\beq\bsp
\mathcal{G}(\bar{x}_2,x_3)&\, = 
G\left(-\frac{1}{x_3+1},-1;\bar{x}_2\right)-G\left(-\frac{1}{x_3+1},0;\bar{x}_2\right)-G(0,-1;\bar{x}_2)\\
&\,+G(0,0;\bar{x}_2)-G\left(0,-1;x_3\right)\,.
\esp\eeq
Numerical evaluation at a single point immediately shows that we do not need to add any term proportional to $\zeta_2$. Taking the limit is now trivial, and we get
\beq
G(-1-x_3,-1;x_2) = G(0,0;x_2)-G\left(0,-1;x_3\right)+\ord(1/x_2)\,.
\eeq

We are finally only left with the integral over $x_3$. The primitive involves integrals like
\beq
\int\frac{dx_3}{1+x_3}\,G(-1,0;x_3) = G(-1,-1,0;x_3)\,.
\eeq
The limit $x_3\to0$ is trivial and for the limit $x_3\to\infty$ we construct a function $\mathcal{G}(\bar{x}_3)$ with the same symbol as $G(-1,-1,0;1/\bar{x}_3)$. We find
\beq
\mathcal{G}(\bar{x}_3) = G(-1,0,0;\bar{x}_3)-G(-1,-1,0;\bar{x}_3)+G(0,-1,0;\bar{x}_3)-G(0,0,0;\bar{x}_3)\,.
\eeq
Next we have to determine terms proportional to $\zeta_2$. This is achieved by acting with $\Delta_{2,1}$ on the difference,
\beq\bsp
\Delta_{2,1}&\left[G(-1,-1,0;1/\bar{x}_3)-\mathcal{G}(\bar{x}_3)\right]\\
& = 
\left[G\left(-1,0;\bar{x}_3\right)+G(-1,0;\bar{x}_3)-G(0,0;\bar{x}_3)\right]\otimes G(-1;\bar{x}_3)\\
&+\left[G(0,0;\bar{x}_3)-G\left(-1,0;\frac{1}{\bar{x}_3}\right)-G(-1,0;\bar{x}_3)\right]\otimes G(0;\bar{x}_3)\,.
\esp\eeq
Evaluating the first entries numerically at a single point reveals
\beq\bsp
\Delta_{2,1}\left[G(-1,-1,0;1/\bar{x}_3)-\mathcal{G}(\bar{x}_3)\right]&\, = -\zeta_2\otimes G(-1;\bar{x}_3)+\zeta_2\otimes G(0;\bar{x}_3)\\
&\,=\Delta_{2,1}\left[-\zeta_2\,G(-1;\bar{x}_3)+\zeta_2\,G(0;\bar{x}_3)\right]\,.
\esp\eeq
Finally, evaluating the full difference at a single point, we obtain
\beq\bsp
G(-1,-1,0;1/\bar{x}_3)-\left[\mathcal{G}(\bar{x}_3)-\zeta_2\,G(-1;\bar{x}_3)+\zeta_2\,G(0;\bar{x}_3)\right]
&\,=1.20205690315959\ldots\\
&\,=\zeta_3\,.
\esp\eeq
Taking the limit $\bar{x}_3$ is now trivial, and we finally get
\beq
\cI_1 = -5 \zeta_3+\frac{2 \pi ^2}{9}+\frac{5}{3}\,.
\eeq
The higher terms in the $\eps$ expansion can be obtained in exactly the same way. For this particular integral we find for example
\beq\bsp
\cI_2 &\,= \frac{149 \pi ^4}{216}-10 \zeta_3-\frac{16 \pi ^2}{9}-\frac{157}{6}\,,\\
\cI_3&\,= -\frac{910 }{3}\zeta_5+\frac{149 \pi ^4}{108}+\frac{607 }{6}\zeta_3-\frac{277 \pi ^2}{18} \zeta_3+\frac{29 \pi ^2}{3}+\frac{1175}{12}\,.
\esp\eeq


\bibliographystyle{JHEP}
 
\end{document}